\documentclass[aps,twocolumn,floatfix]{revtex4}
\usepackage{amsmath}
\usepackage{latexsym}
\usepackage{float}
\usepackage{amssymb}
\usepackage{graphicx}
\usepackage{epsfig}
\usepackage{psfrag}

\begin{document}




\title{One- and two-component bottle-brush polymers: simulations \\compared to theoretical predictions}
\author{Hsiao-Ping Hsu, Wolfgang Paul, and Kurt Binder}
\affiliation{
Institut f\"ur Physik, Johannes Gutenberg-Universit\"at, D-$55099$
Mainz, Germany}


\maketitle

\underline{Summary}: Scaling predictions and results from
self-consistent field calculations for bottle-brush polymers with
a rigid backbone and flexible side chains under good solvent
conditions are summarized and their validity and applicability is
assessed by a comparison with Monte Carlo simulations of a simple
lattice model. It is shown that under typical conditions, as they
are also present in experiments, only a rather weak stretching of
the side chains is realized, and then the scaling predictions based
on the extension of the Daoud-Cotton blob picture are not
applicable.

Also two-component bottle brush polymers are considered, where two
types (A,B) of side chains are grafted, assuming that monomers of
different kind repel each other. In this case, variable solvent
quality is allowed for, such that for poor solvent conditions
rather dense cylinder-like structures result. Theories predict
``Janus Cylinder''-type phase separation along the backbone in
this case. The Monte Carlo simulations, using the
pruned-enriched Rosenbluth method (PERM) then are restricted to
rather short side chain length. Nevertheless, evidence is
obtained that the phase separation between an A-rich part of the
cylindrical molecule and a B-rich part can only occur locally. The
correlation length of this microphase separation can be controlled
by the solvent quality. This lack of a phase transition is
interpreted by an analogy with models for ferromagnets in one
space dimension.

\vskip 1.0truecm
\noindent
{\large \bf  I. Introduction}
\vskip 0.5truecm

Flexible macromolecules can be grafted to various substrates by
special endgroups. Such ``polymer brushes'' find widespread
applications$^\textrm{\cite{1,2,3,4,5}}$ and also pose challenging
theoretical problems, such as an understanding of the
conformational statistics and resulting geometrical structure of
these tethered chain molecules. Only this latter aspect shall be
considered in the present paper, for chains grafted to a straight
line or a very narrow cylinder. This problem is a limiting case of
``bottle brush'' polymers where side chains are grafted to a long
macromolecule that forms the backbone of the bottle brush. When
this backbone chain is also a flexible polymer and the grafting
density is not very high, a ``comb polymer''$^{\textrm {\cite{6}}}$ results,
which is outside of consideration here. Also we shall not discuss
the case where the backbone chain is very short, so the
conformation would resemble a ``star polymer''.$^{\textrm{\cite{7,8,9,10,11}}}$
Here we restrict attention to either stiff backbone chains or high
grafting density of side chains at flexible backbones.
In the lattice case
stiffening of the backbone occurs due to excluded
volume interactions, and a cylindrical shape of the molecule as a
whole results. In fact, many experiments have been carried out
where with an appropriate chemical synthesis bottle brush polymers
with a worm-like cylindrical shape were produced
$^{\textrm {\cite{12,13,14,15,16}}}$. The recent 
papers$^{\textrm {\cite{14,15,16}}}$ contain a
more detailed bibliography on this rapidly expanding field.

On the theoretical side, two aspects of the conformation of bottle
brush polymers where mostly discussed: (i) conformation of a side
chain when the backbone can be treated as a rigid straight line or
thin cylinder$^{\textrm {\cite{10,17,18,19,20,21,22,23,24,25,26,27}}}$ (ii)
conformation of the whole bottle brush when the backbone is
(semi)flexible.$^{\textrm {\cite{28,29,30,31,32,33,34,35,36,37,38,39,40}}}$ The
latter problem is left out of consideration in the present paper.
Problem (i), the stretching of the side chains in the
radial direction in the case of sufficiently high grafting
density, was mostly discussed in terms of a scaling description,
$^{\textrm {\cite{10,17,18,19,20,24,25,26}}}$ extending the Daoud-Cotton
$^{\textrm {\cite{8}}}$ ``blob picture''$^{\textrm {\cite{41,42,43}}}$ from star polymers to
bottle brush polymers. If one uses the Flory exponent$^{\textrm {\cite{44,45}}}$
$\nu = 3/5$ in the scaling relation for the average root mean
square end-to-end distance of a side chain, $R_e\propto \sigma
^{(1-\nu)/(1+\nu)} N ^{2\nu/(1+\nu)}$, where $\sigma$ is the
grafting density and N is the number of effective monomeric units of
a side chain, one obtains $R_e \propto \sigma ^{1/4} N ^{3/4}$.
These  exponents happen to be identical to those which one
would obtain assuming that the chains attain quasi-two-dimensional
configurations, resulting if each chain is confined to a disk of
width $\sigma ^{-1}$.$^{\textrm {\cite{27}}}$ Although this latter picture is a
misconception, in experimental studies (e.g.~\cite{14,15}) this
hypothesis of quasi-two-dimensional chains is discussed as a
serious possibility. Therefore we find it clearly necessary to
first review the correct scaling theory based on the blob picture,
and discuss in detail what quantities need to be recorded in order
to distinguish between these concepts. Thus, in the next section
we shall give a detailed discussion of the scaling concepts for
bottle brush polymers with rigid backbones.

Thereafter we shall describe the Monte Carlo test of these
predictions, that we have recently performed using the pruned
enriched Rosenbluth method (PERM).$^{\textrm {\cite{46,47,48,49}}}$ After a
brief description of the Monte Carlo methodology, we present our
numerical results and compare them to the pertinent theoretical
predictions.

In the second part of this paper, we discuss the extension from
one-component to two-component bottle brush polymers. Just as in a
binary polymer blend (A,B) typically the energetically unfavorable
interaction (described by the Flory-Huggins parameter $\chi$
$^{\textrm {\cite{44,45,50,51,52}}}$) should cause phase separation between
A-rich and B-rich domains. However, just as in block copolymers
where A chains and B-chains are tethered together in a 
point,$^{\textrm {\cite{53,54,55}}}$ no macroscopic phase separation but only
``microphase separation'' is possible: for a binary (A,B) bottle
brush with a rigid backbone one may expect formation of ``Janus
Cylinder'' structures.$^{\textrm {\cite{56,57,58}}}$ This means, phase
separation occurs such that the A-chains assemble in one half of
the cylinder, the B chains in the other half, separated from the
A-chains via a flat interface containing the cylinder axis.
However, it has been argued that the long range order implied by
such a ``Janus cylinder'' type structure has a one-dimensional
character, and therefore true long range order is destroyed by
fluctuations at nonzero temperature.$^{\textrm {\cite{58}}}$ Only local phase
separation over a finite correlation length along the cylinder
axis may persist.$^{\textrm {\cite{58}}}$

We shall again first review the theoretical background on
this problem, and then describe the simulation evidence. We
conclude our paper by a summary and outlook on questions that are
still open, briefly discussing also possible consequences on
experimental work. However, we shall not deal with the related
problems of microphase separation of a bottle brush with only one
kind of side chains induced by deterioration of the solvent
quality$^{\textrm {\cite{59,60}}}$ or by adsorption on flat 
substrates.$^{\textrm {\cite{61,62,63}}}$

\begin{figure*}
\begin{center}
$\begin{array}{c@{\hspace{0.2in}}c}
{\psfig{file=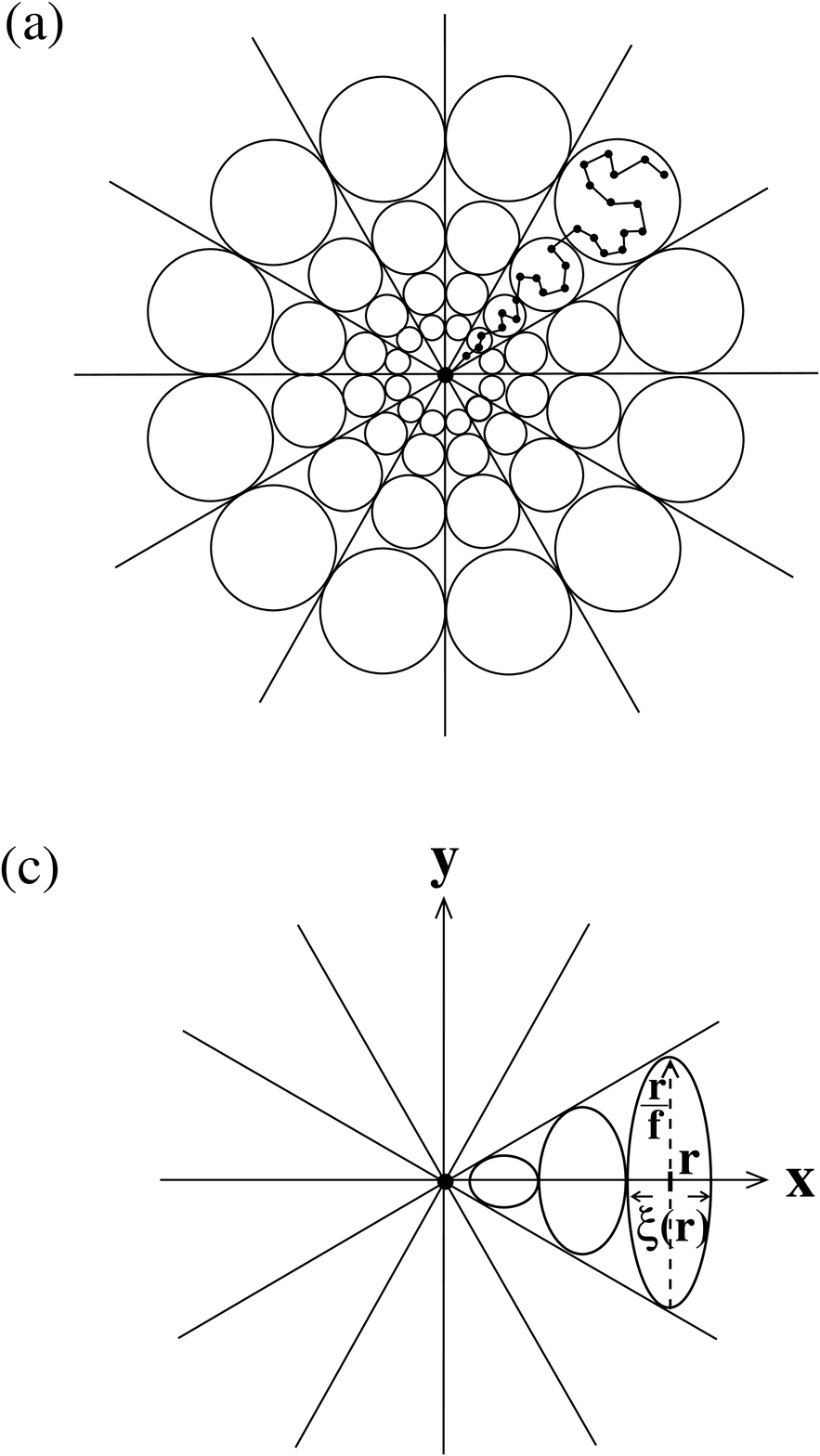, width=7.5cm, angle=0}}
&\psfig{file=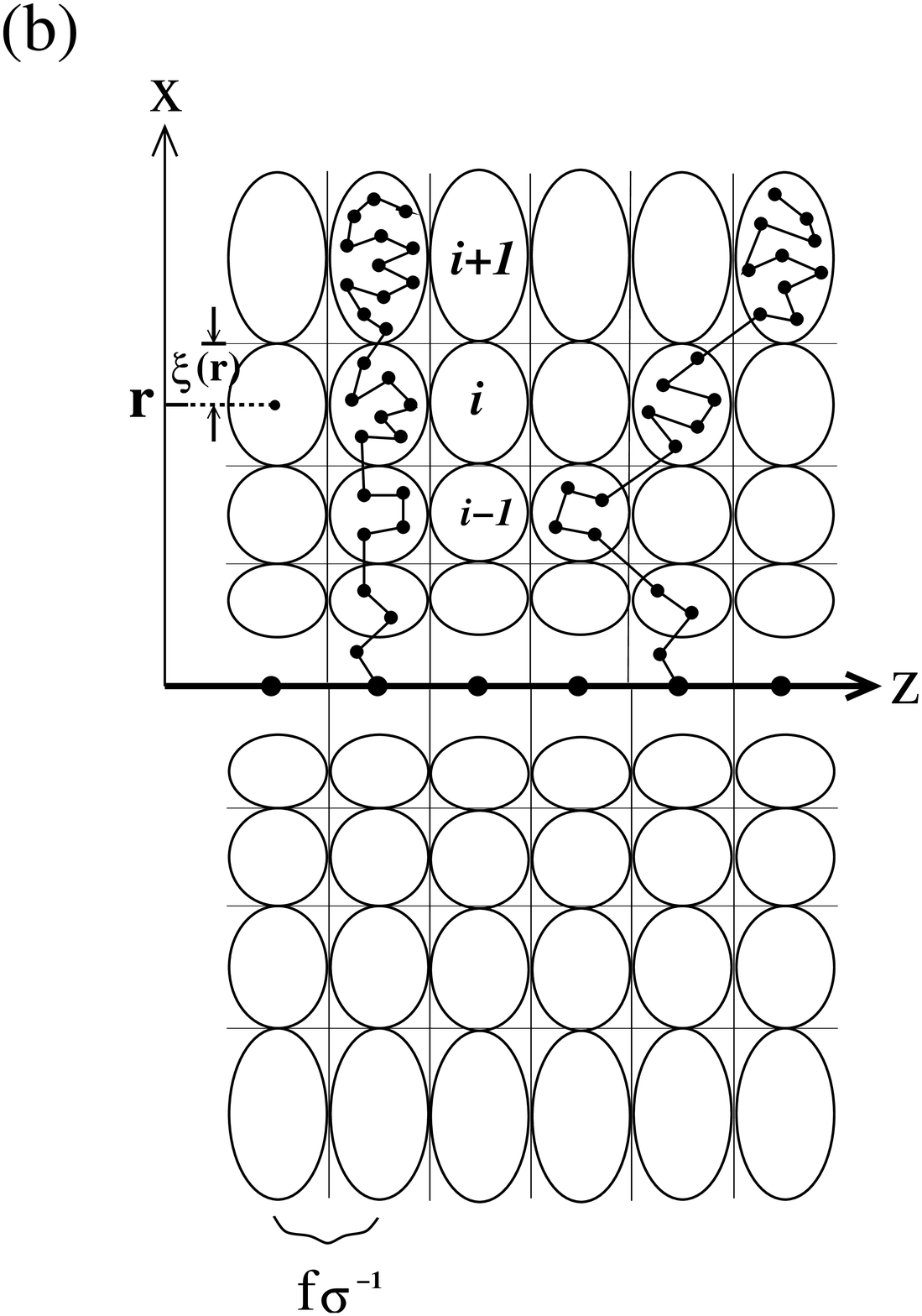,width=7.5cm,angle=0} \\
[0.6cm]
\end{array}$
\caption{Schematic construction of a blob picture for star polymer
(a), and for a bottle brush (b), (c). We assume that along the
rigid backbone there occur at a regular spacing $f\sigma ^{-1}$
grafting sites where at each grafting site $f$ side chains
containing $N$ effective monomeric units are grafted. If the rigid
backbone of the cylindrical bottle brush is oriented along the
z-axis, part (c) shows a view of the blob partitioning in the
xy-plane perpendicular to the backbone, while part (b) shows a cut
through the cylinder along the xz-plane containing the cylinder
axis. A few possible chain configurations are indicated. While for
a star polymer the blobs have a spherical shape, their radius
$\xi(r)$ increasing linearly proportional to the distance $r$ from
the center of the star (a), for the bottle brush the blobs are
ellipsoids with three axes $\xi (r)$ (in x-direction),
proportional to $r/f$ (in y-direction), and $f\sigma ^{-1}$ (in
z-direction), respectively. For a considered chain the x-axis
contains the center of mass of the chain.}
\label{fig1}
\end{center}
\end{figure*}

\vskip 1.0truecm
\noindent
{\large \bf II. Conformation of Side Chains of Bottle Brushes under Good
Solvent Conditions: Theoretical Background}
\vskip 0.5truecm

The most straightforward approach to understand the conformations
of chains in polymer brushes and star polymers under good solvent
conditions uses the concept to partition the space available for
the chains into compartments, called ``blobs''. The idea is that
in each such region there occur only monomers of one chain,
no monomers of any other chains occur in such a blob, and hence
self-avoiding walk statistics holds in each blob. This means, if a
(spherical) blob has a radius $r_B$ and contains $n$ monomers,
these numbers must be related via
\begin{equation}\label{eq1}
r_B =a n^\nu \;, \quad \nu \approx 0.588\;, \quad n \gg 1\; .
\end{equation}
Here $a$ is a length of the order of the size of an effective
monomer, and we emphasize from the start that it is crucial to use
the correct value of the self-avoiding walk exponent $\nu$, as it
is provided from renormalization group calculations$^{\textrm {\cite{64}}}$ or
accurate Monte Carlo simulations.$^{\textrm {\cite{65}}}$ If one would ignore
the small difference between $\nu$ and the Flory estimate 3/5, one
would already miss an important distinction between two different
scaling regimes for a brush on a flat substrate.$^{\textrm {\cite{66}}}$
An almost trivial condition of this ``blobology''$^{\textrm {\cite{43}}}$ is
that each of the $N$ effective monomeric units of a chain must
belong to some blob. So we have
\begin{equation}\label{eq2}
N = nn_B\;,
\end{equation}
where $n_B$ is the number of blobs belonging to one particular
chain.

Finally we note that the space available to the chains must be
densely filled with blobs. It then remains to discuss which
factors control the blob size $r_B$.$^{\textrm {\cite{43}}}$ The simplest case
is a polymer brush on a flat substrate (under good solvent
conditions, as assumed here throughout): if we neglect, for
simplicity, any local fluctuations in the grafting density, the
distance between grafting sites simply is given by $\sigma
^{-1/2}$. Putting $r_B= \sigma^{-1/2}$ in Equation~(\ref{eq1}), we find
$n=(\sigma a ^2)^{-1/2\nu}$, i.e.~each chain is a string of
$n_B=N/n=N(\sigma a^2)^{1/2 \nu}$ blobs. According to the
simple-minded description of polymer brushes due to 
Alexander,$^{\textrm {\cite{41}}}$ this string simply is arranged like a one-dimensional
cigar, and one would conclude that the height of a flat brush is
\begin{equation}\label{eq3}
h= \sigma ^{-1/2}n_B=Na(\sigma a ^2)^{1/(2 \nu)- 1/2}\;.
\end{equation}
The free end of the chain is in the last blob and hence the
end-to-end distance $R_e \approx h$ in this ``Alexander picture''
of polymer brushes.$^{\textrm {\cite{1,41,42,43}}}$ However, a more detailed
theory of polymer brushes, such as the self-consistent field
theory in the strong stretching limit,$^{\textrm {\cite{67,68,69}}}$ yields a
somewhat different behavior: the end monomer is not localized at
the outer edge of the brush, but rather can be located anywhere in
the brush, according to a broad distribution; also the monomer
density in polymer brushes at flat substrates is not constant up
to the brush height $h$, but rather decreases according to a
parabolic profile. So even for a polymer brush at a flat substrate
already a description in terms of non-uniform blob sizes, that
increase with increasing distance $z$ from the substrate, is
required.$^{\textrm {\cite{70}}}$ However, in the following we shall disregard
all these caveats about the Alexander picture for flat brushes,
and consider its generalization to the bottle brush geometry,
where polymer chains are tethered to a line rather than a flat
surface. Then we have to partition space into blobs of nonuniform
size and shape in order to respect the cylindrical geometry
(Figure~\ref{fig1}).

In the discussion of brushes in cylindrical geometry in terms of
blobs in the literature$^{\textrm {\cite{19,20,26}}}$ the non-spherical
character of the blob shape is not explicitly accounted for, and
it rather is argued that one can characterize the blobs by one
effective radius $\xi(r)$ depending on the radial distance $r$
from the cylinder axis. One considers a segment of the array of
length $L$ containing $p$ polymer chains.$^{\textrm {\cite{19}}}$ On a surface
of a cylinder of radius $r$ and length $L$ there should then be
$p$ blobs, each of cross-sectional area $\xi^2(r)$; geometrical
factors of order unity are ignored throughout. Since the surface
area of the cylindrical segment is $Lr$, we must have$^{\textrm {\cite{19,20,26}}}$
\begin{equation}\label{eq4}
p \xi ^2(r) = Lr\;, \quad \xi(r)=(Lr/p)^{1/2}=(r/\sigma)^{1/2}\;.
\end{equation}
If the actual non-spherical shape of the blobs (Figure~\ref{fig1})
is neglected, the blob volume clearly is of the order of $\xi
^3(r)=(r/\sigma)^{3/2}$. Invoking again the principle that inside
a blob self-avoiding walk statistics hold, we have, in analogy
with Equation~(\ref{eq1}),
\begin{equation}\label{eq5}
\xi (r) = a [n(r)]^\nu\;,\enspace n(r)= [\xi(r)/a]^{1/\nu}=[r/(\sigma
a^2)]^{1/2 \nu}\;.
\end{equation}
From this result one immediately derives the power law decay for
the density profile $\rho(r)$ as follows$^{\textrm {\cite{3,19,20,26}}}$
\begin{eqnarray}\label{eq6}
\rho (r) &=& n(r)/\xi ^3(r)= a^{-3}[r/(\sigma
a^2)]^{-(3\nu-1)/2\nu} \nonumber \\ &\approx& a ^{-3}[r/(\sigma
a^2)]^{-0.65}
\end{eqnarray}
Using the Flory approximation$^{\textrm {\cite{44,45}}}$ $\nu = 3/5$ one would
find $\rho(r)\propto r^{-2/3}$ instead.

Now the average height $h$ of the bottle brush is estimated by
requiring that we obtain all $\sigma N$ monomers per unit length
in the z-direction along the axis of the bottle brush (cf.
Figure~\ref{fig1}) when we integrate $\rho(r)$ from $r=0$ to $r=h$,
\begin{equation}\label{eq7}
\sigma N = \int \limits _0^h \rho (r) r dr = a ^{-3}(\sigma a^2)
^{(3\nu - 1)/2\nu} \int \limits _0^h r ^{(1-\nu)/2\nu} dr\;.
\end{equation}
This yields, again ignoring factors of order unity,
\begin{subequations}
\begin{equation}\label{eq8a}
N = (\sigma a)^{(\nu-1)/(2\nu)} (h/a)^{(\nu +1)/(2\nu)},
\end{equation}
\begin{equation}\label{eq8b}
h/a = (\sigma a)^{(1-\nu)/(1+\nu)}N^{2\nu/(1+\nu)} = (\sigma
a)^{0.259}N^{0.74}\; .
\end{equation}
\end{subequations}
Note that the use of the Flory estimate $\nu = 3/5$ would simply
yield $h \propto \sigma ^{1/4}N^{3/4}$, which happens to be
identical to the relation that one obtains when one partitions the
cylinder of height $L$ and radius $h$ into disks of height $\sigma
^{-1}$, requiring hence that each chain is confined strictly into
one such disk. Then each chain would form a quasi-two-dimensional
self-avoiding walk formed from $n'_{\textrm{blob}}=N/n'$ blobs of
diameter $\sigma ^{-1}$. Since we have $\sigma ^{-1}= an'^\nu$
again, we would conclude that the end-to-end distance $R$ of such
a quasi-two-dimensional chain is
\begin{eqnarray}\label{eq9}
R &=& \sigma ^{-1}(n'_{\textrm{blob}})^{3/4} = \sigma
^{-1}n'^{-3/4}N^{3/4} \nonumber \\
&=& a (\sigma a)^{-1+3/4\nu}N^{3/4} 
= a (\sigma a )^{1/4} N^{3/4}\;.
\end{eqnarray}
Putting then $R=h$, Equation~(\ref{eq8b}) results when we use there
$\nu = 3/5$. However, this similarity between Equations~(\ref{eq8b}),
(\ref{eq9}) is a coincidence: in fact, the assumption of a
quasi-two-dimensional conformation does not imply a stretching of
the chain in radial direction. In fact, when we put the x-axis of
our coordinate system in the direction of the end-to-end vector
$\vec{R}$ of the chain, we would predict that the y-component of
the gyration radius scales as
\begin{equation}\label{eq10}
R_{gy}=a(\sigma a)^{1/4} N^{3/4}\;,
\end{equation}
since for quasi-two-dimensional chains we have $R_{gx}\propto
R_{gy} \propto R$, all these linear dimensions scale with the same
power laws. On the other hand, if the Daoud-Cotton-like$^{\textrm {\cite{8}}}$
picture \{Figure~\ref{fig1}a\} holds, in a strict sense, one would
conclude that $R_{gy}$ is of the same order as the size of the
last blob for $r=h$,
\begin{eqnarray}\label{eq11}
R_{gy}= \xi(r=h) &=& (h/\sigma)^{1/2} \nonumber \\
&=& a (\sigma a) ^{-\nu /(1+\nu)} N^{\nu/(1+\nu)}
\end{eqnarray}
Clearly this prediction is very different from Equation~(\ref{eq10}),
even with the Flory exponent $\nu = 3/5$ we find from
Equation~(\ref{eq11}) that $R_{gy} \propto \sigma ^{-3/8}N^{3/8}$.
Hence it is clear that an analysis of all three components of the
gyration radius of a polymer chain in a bottle brush is very
suitable to distinguish between the different versions of scaling
concepts discussed in the literature.

However, at this point we return to the geometrical construction
of the filling of a cylinder with blobs, Figure~\ref{fig1}. We
explore the consequences of the obvious fact that the blobs cannot
be simple spheres, when we require that each blob contains
monomers from a single chain only, and the available volume is
densely packed with blobs touching each other.

As mentioned above, $\xi(r)$ was defined from the available
surface area of a cylinder at radius $r$, cf.~Equation~(\ref{eq4}). It
was argued that for each chain in the surface area $Lr$ of such a
cylinder a surface area $\xi^2(r)$ is available. However,
consideration of Figure~\ref{fig1} shows that these surfaces are not
circles but rather ellipses, with axes $f\sigma ^{-1}$ and $r/f$.
The blobs hence are not spheres but rather ellipsoids, with three
different axes: $f\sigma ^{-1}$ in z-direction along the cylinder
axes, $r/f$ in the y-direction tangential on the cylinder surface,
normal to both the z-direction and the radial direction, and the
geometric mean of these two lengths $(r\sigma ^{-1})^{1/2}$, in
the radial x-direction. Since the physical meaning of a blob is
that of a volume region in which the excluded volume interaction
is not screened, this result implies that the screening of
excluded volume in a bottle brush happens in a very anisotropic
way: there are three different screening lengths, $f \sigma ^{-1}$
in the axial z-direction, $(r \sigma ^{-1})^{1/2}$ in the radial
x-direction, and $r/f$ in the third, tangential, y-direction. Of
course, it remains to be shown by a more microscopic theory that
such an anisotropic screening actually occurs.

However, the volume of the ellipsoid with these three axes still
is given by the formula
\begin{equation}\label{eq12}
V_{\textrm{ellipsoid}} = (\sigma ^{-1}f)(r/f)(r\sigma
^{-1})^{1/2}=(r\sigma ^{-1})^{3/2}= [\xi (r)]^3
\end{equation}
with $\xi(r)$ given by Equation~(\ref{eq4}), and hence the volume of
the blob has been correctly estimated by the spherical
approximation. As a consequence, the estimations of the density
profile $\rho(r)$, Equation~(\ref{eq6}), and resulting brush height
$h$, Equation~(\ref{eq8b}), remain unchanged.

More care is required when we now estimate the linear dimensions
of the chain in the bottle brush. We now orient the x-axis such
that the xz-plane contains the center of mass of the considered
chain. As Figure~\ref{fig1}b indicates, we can estimate $R_{gz}$
assuming a random walk picture in terms of blobs. When we go along
the chain from the grafting site towards the outer boundary of the
bottle brush, the blobs can make excursions with $\Delta z=\pm f
\sigma ^{-1}$, independent of $r$. Hence we conclude, assuming
that the excursions of the $n_{\textrm{blob}}$ steps add up
randomly
\begin{eqnarray}\label{eq13}
R_{gz}^2 &=& \sum \limits _{i=1}^{n_{\textrm{blob}}} (f \sigma
^{-1})^2= n_{\textrm{blob}} (f \sigma^{-1})^2\;, \nonumber \\
R_{gz}&=& f
\sigma^{-1} \sqrt{n_{\textrm{blob}}}\;.
\end{eqnarray}
Hence we must estimate the number of blobs $n_{\textrm{blob}}$ per
chain in a bottle brush. We must have
\begin{equation}\label{eq14}
n_{\textrm{blob}} = \sum \limits _{i=1}^{n_{\textrm{blob}}} 1 =
\int \limits _0^h [\xi(r)]^{-1}dr
\end{equation}
Note from Figure~\ref{fig1} that we add an increment $2\xi(r_i)$ to
$r$ when we go from the shell $i$ to shell $i+1$ in the cylindrical
bottle brush. So the discretization of the integral in
Equation~(\ref{eq14}) is equivalent to the sum. From Equation~(\ref{eq14}) we
then find, again ignoring factors of order unity
\begin{equation}\label{eq15}
n_{\textrm{blob}}= \sigma^{1/2} h^{1/2} = (\sigma a)^{1/(1+\nu)}
N^{\nu/(1+\nu)}\;,
\end{equation}
and consequently
\begin{equation}\label{eq16}
R_{gz}=fa(\sigma a)^{- \frac {2 \nu +1}{2\nu +2}} N^{\nu
/[2(1+\nu)]} \propto \sigma ^{-0.685}N^{0.185}
\end{equation}
With Flory exponents we hence conclude $n_{\textrm{blob}} \propto
\sigma ^{5/8}N^{3/8}$, and thus $R_{gz} \propto f \sigma
^{-11/16}N^{3/16}$.
The estimation of $R_{gy}$ is most delicate, of course, because
when we consider random excursions away from the radial
directions, the excursions $\Delta y= \pm \xi (r)$ are
non-uniform. So we have instead of Equation~(\ref{eq13}),
\begin{equation}\label{eq17}
R^2_{gy}= \sum _{i=1}^{n_{\textrm{blob}}} \xi ^2 (r_i) = \int
\limits _0^h\xi(r)dr ,
\end{equation}
in analogy with Equation~(\ref{eq14}). This yields $R^2_{gy}= \sigma
^{-1/2} h^{3/2}$ and hence
\begin{eqnarray}\label{eq18}
R_{gy}=\sigma ^{-1/4} h^{3/4} &=& a (\sigma a)^{-(2\nu-1)/(2\nu+2)}
N^{3\nu/(2\nu +2)}  \nonumber \\
&\propto& \sigma ^{-0.055}N^{0.555}
\end{eqnarray}
while the corresponding result with Flory exponents is $R_{gy}\
\propto \sigma^{-1/16}N^{9/16}$. These different results for
$R_{gx}\propto h$ \{Equation~(\ref{eq8b})\}, $R_{gy}$
\{Equation~(\ref{eq18})\} and $R_{gz}$ \{Equation~(\ref{eq16})\} clearly
reflect the anisotropic structure of a chain in a bottle brush.
Note that the result for $R_{gy}$ according to Equation~(\ref{eq18}) is
much larger than the simple Daoud-Cotton-like prediction,
Equation~(\ref{eq11}), but is clearly smaller than the result of the
quasi-two-dimensional picture, Equation~(\ref{eq10}).

As a consistency check of our treatment, we note that also
Equation~(\ref{eq7}) can be formulated as a discrete sum over blobs
\begin{equation}\label{eq19}
N=\sum \limits _{i=1}^{n_{\textrm{blob}}} n(r)= \int \limits _0^h
[n(r)/\xi(r)]dr= \int \limits _0^h[r/(\sigma a)]^{1/2\nu -
1/2}dr\;,
\end{equation}
which yields Equation~(\ref{eq8a}), as it should be.

Finally we discuss the crossover towards mushroom behavior.
Physically, this must occur when the distance between grafting
points along the axis, $f\sigma^{-1}$, becomes equal to $aN^\nu$.
Thus we can write
\begin{equation}\label{eq20}
h=aN^\nu \tilde{h}(\sigma a N^\nu),
\end{equation}
where we have absorbed the extra factor $f$ in the scaling
function $\tilde{h}(\zeta)$. We note that Equation~(\ref{eq8b}) results
from Equation~(\ref{eq20}) when we request that
\begin{equation}\label{eq21}
\tilde{h}(\zeta \gg1)\propto \zeta ^{(1-\nu)/(1+\nu)},
\end{equation}
and hence a smooth crossover between mushroom behavior and
radially stretched polymer conformations occurs for $\sigma a
N^\nu$ of order unity, as it should be. Analogous crossover
relations can be written for the other linear dimensions, too:
\begin{equation}\label{eq22}
R_{gz}= aN^\nu \tilde{R}_{gz}(\zeta),\enspace \tilde{R}_{gz}(\zeta
\gg 1)\propto \zeta ^{-(2\nu +1)/(2\nu +2)},
\end{equation}
\begin{equation}\label{eq23}
R_{gy}= aN^\nu \tilde{R}_{gy}(\zeta),\enspace \tilde{R}_{gy}(\zeta
\gg 1)\propto \zeta ^{-(2\nu -1)/(2\nu +2)},
\end{equation}
The agreement between Equation~(\ref{eq16}) and Equation~(\ref{eq22}) or
Equation~(\ref{eq18}) and Equation~(\ref{eq23}), respectively, provides a
check on the self-consistency of our scaling arguments.

However, it is important to bear in mind that the blob picture of
polymer brushes, as developed by Alexander,$^{\textrm {\cite{41}}}$ Daoud and
Cotton,$^{\textrm {\cite{8}}}$ Halperin$^{\textrm {\cite{1,43}}}$ and many others, is a severe
simplification of reality, since its basic assumptions that (i)
all chains in a polymer brush are stretched in the same way, and
(ii) the chain ends are all at a distance $h$ from the grafting
sites, are not true. Treatments based on the self-consistent field
theory$^{\textrm {\cite{67,68,69,71,72,73,74,75,76}}}$ and computer 
simulations$^{\textrm {\cite{3,22,66,70,77,78,79,80,81,82,83}}}$ have shown that chain ends
are not confined at the outer boundary of the brush, and the
monomer density distribution is a nontrivial function, that cannot
be described by the blob model. However, it is widely believed
that the blob model yields correctly the power laws of chain
linear dimensions in terms of grafting density and chain length,
so the shortcomings mentioned above affect the pre-factors in
these power laws only. In this spirit, we have extended the blob
model for brushes in cylindrical geometry in the present section,
taking the anisotropy in the shape of the blobs into account to
predict the scaling behavior of both the brush height $h$ (which
corresponds to the chain end-to-end distance $R$ and the
x-component $R_{gx}$ of the gyration radius) and of the transverse
gyration radius components $R_{gy},\; R_{gz}$. To our knowledge,
the latter have not been considered in the previous literature.

\vskip 1.0truecm
\noindent
{\large \bf III. Monte Carlo Methodology}
\vskip 0.5truecm

\begin{figure}
\begin{center}
\psfig{file=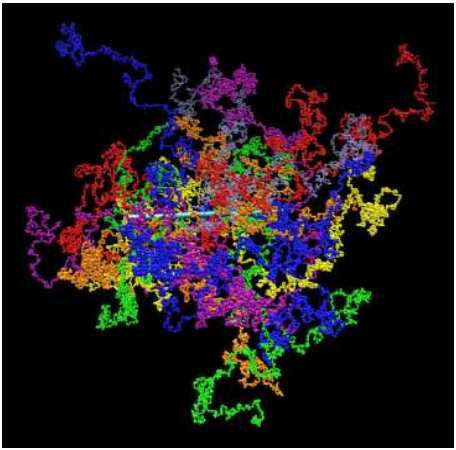,width=6.0cm,angle=0}
\caption{Snapshot
picture of a bottle brush polymer with $L_b= 128\; \sigma = 1/4$,
and $N=2000$ on the simple cubic lattice. Note that different
chains are displayed in different colors to distinguish them, and
the periodic boundary condition is undone for the sake of a better
visibility in the visualization of this configuration.}
\label{fig2}
\end{center}
\end{figure}

We consider here the simplest lattice model of polymers under good
solvent conditions, namely, the self-avoiding walk on the simple
cubic lattice. The backbone of the bottle brush is just taken to
be the z-axis of the lattice, and in order to avoid any effects
due to the ends of the backbone, we choose periodic boundary
conditions in the z-direction. The length of the backbone is taken
to be $L_b$ lattice spacings, and the lattice spacing henceforth
is taken as one unit of length. Note that in order to avoid finite
size effects $L_b$ has to be chosen large enough so that no side
chain can interact with its own ``images'' created by the periodic
boundary condition, i.e. $R_{gz}\ll L_b$.

We create configurations of the bottle brush polymers applying the
pruned-enriched-Rosenbluth method 
(PERM).$^{\textrm {\cite{46,47,48,49,87,88,89,90}}}$ This algorithm is based on the
biased chain growth algorithm of Rosenbluth and Rosenbluth,$^{\textrm {\cite{91}}}$ 
and extends it by re-sampling technique (``population
control''$^{\textrm {\cite{46}}}$) with depth-first implementation.
Similar to a recent study of star
polymers$^{\textrm {\cite{47,48}}}$ all side chains of the bottle brush are
grown simultaneously, adding one monomer to each side chain one
after the other before the growth process of the first chain
continues by the next step.

When a monomer is added to a chain of length $n-1$
(containing $n$ monomers) at the $n$th step,
one scans the neighborhood of the chain end to identify
the free neighboring sites of the chain end, to which a
monomer could be added. Out of these
$n_{\textrm{free}}$ sites available for the addition of a monomer
one site is chosen with the probability $p_{n,i}$ for the
$i$th direction. One has the freedom to sample these
steps from a wide range of possible distributions, e.g.
$p_{n,i}=1/n_{\textrm{free}}$, if one site is
chosen at random, and this additional bias is taken into
account by suitable weight factors.
The total weight $W_n$ of a chain of length $n$ with an
unbiased sampling is determined recursively by
$W_n=\Pi_{k=1}^{n} w_k=W_{n-1}w_n$. While the weight $w_n$ is gained
at the $n$th step with a probability $p_{n,i}$, one has to
use $w_n/p_{n,i}$ instead of $w_n$.
The partition sum of a chain of length $n$ (at the $n$th step) is
approximated as
\begin{equation}\label{eq24}
Z_n \approx \hat{Z}_n \equiv M^{-1}_n \sum \limits _{\alpha
=1}^{M_n}W_n(\alpha),
\end{equation}
where $M_n$ is the total number of configurations $\{\alpha \}$,
and averages of any chain property (e.g. its end-to-end distance,
gyration radius components, etc.) $A(\alpha)$ are obtained as
\begin{equation}\label{eq25}
\bar{A}_n=\frac{1}{M_n}\frac{\sum \limits _{\alpha =1}^{M_n} A(\alpha)W_n
(\alpha)}{Z_n}
\end{equation}
As is well-known from Ref.~\cite{84}, for this original biased 
sampling$^{\textrm {\cite{91}}}$ the statistical errors for large 
$n$ are very hard to control. This problem is alleviated by 
population control.$^{\textrm {\cite{87}}}$ One introduces two thresholds
\begin{equation}\label{eq26}
W_n^+= C_+ \hat{Z}_n,\quad W^-_n=C_- \hat{Z}_n\quad \;,
\end{equation}
where $\hat{Z}_n$ is the current estimate of the partition sum,
and $C_+$ and $C_-$ are constants of order unity.
The optimal ratio between $C_+$ and $C_-$ is found to
be $C_+/C_- \sim 10$ in general.
If $W_n$ exceeds $W_n^+$ for the configuration $\alpha$, one
produces $k$ identical copies of this configuration, replaces
their weight $W_n$ by $W_n/k$, and uses these copies as members of
the population for the next step, where one adds a monomer to go
from chain length $n$ to $n+1$. However, if $W_n < W_n^-$, one
draws a random number $\eta$, uniformly distributed between zero
and one. If $\eta < 1/2$, the configuration is eliminated from the
population when one grows the chains from length $n$ to $n+1$.
This ``pruning'' or ``enriching'' step has given the PERM algorithm
its name. On the other hand, if $\eta \geq 1/2$, one keeps the configuration
and replaces its weight $W_n$ by $2W_n$.
In a depth-first implementation, at each time one deals with only a single
configuration until a chain has been grown either to the end of
reaching the maximum length or to be killed in between, and
handles the copies by recursion. Since only a single configuration
has to be remembered during the run, it requires much less memory.

In our implementation, we used $W_n^+=\infty$ and $W_n^-=0$ for the first
configuration hitting chain length $n$.
For the following configurations,
we used $W_n^+=C\hat{Z}_n(c_n/c_0)$ and $W_n^-=0.15W_n^+$,
here $C=3.0$ is a positive constant, and $c_n$ is the total
number of configurations of length $n$ already created during the run.
The bias of growing
side chains was used by giving higher probabilities in the direction
where there are more free next neighbor sites and in the outward
directions perpendicular to the backbone, where the second part of bias
decreases with the length of side chains and increases with the grafting
density. Totally $10^5 \sim 10^6$ independent configurations were
obtained in most cases we simulated.

Typical simulations used backbone lengths $L_b=32,64,$ and 128,
$f=1$ (one chain per grafting site of the backbone, although
occasionally also $f=2$ and $f=4$ were used), and grafting
densities $\sigma = 1/32$, $1/16$, $1/8$, $1/4$, $1/2$ and $1$. 
The side chain
length $N$ was varied up to $N=2000$. So a typical bottle brush
with $L_b = 128, \sigma = 1/4$ (i.e., $n_c =32$ side chains) and
$N=2000$ contains a total number of monomers
$N_{\textrm{tot}}=L_b+n_cN=64128$. Figure~\ref{fig2} shows a
snapshot configuration of such a bottle brush polymer. Note that
most other simulation algorithms for polymers$^{\textrm {\cite{65,84,85,86}}}$
would fail to produce a large sample of well-equilibrated
configurations of bottle brush polymers of this size: dynamic Monte
Carlo algorithms using local moves involve
a relaxation time (in units of Monte Carlo steps per monomer
[MCS]) of order $N^z$ where $z=2 \nu +1$ if one assumes that the
side chains relax independently of each other and one applies the
Rouse model$^{\textrm {\cite{92}}}$ in the good solvent 
regime.$^{\textrm {\cite{85}}}$ For
$N=2000$ such an estimate would imply that the time between
subsequent statistically independent configurations is of the
order of 10$^7$ MCS, which clearly is impractical. While the pivot
algorithm$^{\textrm {\cite{65}}}$ would provide a significantly faster
relaxation in the mushroom regime, the acceptance rate of the
pivot moves quickly deteriorates when the monomer density
increases. This algorithm could equilibrate the outer region of
the bottle brush rather efficiently, but would fail to equilibrate
the chain configurations near the backbone. The configurational
bias algorithm$^{\textrm {\cite{86,93}}}$ would be an interesting alternative
when the monomer density is high, but it is not expected to work
for very long chains, such as $N=2000$. Also, while the simple
enrichment technique is useful to study both star polymers$^{\textrm {\cite{94}}}$ 
and polymer brushes on flat substrates$^{\textrm {\cite{95}}}$ it also
works only for chain lengths up to about $N=100$. Thus, existing
Monte Carlo simulations of one-component bottle brushes under good
solvent conditions either used the bond fluctuation 
model$^{\textrm {\cite{85,96}}}$ 
on the simple cubic lattice applying local moves with
side chain lengths up to $N=41$$^{\textrm {\cite{30}}}$ 
or $N=64$$^{\textrm {\cite{36}}}$ or
the pivot algorithm$^{\textrm {\cite{65}}}$ with side chain lengths up to 
$N=80$,$^{\textrm {\cite{34}}}$ but considering flexible backbone of 
length $L=800$. An alternative approach was followed by Yethiraj$^{\textrm {\cite{31}}}$ who
studied an off-lattice tangent hard sphere model by a pivot algorithm
in the continuum, for $N \leq 50$. All these studies addressed the
question of the overall conformation of the bottle brush
for a flexible backbone, and did not address in detail the
conformations of the side chains. Only the total mean square
radius of gyration of the side chains was estimated occasionally,
obtaining$^{\textrm {\cite{30,36}}}$ $R^2_g \propto N^{1.2}$ or$^{\textrm {\cite{31}}}$ $R_g^2
\propto N^{1.36} $ and$^{\textrm {\cite{34}}}$ $R_g^2 \propto N^{1.4}$. However,
due to the smallness of the side chain lengths used in these
studies, as quoted above, these results have to be considered as
somewhat preliminary, and also a systematic study of the
dependence on both $N$ and $\sigma$ was not presented. We also
note that the conclusions of the quoted papers are somewhat
contradictory.

An interesting alternative simulation method to the Monte Carlo
study of polymeric systems is Molecular Dynamics,$^{\textrm {\cite{85}}}$ of
course. While in corresponding studies of a bead-spring model of
polymer chains for flat brushes$^{\textrm {\cite{81}}}$ chain lengths $N$ up to
$N=200$ were used, for chains grafted to thin cylinders$^{\textrm {\cite{22}}}$
the three chain lengths $N=50, 100$ and 150 were used. For $N=50$,
also four values of grafting density were studied.$^{\textrm {\cite{22}}}$ Murat
and Grest$^{\textrm {\cite{22}}}$ used these data to test the scaling prediction
for the density profile, Equation~(\ref{eq6}), but found that $\rho(r)$
is better compatible with $\rho(r) \propto r^{-0.5}$ rather than
$\rho(r) \propto r^{-0.65}$. However, for $N=50$ the range where
the power law is supposed to apply is very restricted, and hence
this discrepancy was not considered to be a problem for the theory.$^{\textrm {\cite{22}}}$

Thus, we conclude that only due to the use of the
PERM algorithm has it become possible to study such large bottle
brush polymers as depicted in Figure~\ref{fig2}. Nevertheless, as we
shall see in the next section, even for such large side chain
lengths one cannot yet reach the asymptotic region where the power
laws derived in the previous section are strictly valid.

\begin{figure}
\begin{center}
$\begin{array}{c@{\hspace{0.2in}}c}
\multicolumn{2}{l}{\mbox{(a)}} \\ [-1.5cm] \\
&\psfig{file=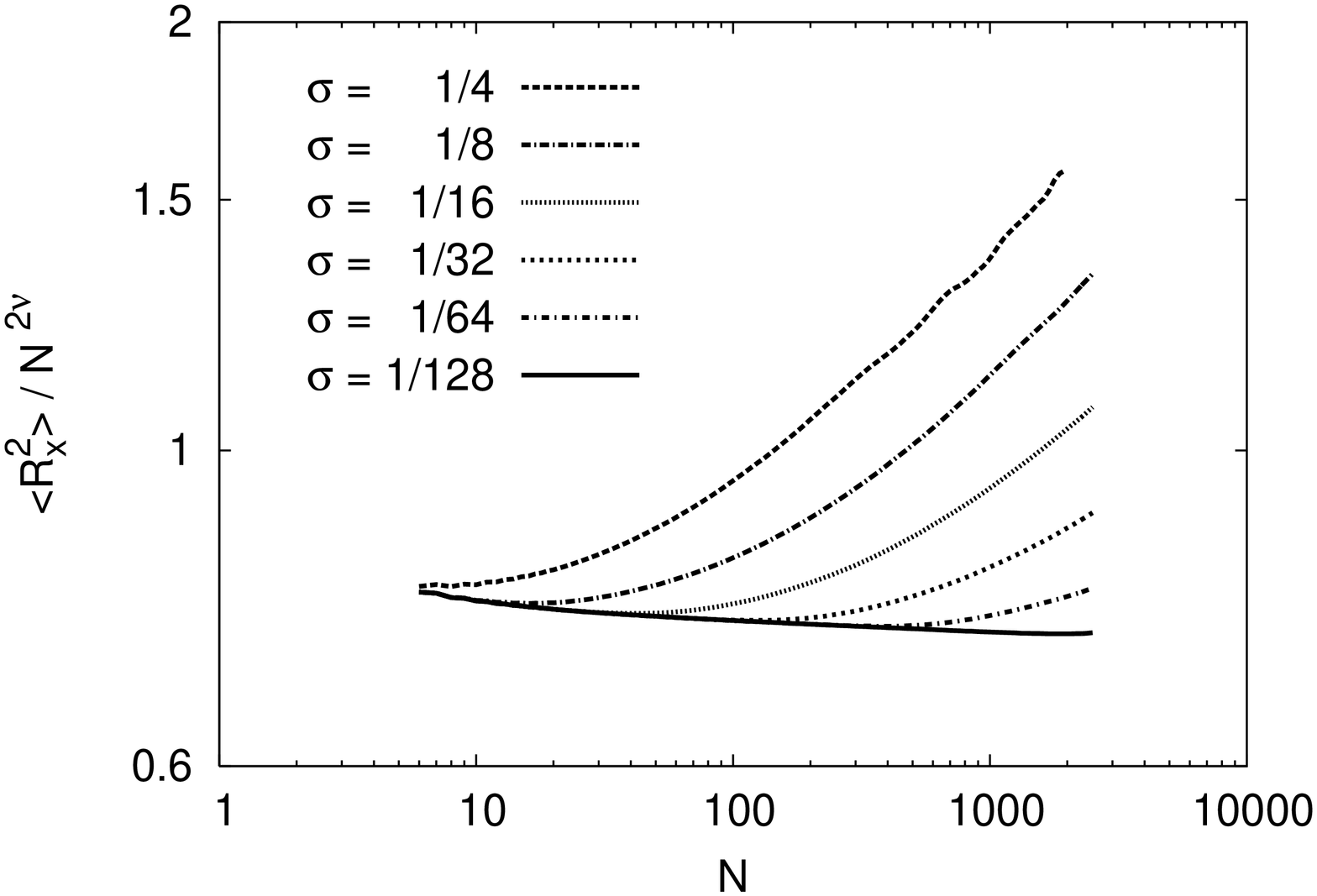,width=5.0cm,angle=270} \\
[0.05cm]\\
\multicolumn{2}{l}{\mbox{(b)}} \\ [-1.5cm] \\
&\psfig{file=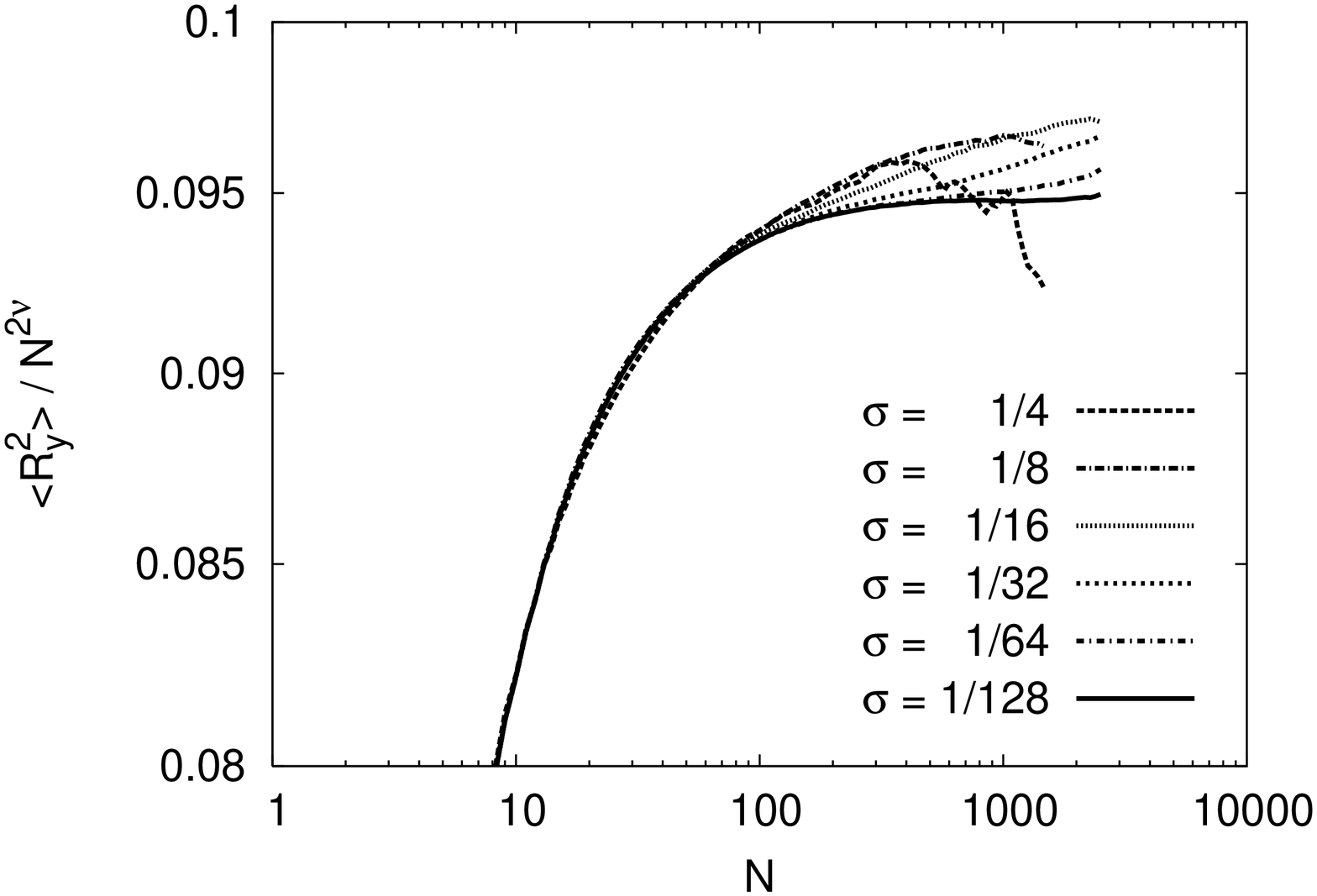, width=5.0cm, angle=270} \\
[0.05cm]\\
\multicolumn{2}{l}{\mbox{(c)}} \\ [-1.5cm] \\
&\psfig{file=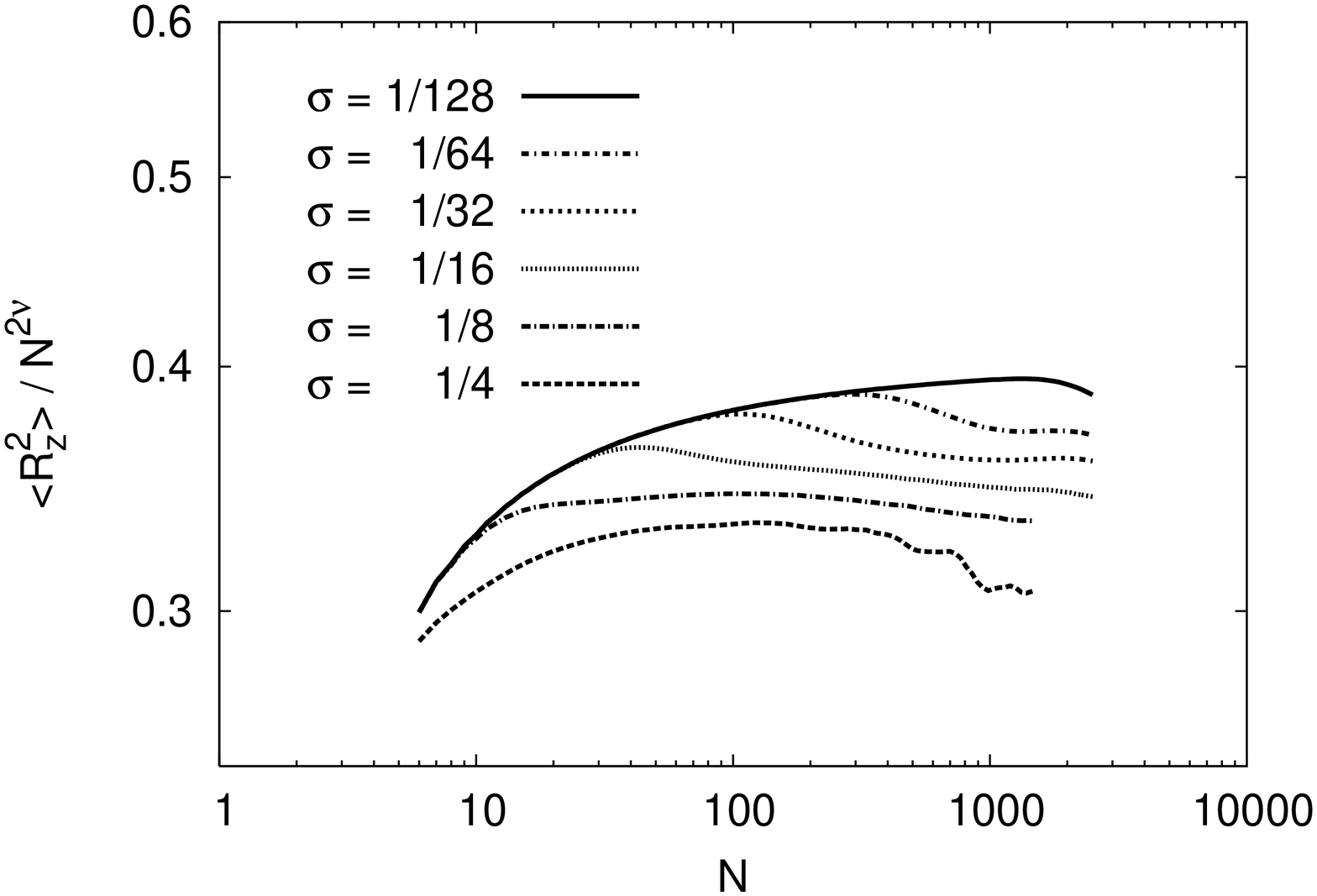, width=5.0cm, angle=270} \\
[1.0cm]
\end{array}$
\caption{Log-log plot of the mean square end-to-end distance
components $\langle R_x^2 \rangle $ (a), $\langle R_y^2 \rangle $
(b), and $\langle R_z^2 \rangle$ (c) versus side chain length $N$,
for various choices of the grafting density $\sigma$ as indicated.
All data refer to $f=1$ (one chain per possible grafting site)
and $N>5$.
Note that the x-direction for every chain is the normal direction
from its center of mass position to the bottle brush backbone. All
data are for $L_b=128$. The chain mean square linear dimensions
are all normalized by $N^{2 \nu}$.}
\label{fig3}
\end{center}
\end{figure}

\begin{figure}
\begin{center}
$\begin{array}{c@{\hspace{0.2in}}c}
\multicolumn{2}{l}{\mbox{(a)}} \\ [-1.5cm] \\
&\psfig{file=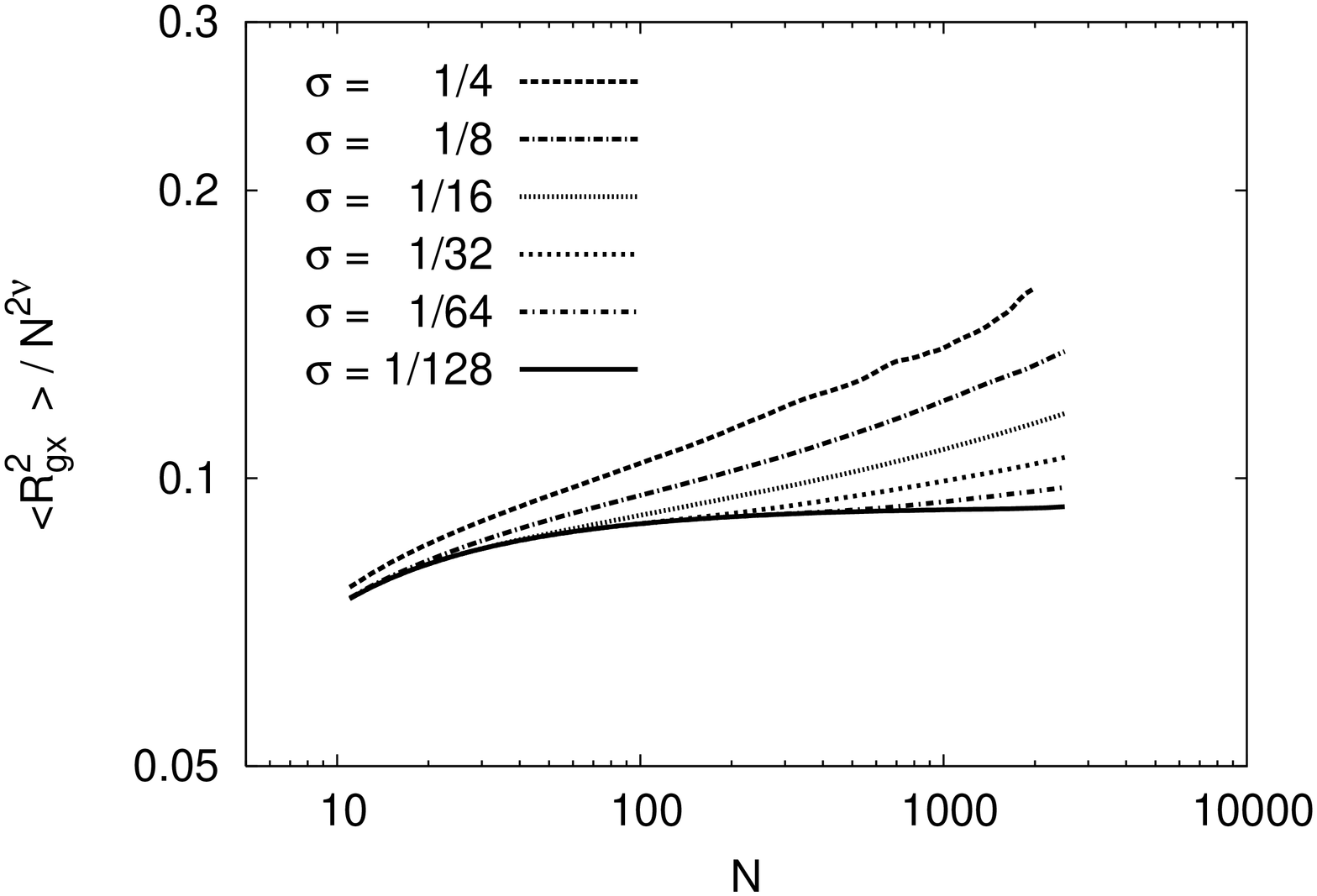,width=5.0cm,angle=270} \\
[0.05cm]\\
\multicolumn{2}{l}{\mbox{(b)}} \\ [-1.5cm] \\
&\psfig{file=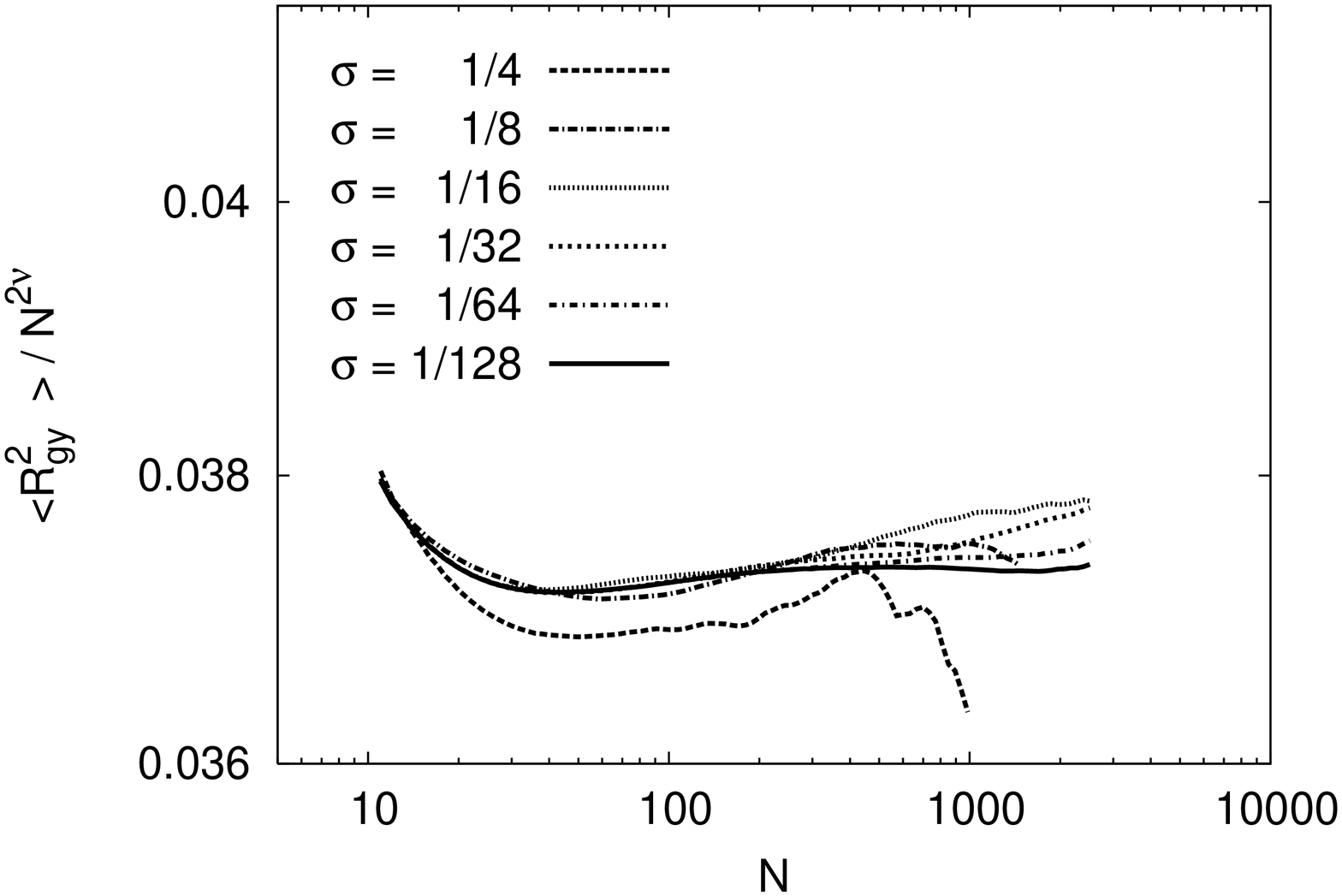, width=5.0cm, angle=270} \\
[0.05cm]\\
\multicolumn{2}{l}{\mbox{(c)}} \\ [-1.5cm] \\
&\psfig{file=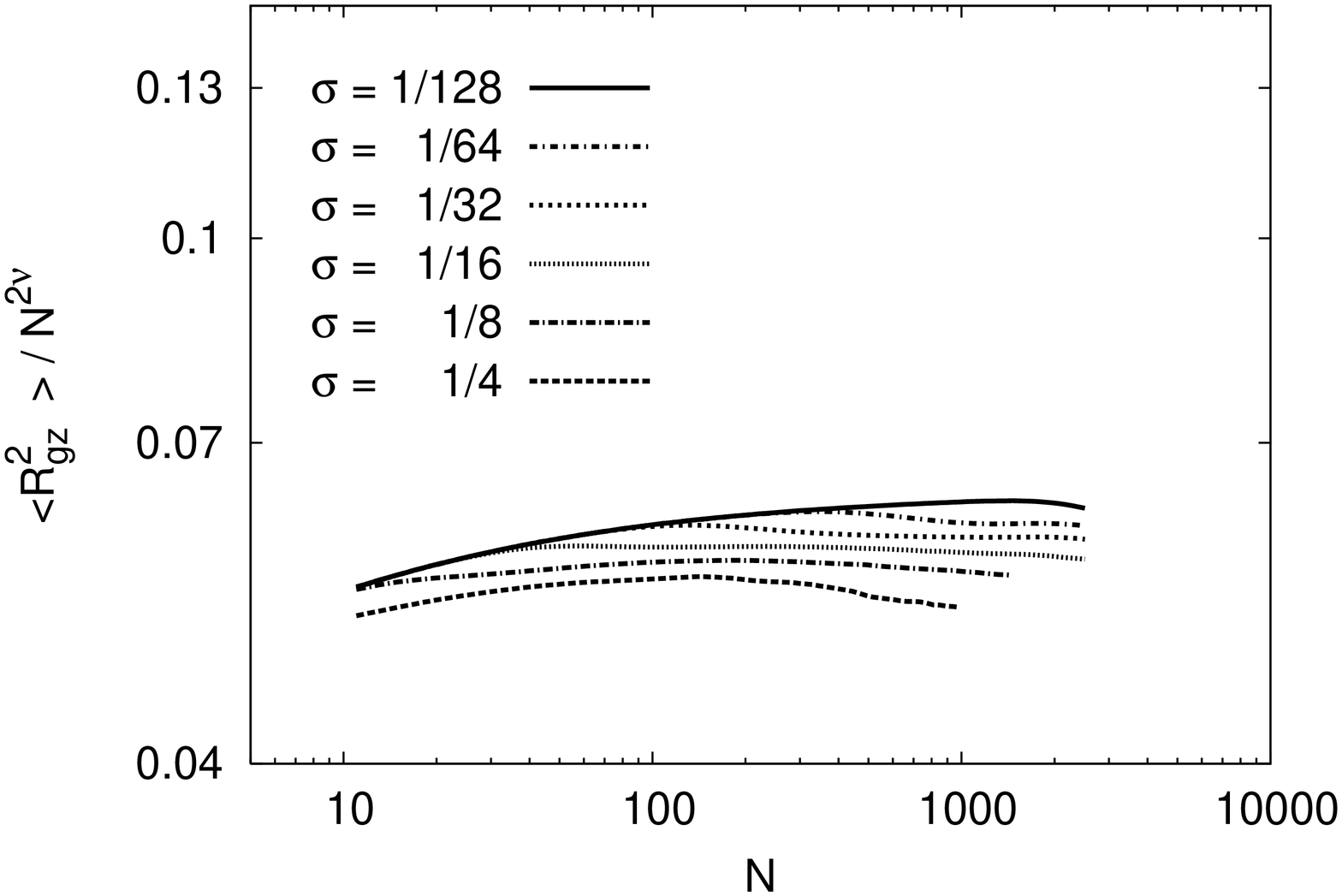, width=5.0cm, angle=270} \\
[1.0cm]
\end{array}$
\caption{Log-log plot of the mean square gyration radius
components $\langle R_{gx}^2 \rangle$ (a), $\langle R_{gy}^2
\rangle $ (b), and $\langle R_{gz}^2 \rangle $ (c), versus side
chain length $N$. Only data for $N > 10$ are included.
All data are for $f=1, L_b=128$, and various choices
of $\sigma$. All chain mean square linear dimensions are
normalized by $N^{2 \nu}$ with $\nu=\nu_3 \approx 0.588$.}
\label{fig4}
\end{center}
\end{figure}

\begin{figure}
\begin{center}
$\begin{array}{c@{\hspace{0.2in}}c}
\multicolumn{2}{l}{\mbox{(a)}} \\ [-1.5cm] \\
&\psfig{file=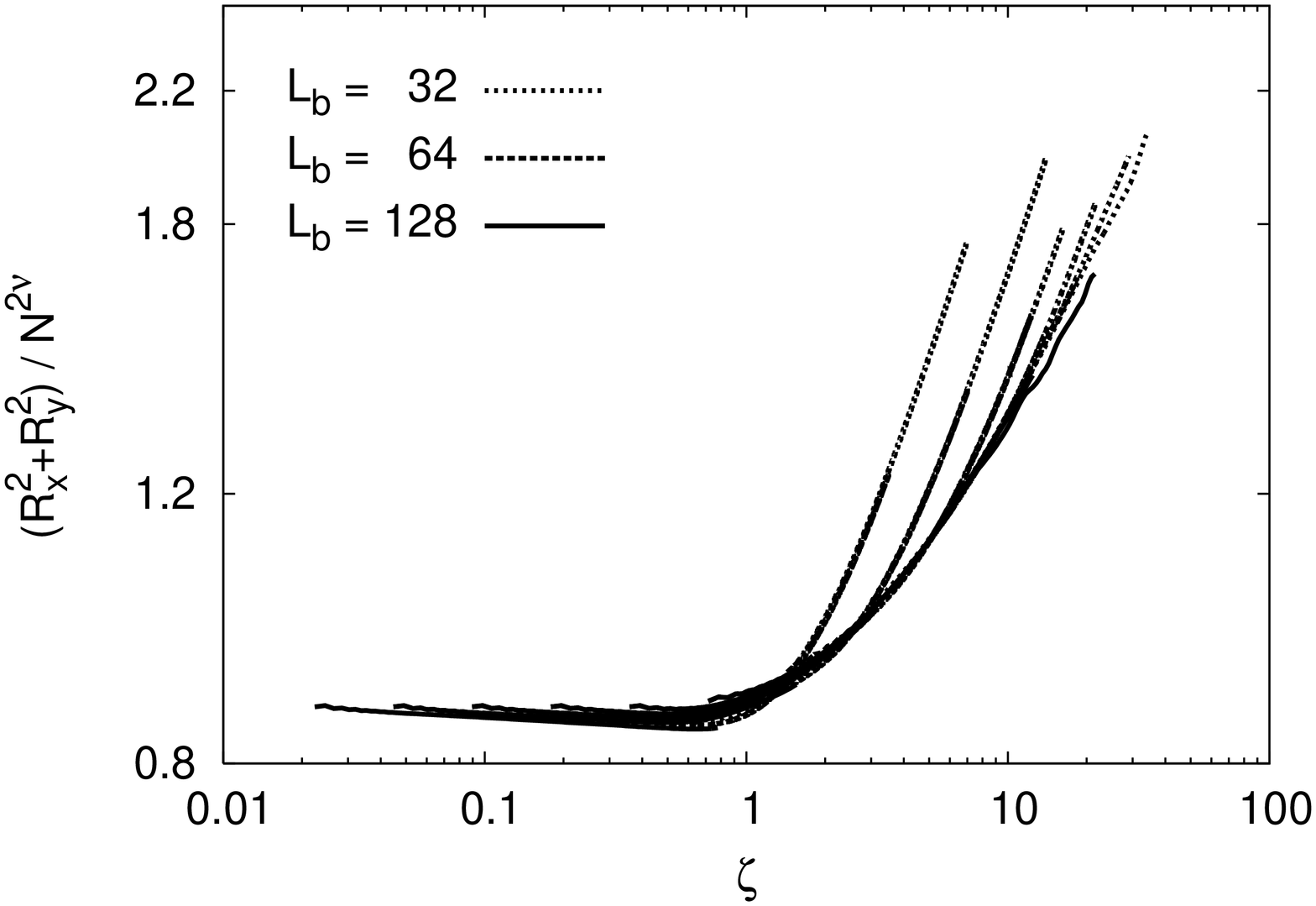,width=5.0cm,angle=270} \\
[0.05cm]\\
\multicolumn{2}{l}{\mbox{(b)}} \\ [-1.5cm] \\
&\psfig{file=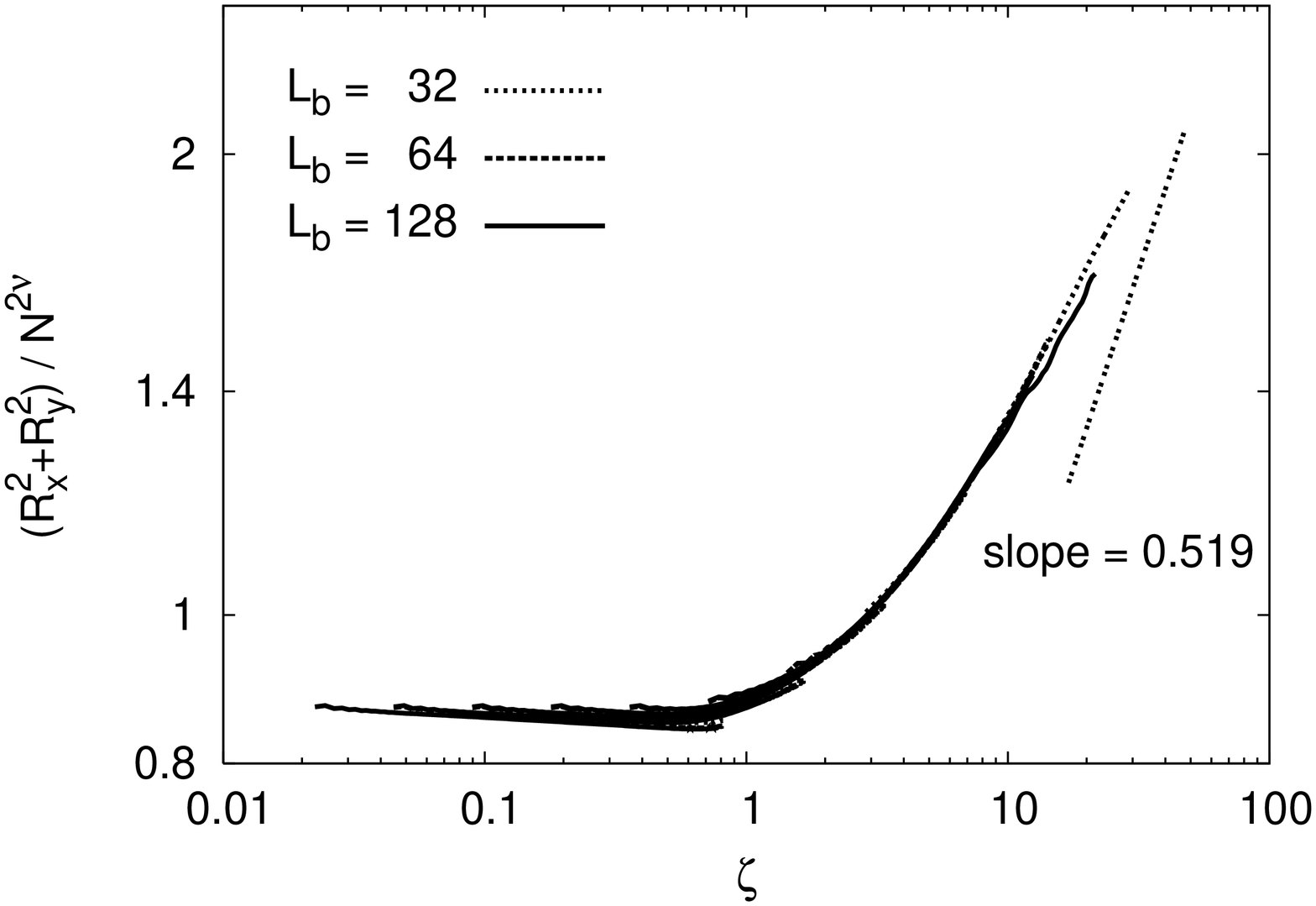, width=5.0cm, angle=270} \\
[1.0cm]
\end{array}$
\caption{Log-log plot of $R^2_\bot = R^2_x+R_y^2$ divided by $N^{2 \nu}$
vs.~$\zeta = \sigma N ^\nu$, including all data for $N > 5 $, and three choices of
$L_b$ as indicated (a), or alternatively removing data affected by
the finite size of the backbone length via the periodic boundary
condition (b). The slope indicated in (b) by the dotted straight
line corresponds to the scaling estimate from Equation~(\ref{eq21}),
$2(1-\nu)/(1+\nu) \approx 0.519.$}
\label{fig5}
\end{center}
\end{figure}

\begin{figure}
\begin{center}
$\begin{array}{c@{\hspace{0.2in}}c}
\multicolumn{2}{l}{\mbox{(a)}} \\ [-1.5cm] \\
&\psfig{file=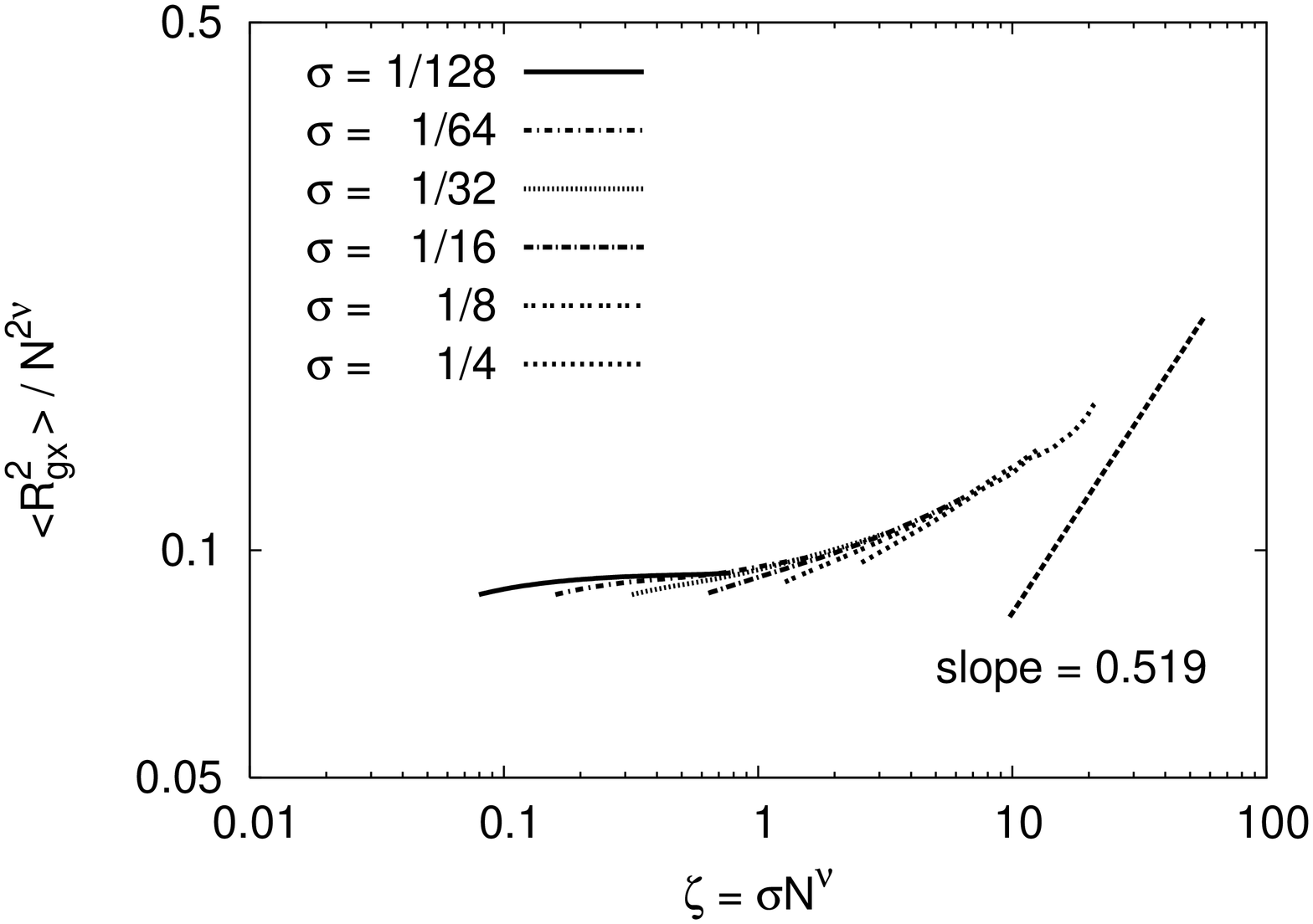,width=5.0cm,angle=270} \\
[0.05cm]\\
\multicolumn{2}{l}{\mbox{(b)}} \\ [-1.5cm] \\
&\psfig{file=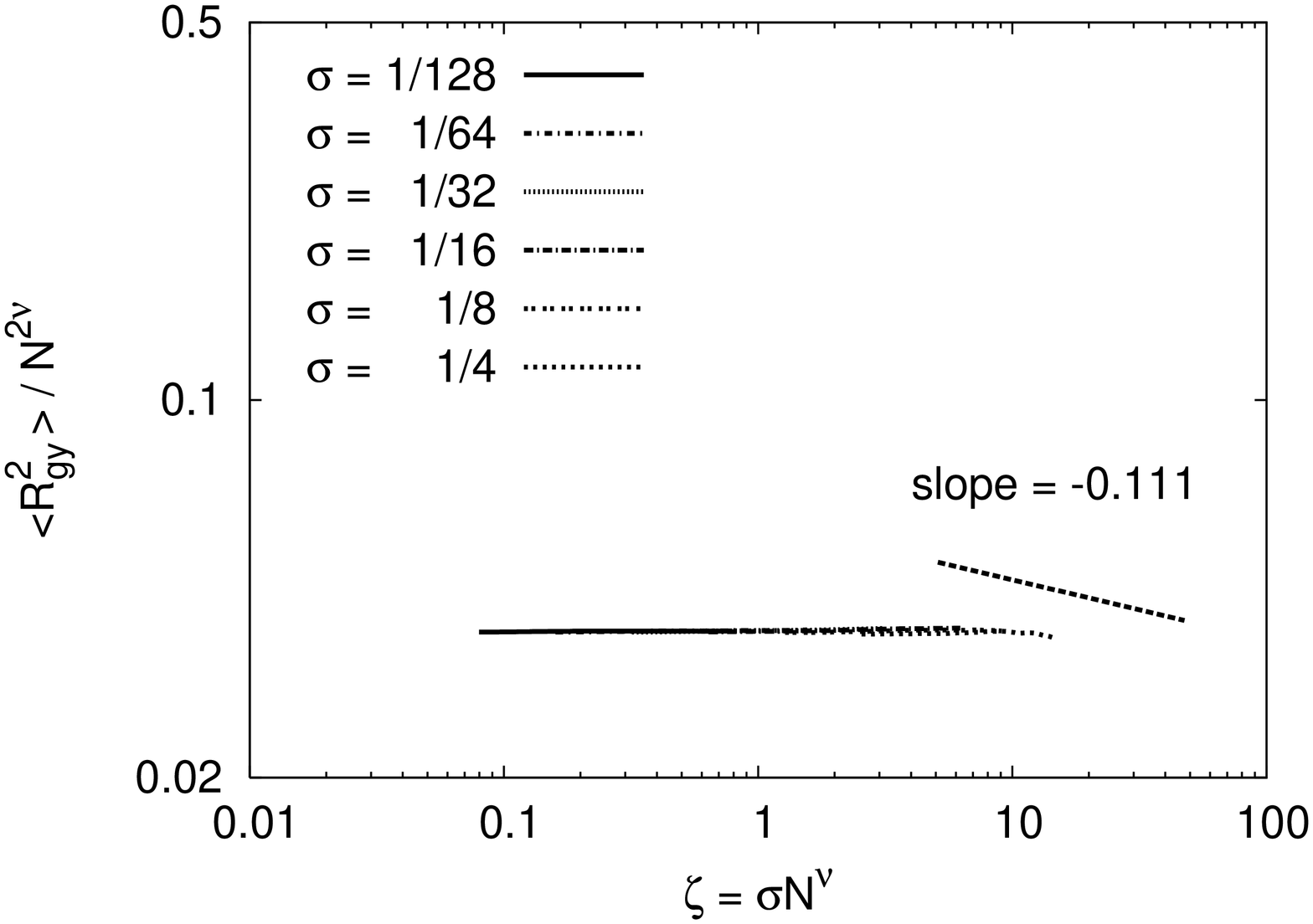, width=5.0cm, angle=270} \\
[0.05cm]\\
\multicolumn{2}{l}{\mbox{(c)}} \\ [-1.5cm] \\
&\psfig{file=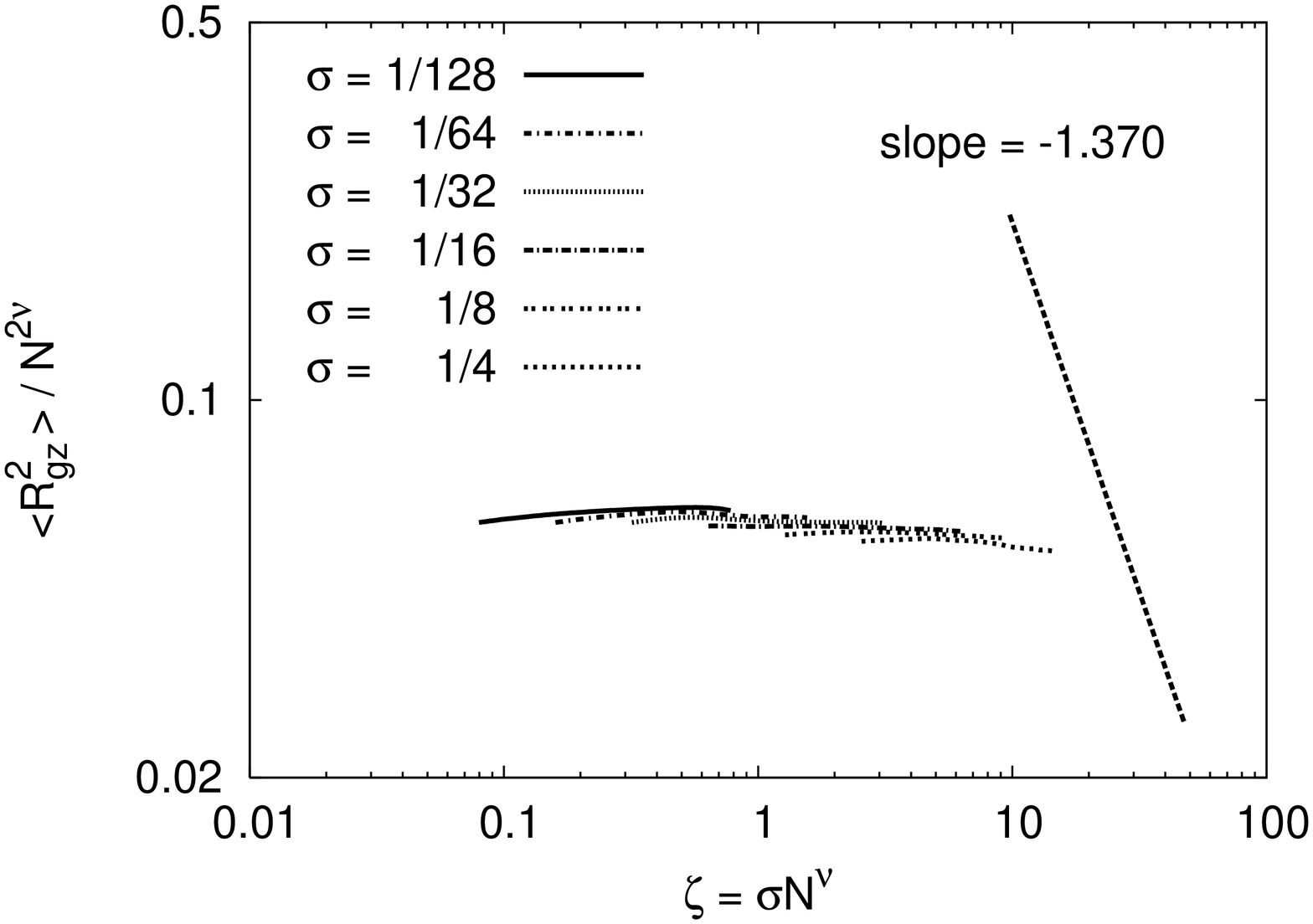, width=5.0cm, angle=270} \\
[1.0cm]
\end{array}$
\caption{Log-log plot of $\langle R_{gx}^2\rangle /N^{2 \nu}$ (a),
$\langle R_{gy}^2 \rangle /N^{2 \nu}$ (b) and $\langle
R_{gz}^2 \rangle /N^{2 \nu}$ (c) versus the scaling variable $\zeta =
\sigma N ^\nu$, using the data in Figure~\ref{fig4} but for $N>50$.
The slopes indicated by
dashed straight lines illustrate the scaling estimates from
Equation~(\ref{eq21}), $2(1-\nu)/(1+\nu) \approx 0.519$ (a), from Equation~
(\ref{eq23}), $-(2\nu - 1)/(\nu + 1) \approx - 0.111$ (b), and from
Equation~(\ref{eq22}), $-(2\nu+1))/(\nu+1) \approx - 1.370$ (c).}
\label{fig6}
\end{center}
\end{figure}

\begin{figure}
\begin{center}
$\begin{array}{c@{\hspace{0.2in}}c}
\multicolumn{2}{l}{\mbox{(a)}} \\ [-1.5cm] \\
&\psfig{file=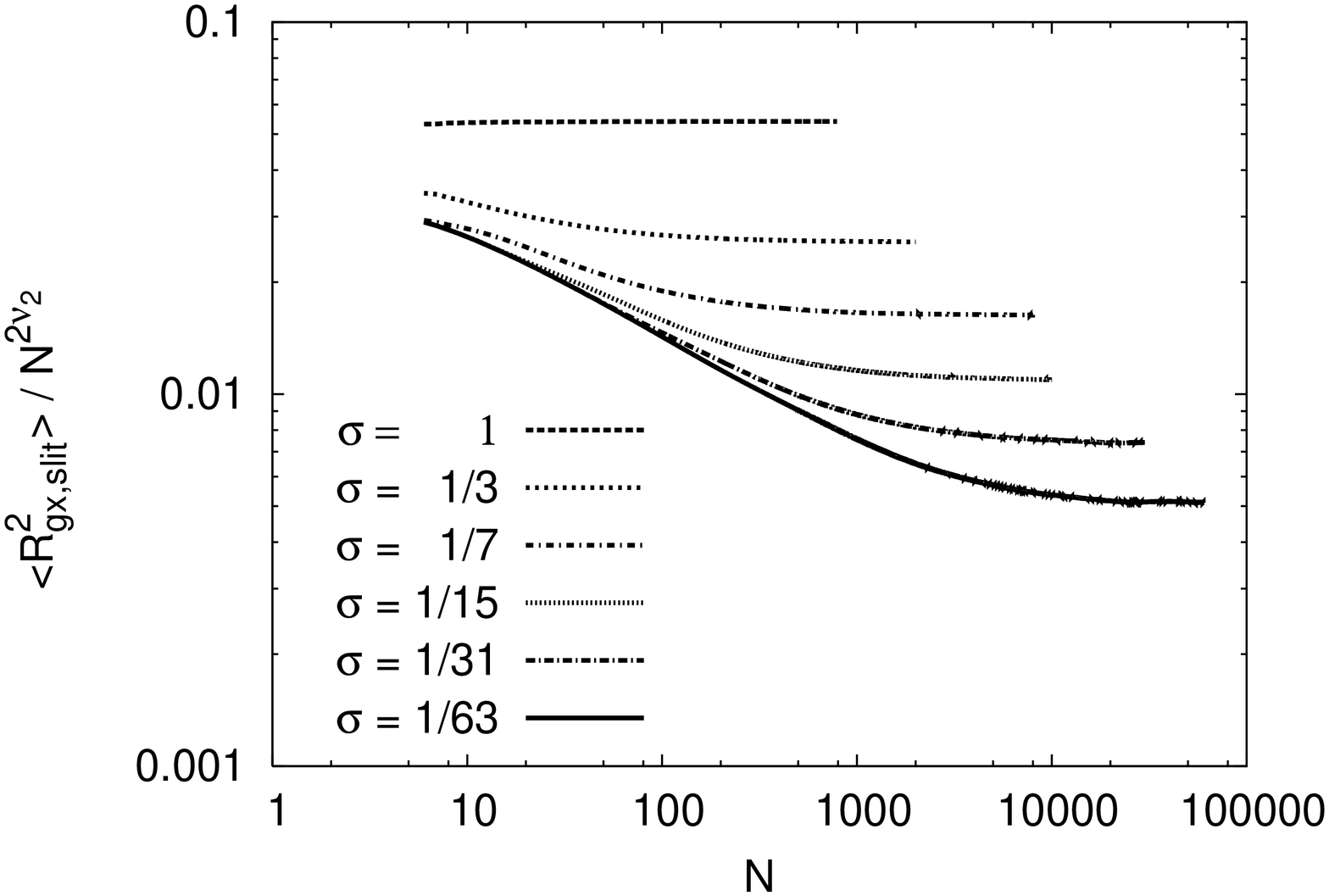,width=5.0cm,angle=270} \\
[0.05cm]\\
\multicolumn{2}{l}{\mbox{(b)}} \\ [-1.5cm] \\
&\psfig{file=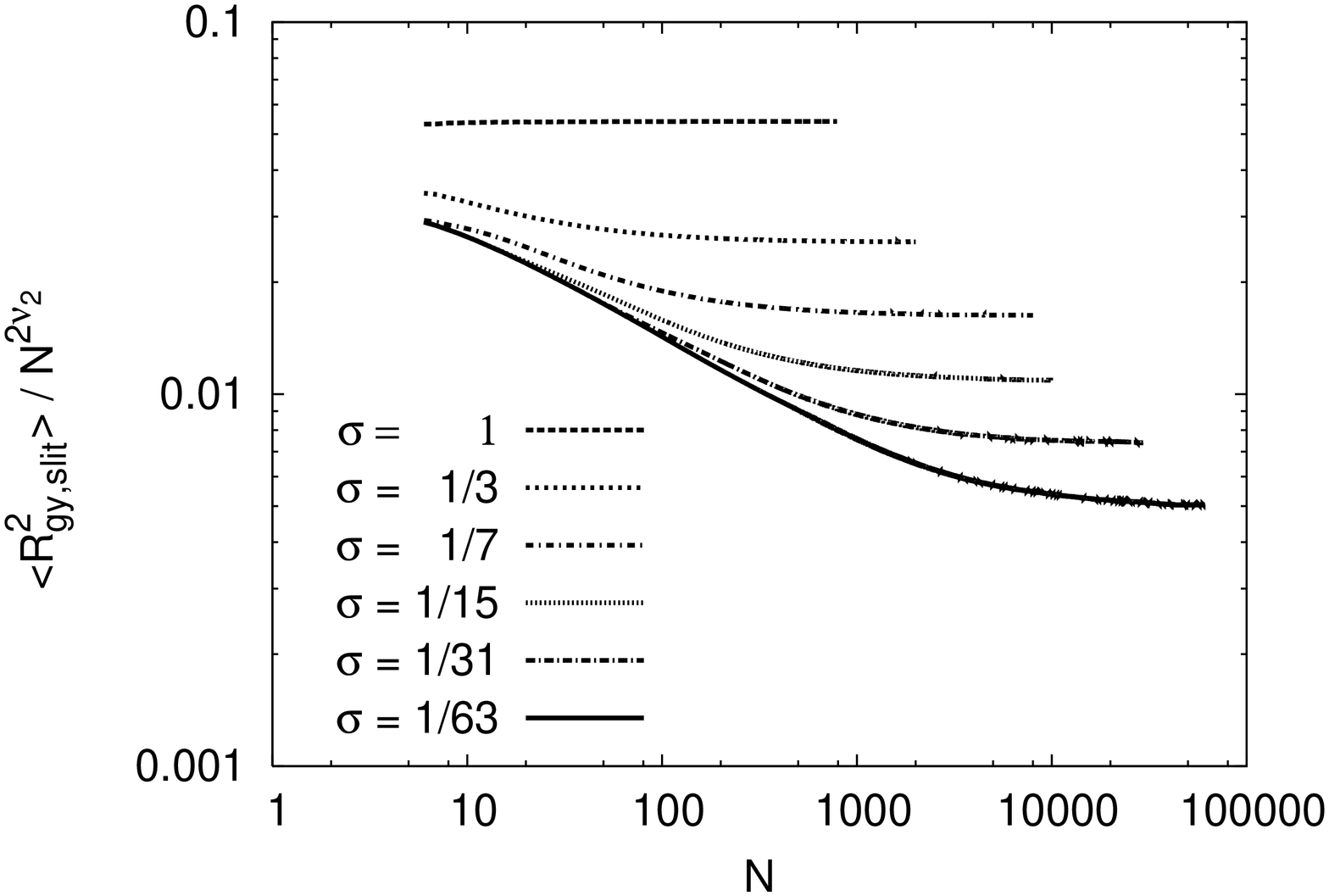, width=5.0cm, angle=270} \\
[0.05cm]\\
\multicolumn{2}{l}{\mbox{(c)}} \\ [-1.5cm] \\
&\psfig{file=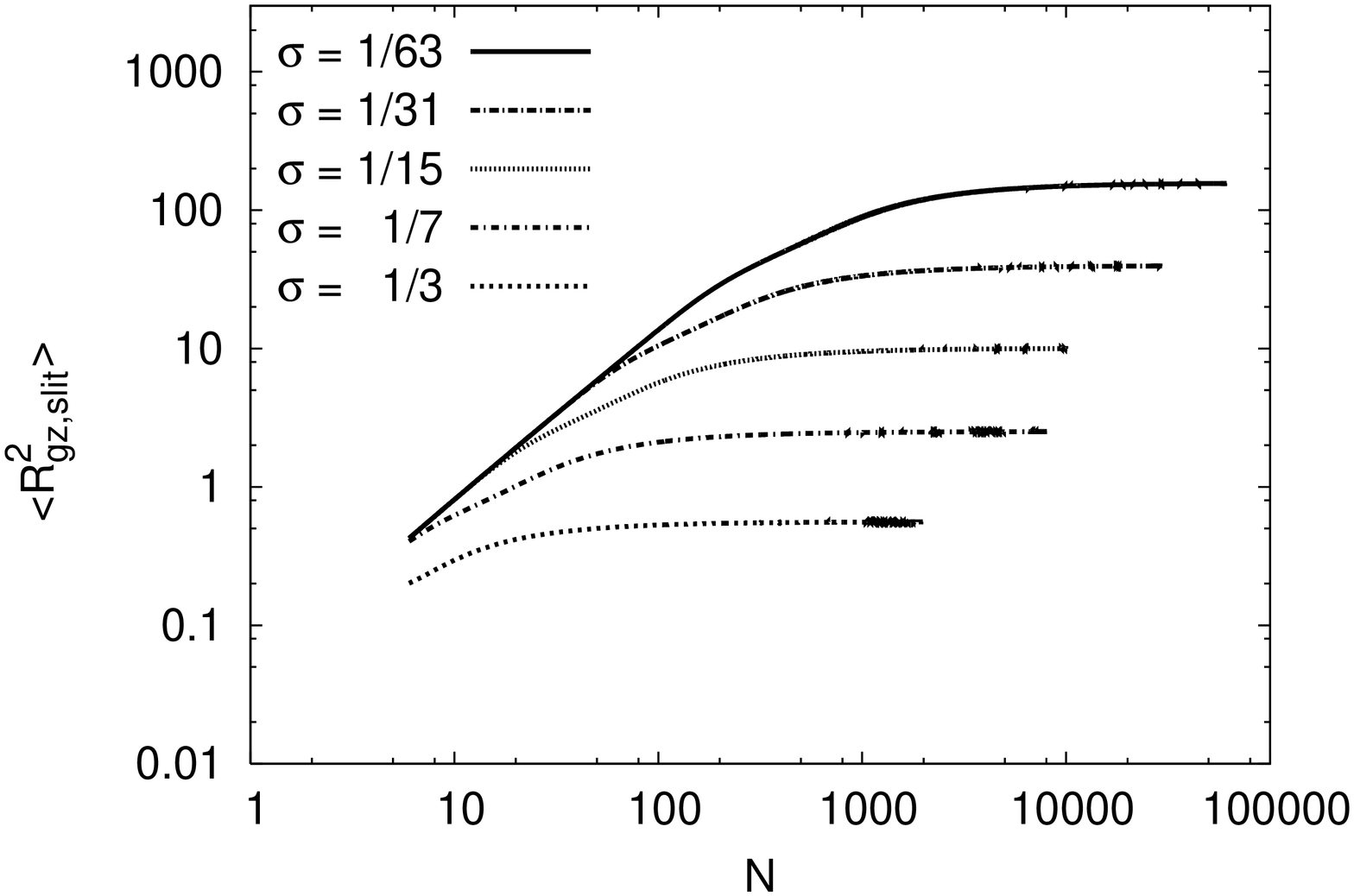, width=5.0cm, angle=270} \\
[1.0cm]
\end{array}$
\caption{Log-log plot of the mean-square gyration radius
components $\langle R_{gx, \textrm{slit}}^2 \rangle /N^{2 \nu_2}$ (a),
$\langle R_{gy, \textrm{slit}}^2 \rangle / N^{2 \nu_2}$ (b) and $\langle
R^2_{gz,\textrm{slit}} \rangle$ (c) versus side chain length $N$
with $\nu_2=3/4$.
As in Figure~\ref{fig4}, only data for $N > 5$ are included.
Various choices of $\sigma$ are included. Note that the walls of
the slit are located at $z = \pm (\sigma ^{-1}+1)/2$ and are
strictly repulsive hard walls. One chain end is fixed at the
x-axis at $z=0$, and since the x-axis represents the backbone if
this polymer in a slit is taken as a model of a disk-shaped
section of a bottle brush, all sites of the x-axis are excluded
from occupation of the monomers of the (side) chain as well.}
\label{fig7}
\end{center}
\end{figure}

\begin{figure}
\begin{center}
$\begin{array}{c@{\hspace{0.2in}}c}
\multicolumn{2}{l}{\mbox{(a)}} \\ [-1.5cm] \\
&\psfig{file=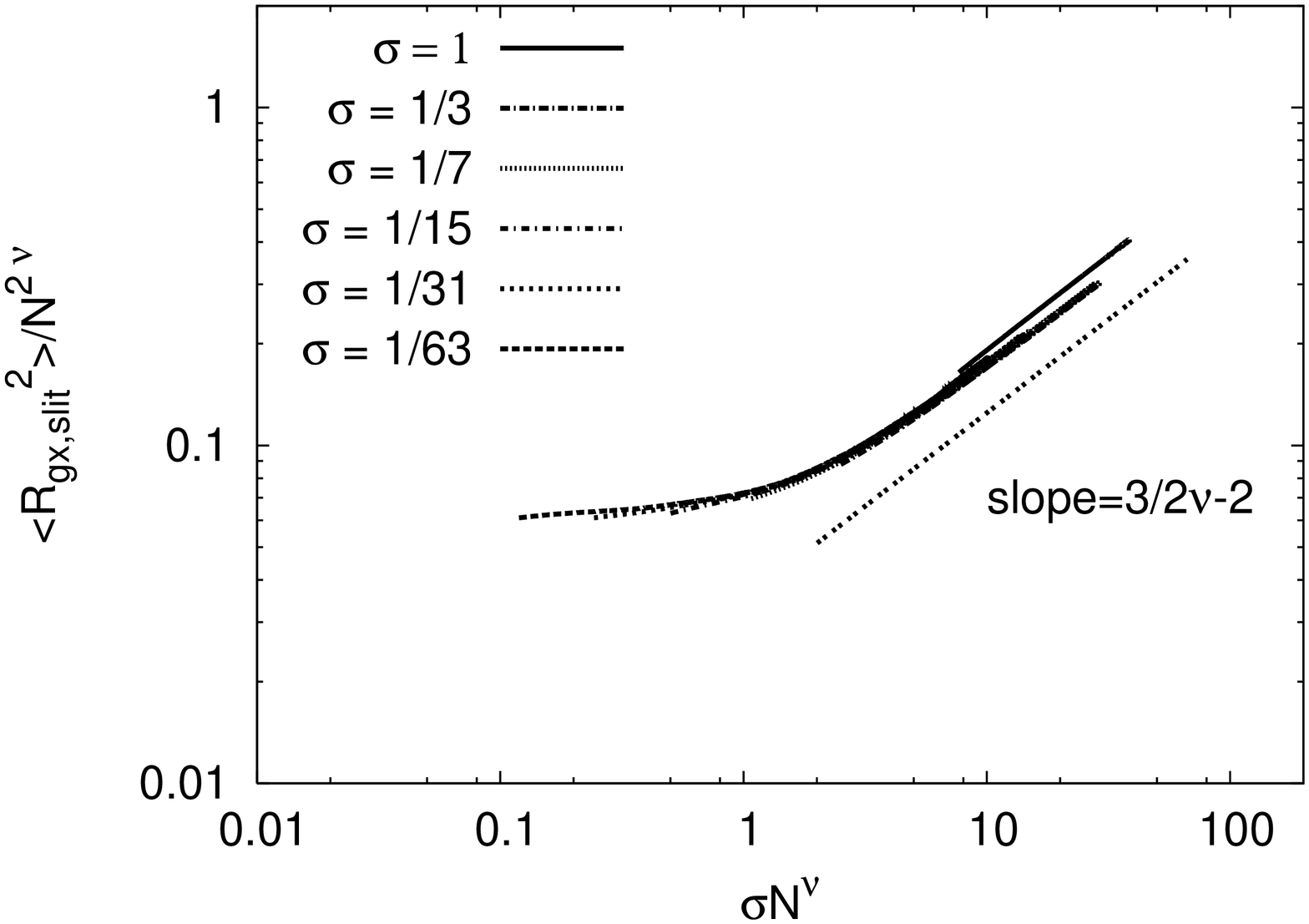,width=5.0cm,angle=270} \\
[0.05cm]\\
\multicolumn{2}{l}{\mbox{(b)}} \\ [-1.5cm] \\
&\psfig{file=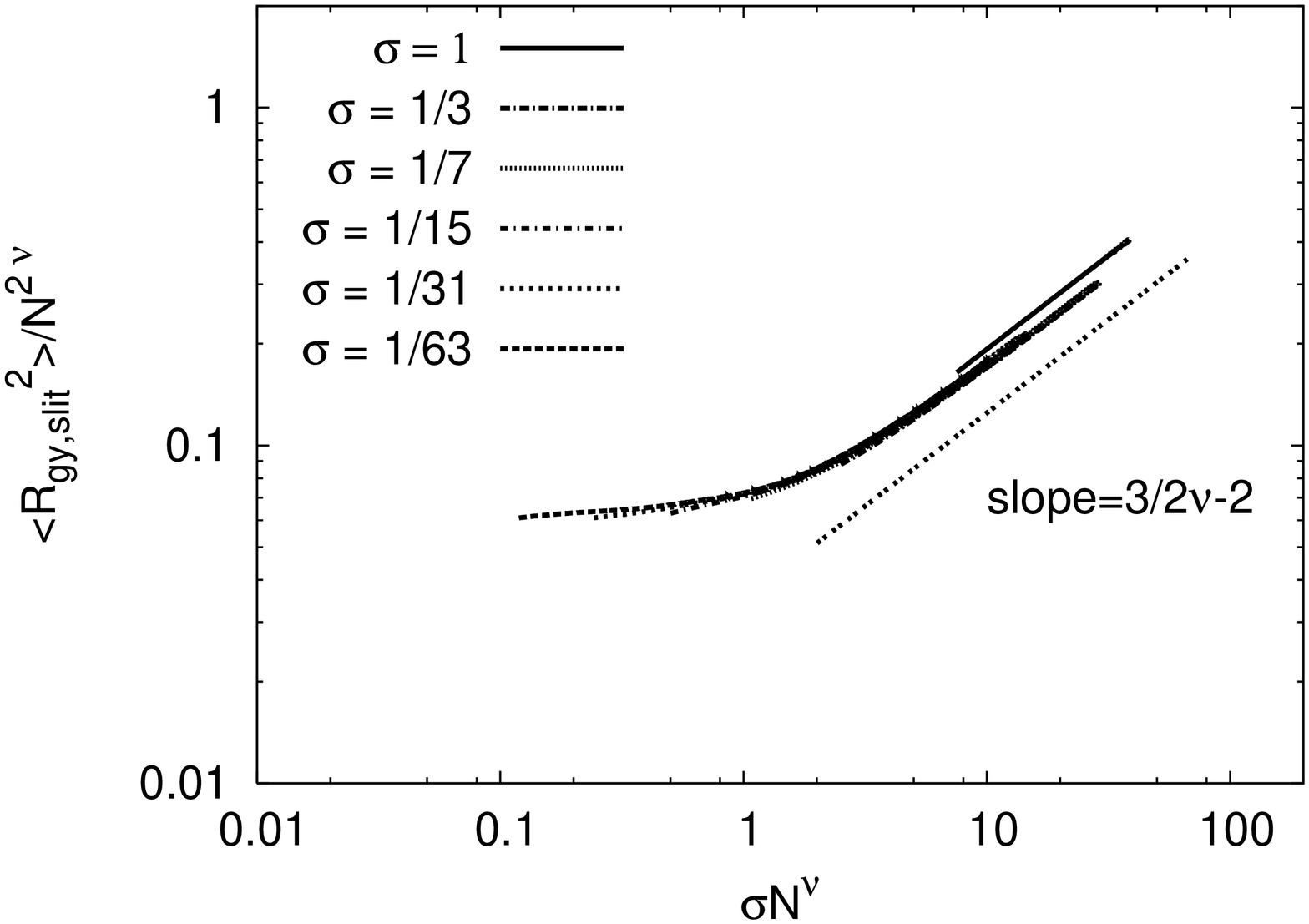, width=5.0cm, angle=270} \\
[0.05cm]\\
\multicolumn{2}{l}{\mbox{(c)}} \\ [-1.5cm] \\
&\psfig{file=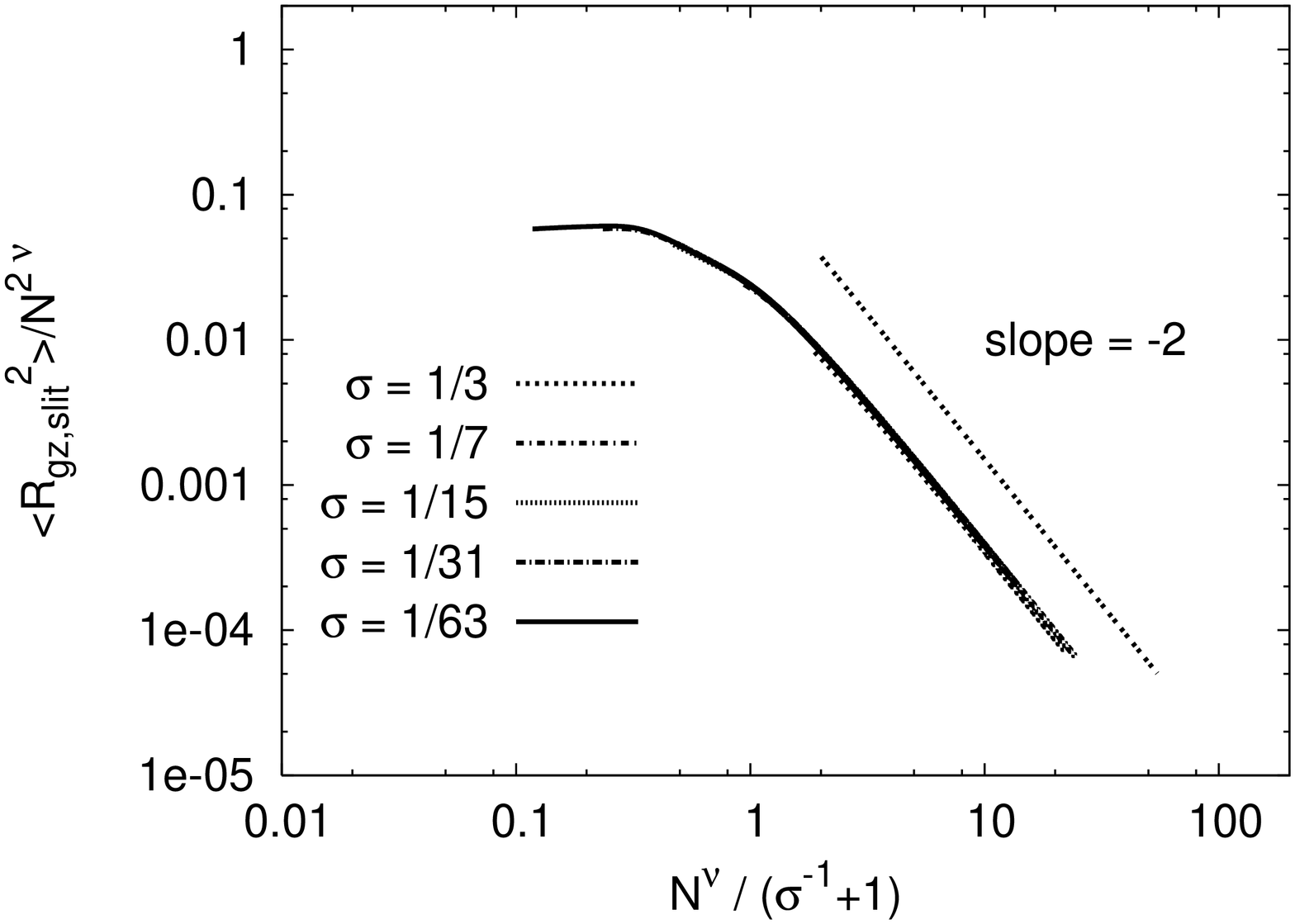, width=5.0cm, angle=270} \\
[1.0cm]
\end{array}$
\caption{Log-log plot of $\langle R_{gx,\textrm{slit}}\rangle
/N^{2 \nu}$ (a), $\langle R_{gy,\textrm{slit}}^2\rangle
/N^{2\nu}$, and $\langle R_{gz \textrm{slit}}^2 \rangle /N^{2 \nu}$
(c) versus the scaling variable $\zeta = \sigma N ^\nu$.
Only data for $N>10$ are included. The
slopes indicated by dashed straight lines illustrate the scaling
exponent implied by Equation~(\ref{eq27}), namely $3/2\nu-2$ (a,b) and
$-2$ (c), respectively. In (c) $\sigma^{-1}$ is replaced by $\sigma^{-1}+1$
to remove a finite-size effect. }
\label{fig8}
\end{center}
\end{figure}

\begin{figure}
\begin{center}
$\begin{array}{c@{\hspace{0.2in}}c}
\multicolumn{2}{l}{\mbox{(a)}} \\ [-1.5cm] \\
&\psfig{file=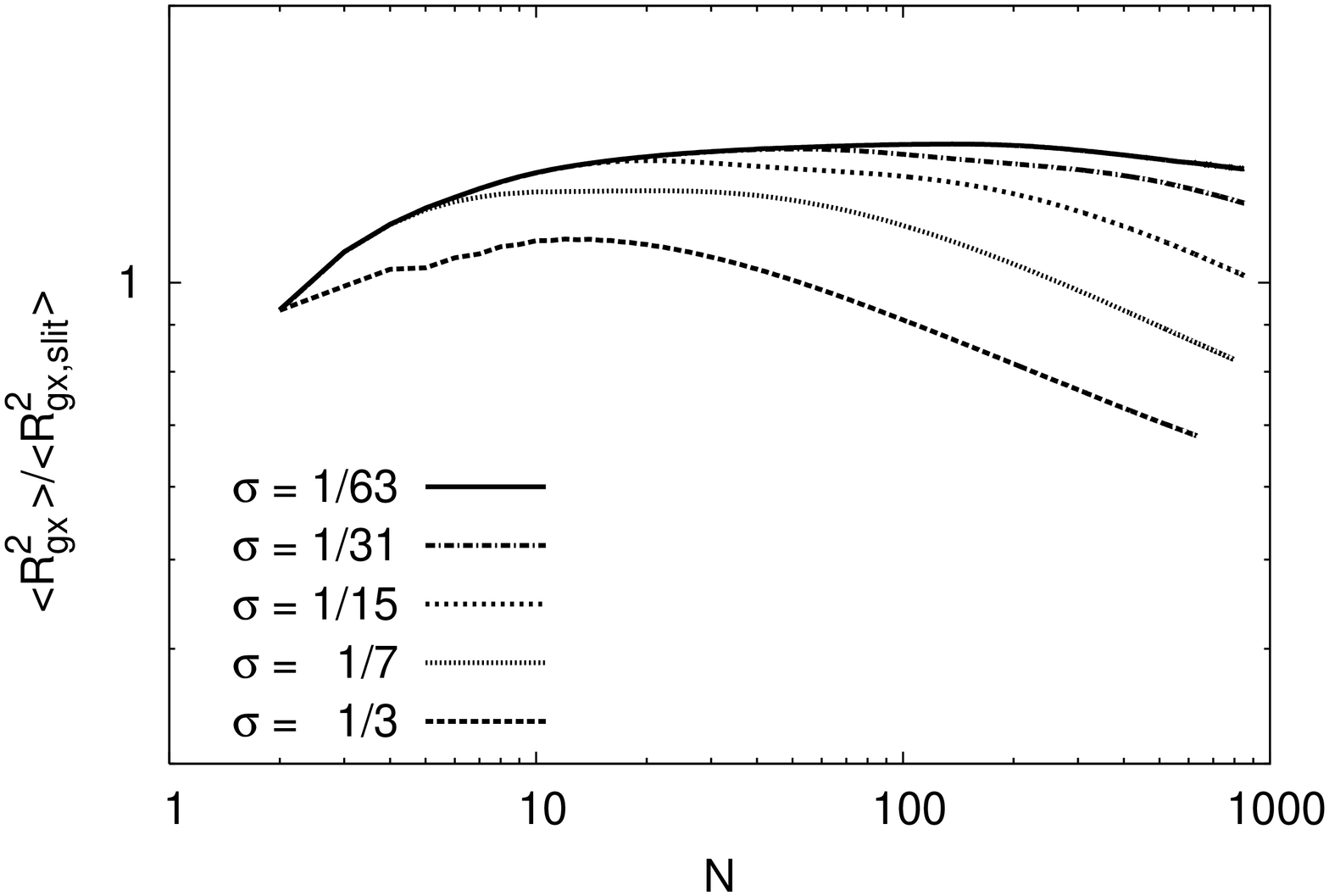,width=5.0cm,angle=270} \\
[0.05cm]\\
\multicolumn{2}{l}{\mbox{(b)}} \\ [-1.5cm] \\
&\psfig{file=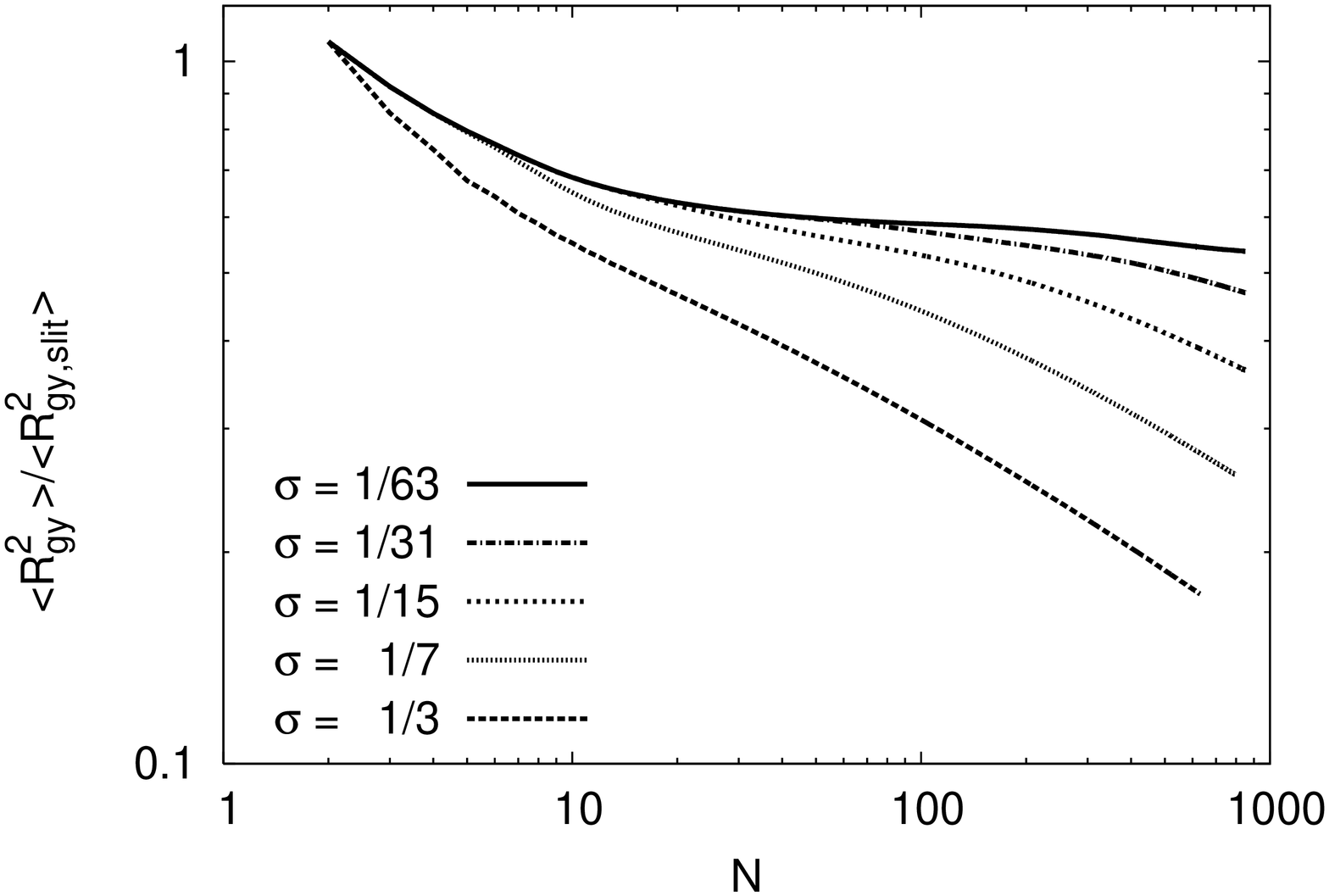, width=5.0cm, angle=270} \\
[0.05cm]\\
\multicolumn{2}{l}{\mbox{(c)}} \\ [-1.5cm] \\
&\psfig{file=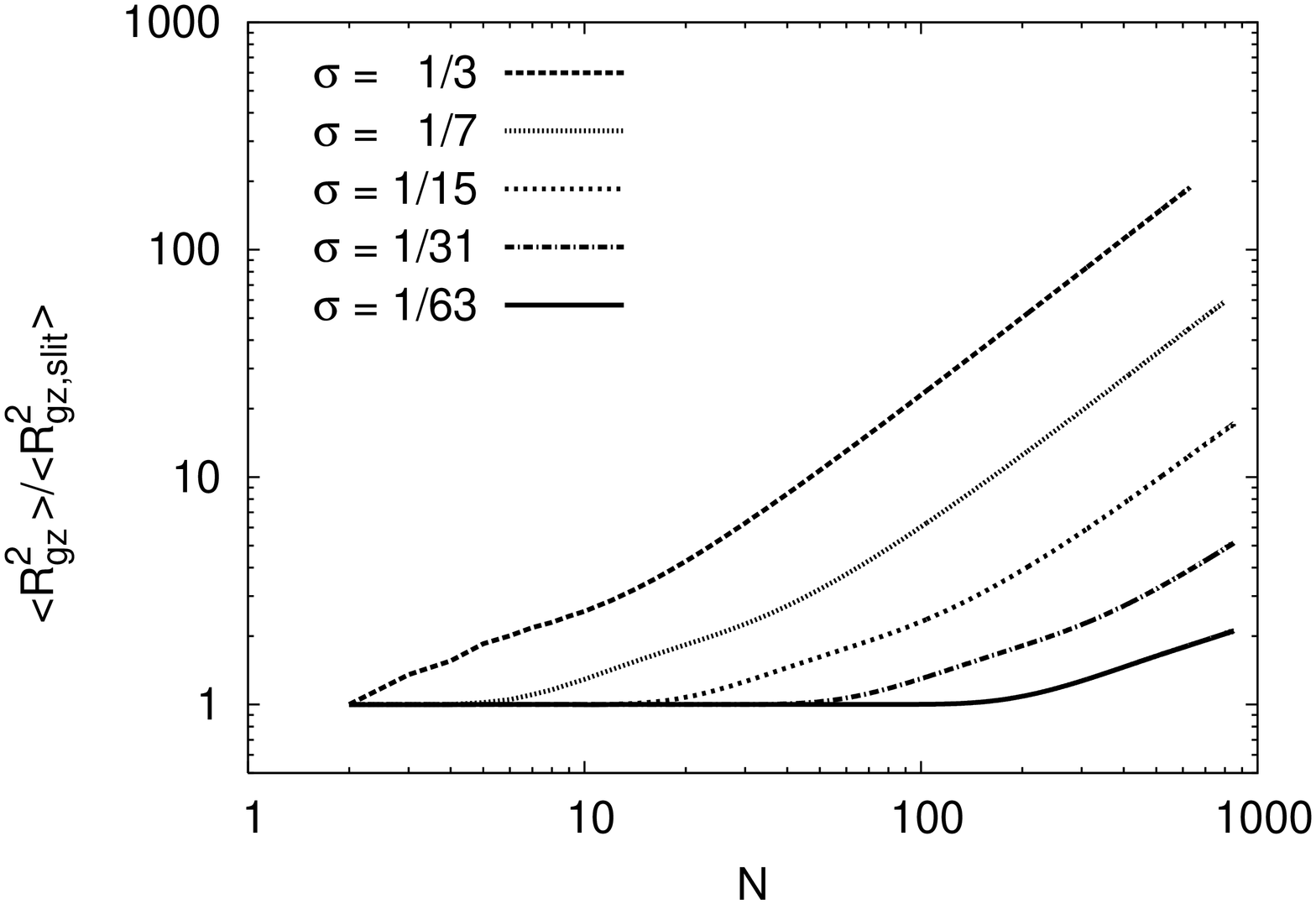, width=5.0cm, angle=270} \\
[1.0cm]
\end{array}$
\caption{Log-log plot of $\langle R_{gx}^2\rangle/\langle
R_{gx,\textrm{slit}}^2\rangle$ (a), $\langle R_{gy}^2\rangle
/\langle R_{gy,\textrm{slit}}^2\rangle $ (b) and $\langle
R_{gz}^2\rangle /\langle R_{gz,\textrm{slit}}^2 \rangle $ versus
$N$, for values of $\sigma$ strictly corresponding to each other,
as indicated. }
\label{fig9}
\end{center}
\end{figure}

\begin{figure}
\begin{center}
$\begin{array}{c@{\hspace{0.2in}}c}
\multicolumn{2}{l}{\mbox{(a)}} \\ [-1.5cm] \\
&\psfig{file=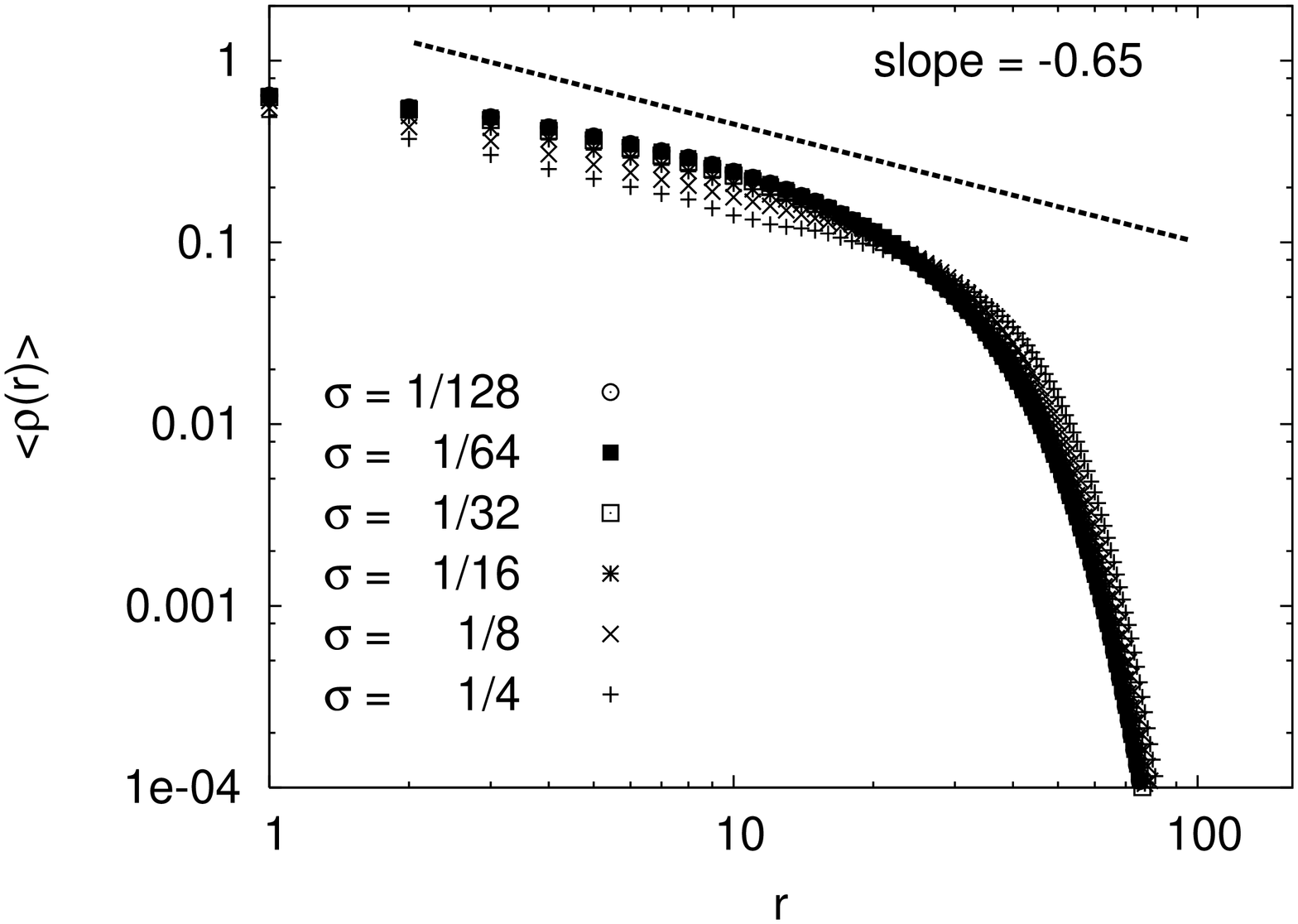,width=5.0cm,angle=270} \\
[0.05cm]\\
\multicolumn{2}{l}{\mbox{(b)}} \\ [-1.5cm] \\
&\psfig{file=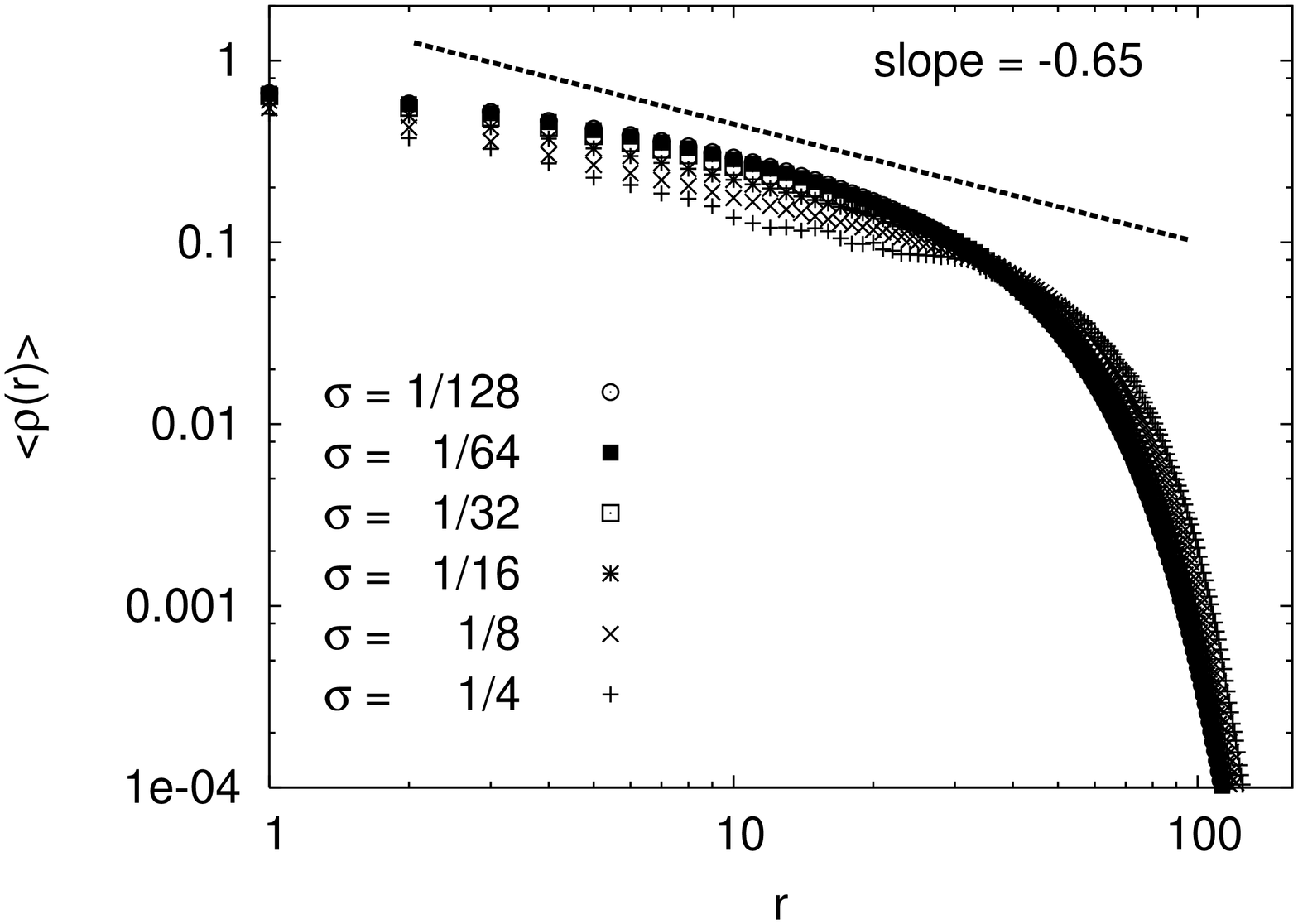, width=5.0cm, angle=270} \\
[0.05cm]\\
\multicolumn{2}{l}{\mbox{(c)}} \\ [-1.5cm] \\
&\psfig{file=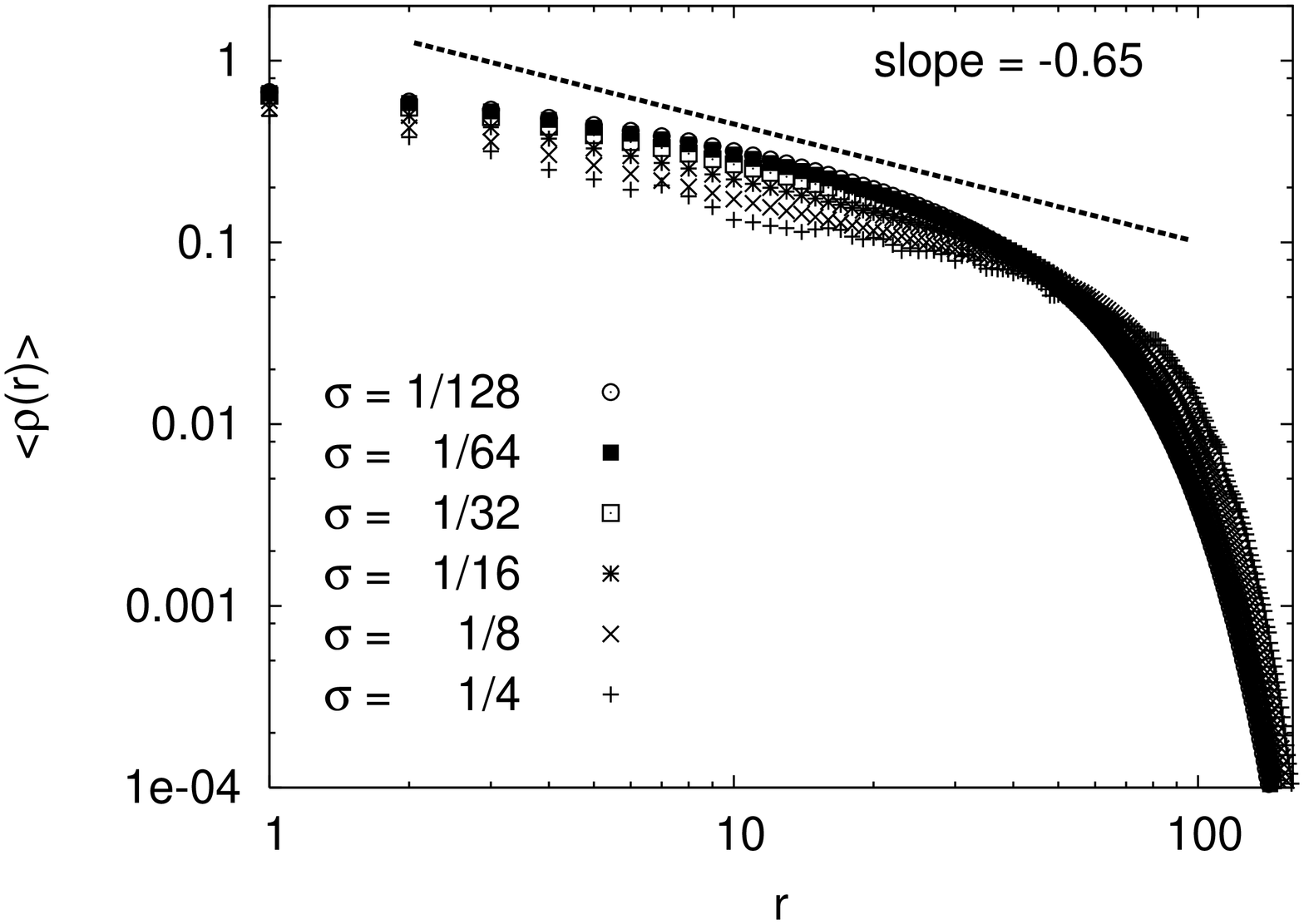, width=5.0cm, angle=270} \\
[1.0cm]
\end{array}$
\caption{Radial distribution function $\rho (r)$ plotted vs.~$r$,
for $N=500 $ (a), $N=1000$ (b), and $N=1500$ (c), for various choices of
$\sigma$, as indicated. Note that due to the discreteness of the
lattice, the number $N(r)$ of monomers in the interval $[r,r+dr]$
is not normalized by the factor $\pi r$ that applies in the
continuum limit, but by the number $N_r$ of lattice sites (x,y)
satisfying the constraint $r^2= x^2+y^2$, i.e.
$\sum_r N(r)=N$ and $\rho(r)=N(r)/N_r$. All data refer to
$L_b = 128$ and $f=1$.}
\label{fig10}
\end{center}
\end{figure}

\begin{figure}
\begin{center}
$\begin{array}{c@{\hspace{0.2in}}c}
\multicolumn{2}{l}{\mbox{(a)}} \\ [-1.5cm] \\
&\psfig{file=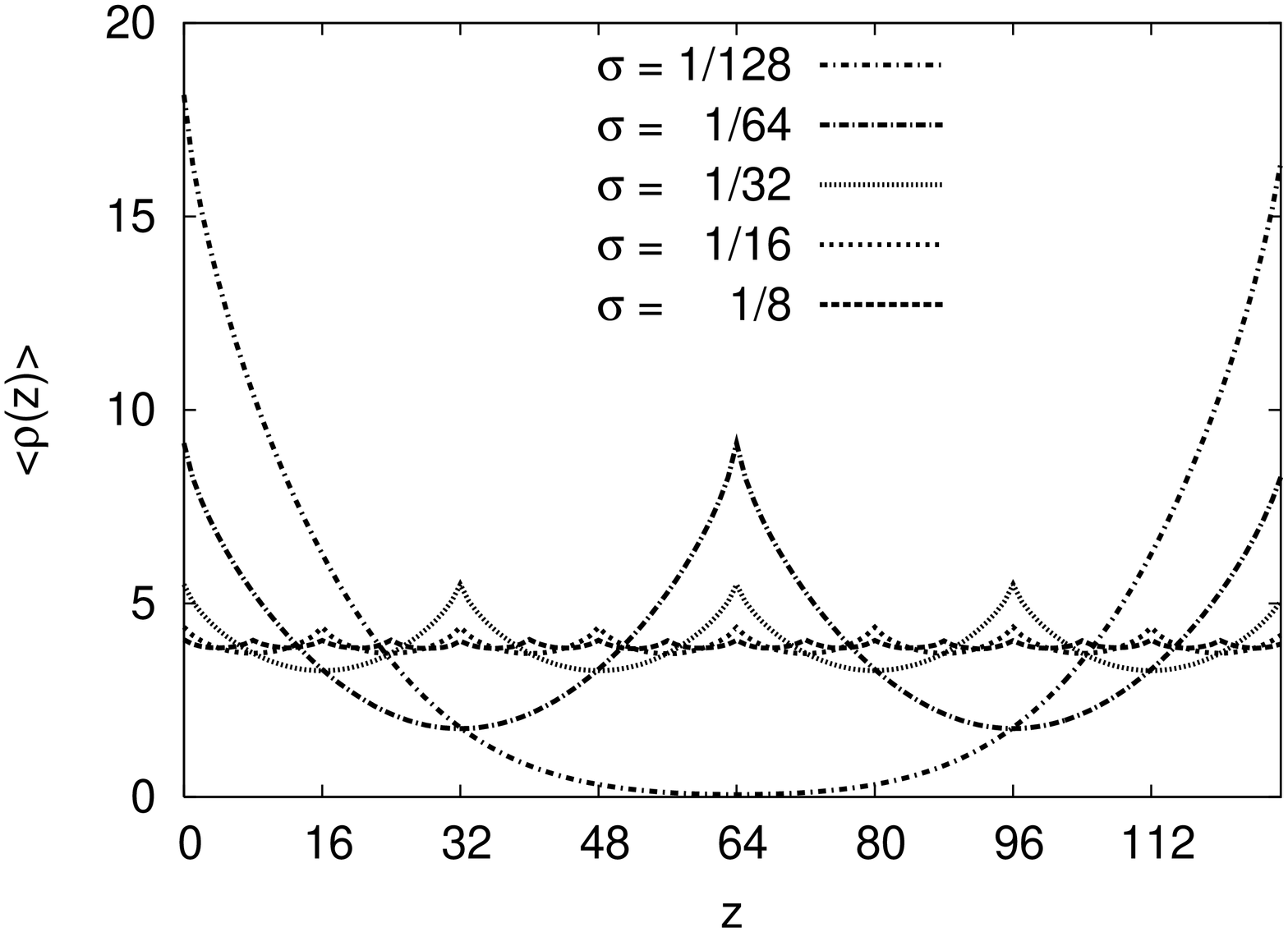,width=5.0cm,angle=270} \\
[0.05cm]\\
\multicolumn{2}{l}{\mbox{(b)}} \\ [-1.5cm] \\
&\psfig{file=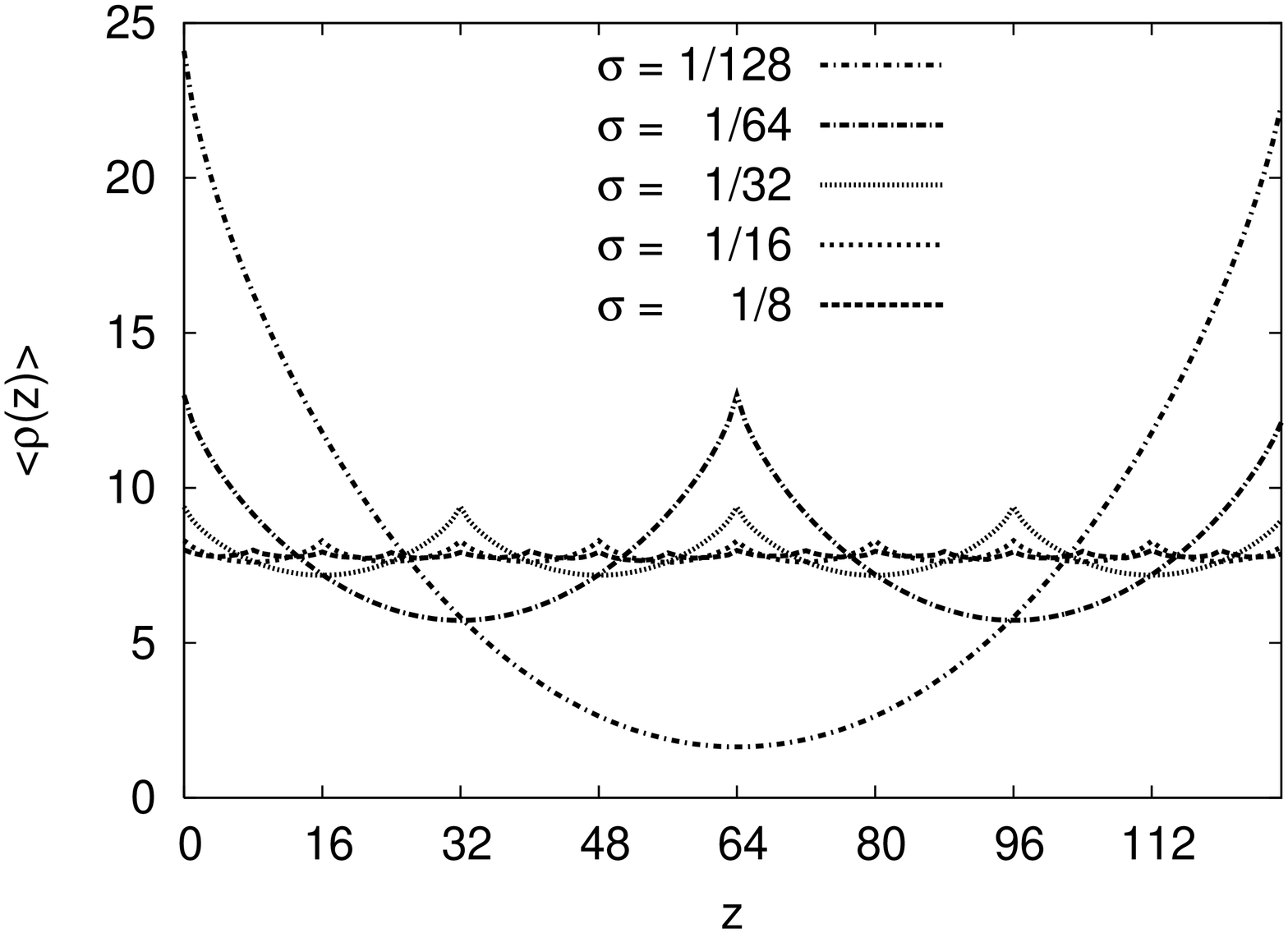, width=5.0cm, angle=270} \\
[0.05cm]\\
\multicolumn{2}{l}{\mbox{(c)}} \\ [-1.5cm] \\
&\psfig{file=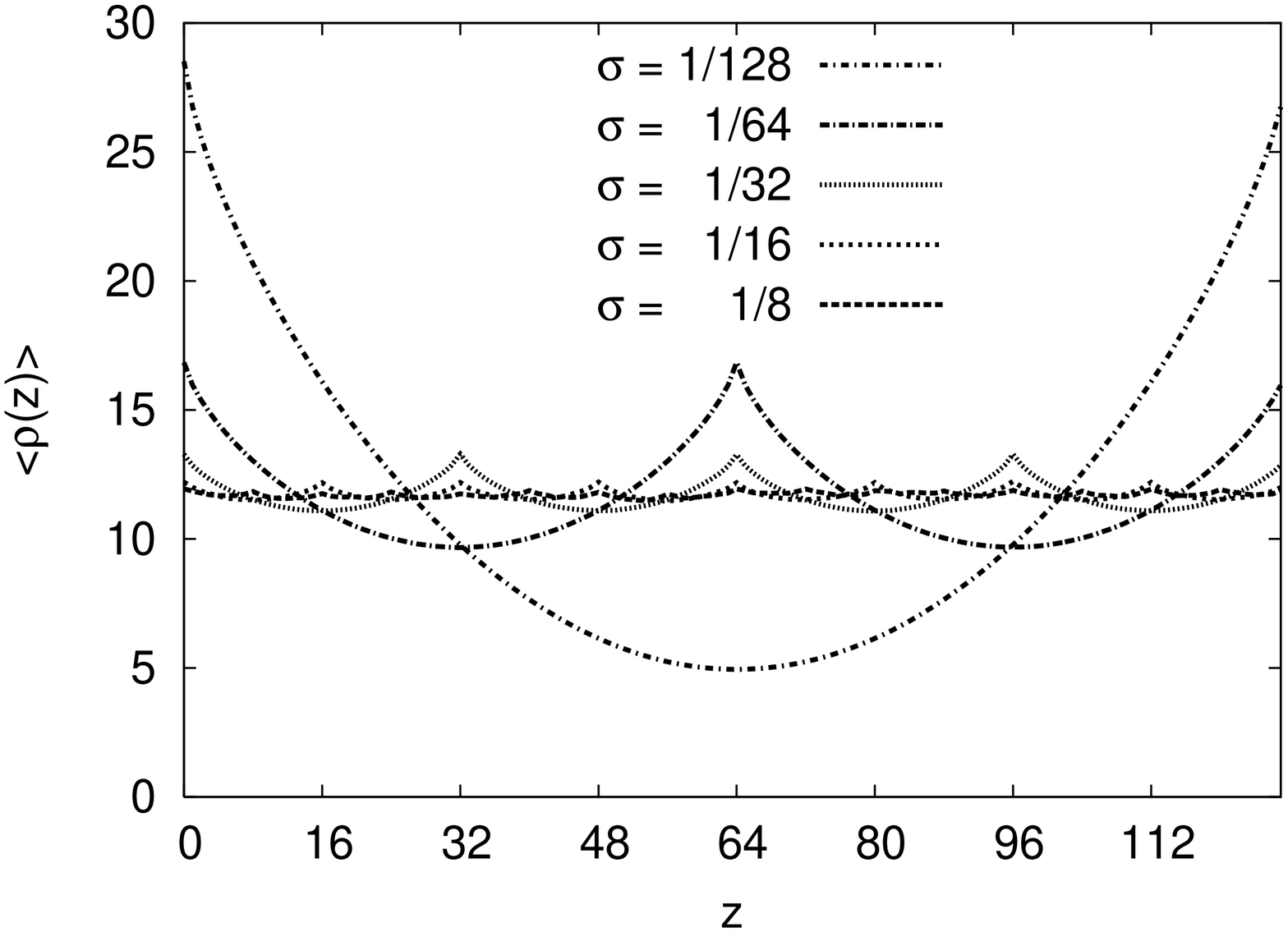, width=5.0cm, angle=270} \\
[1.0cm]
\end{array}$
\caption{Density distribution function $\rho(z)$ plotted vs.~the
coordinate z along the backbones, for $N = 500$ (a), $N=1000$ (b), and
$N=1500$ (c), and various choices of $\sigma$, as indicated. This
distribution is normalized by choosing $\sum_z \rho(z)=1$}
\label{fig11}
\end{center}
\end{figure}

\begin{figure}
\begin{center}
$\begin{array}{c@{\hspace{0.2in}}c}
\multicolumn{2}{l}{\mbox{(a)}} \\ [-1.5cm] \\
&\psfig{file=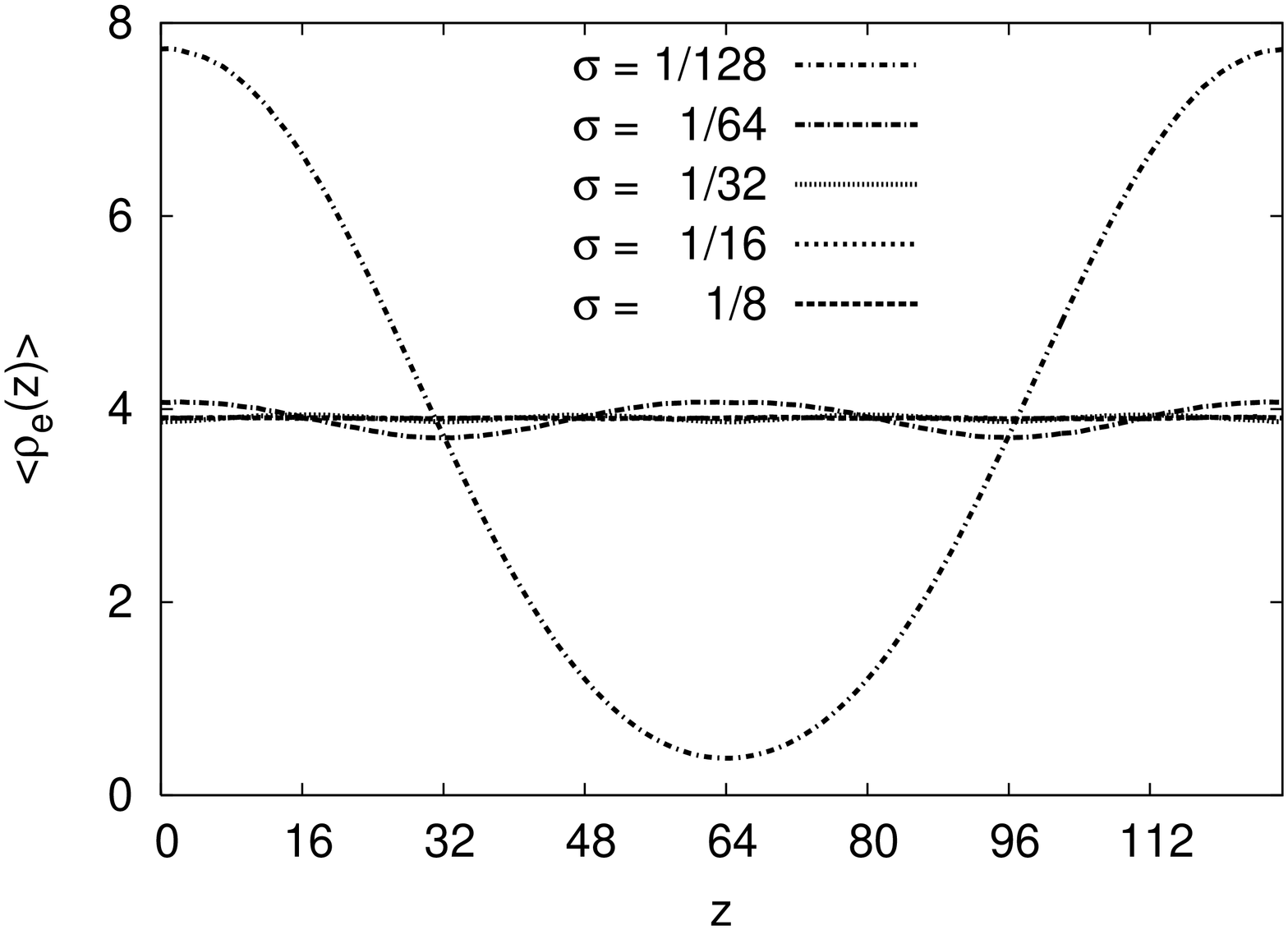,width=5.0cm,angle=270} \\
[0.05cm]\\
\multicolumn{2}{l}{\mbox{(b)}} \\ [-1.5cm] \\
&\psfig{file=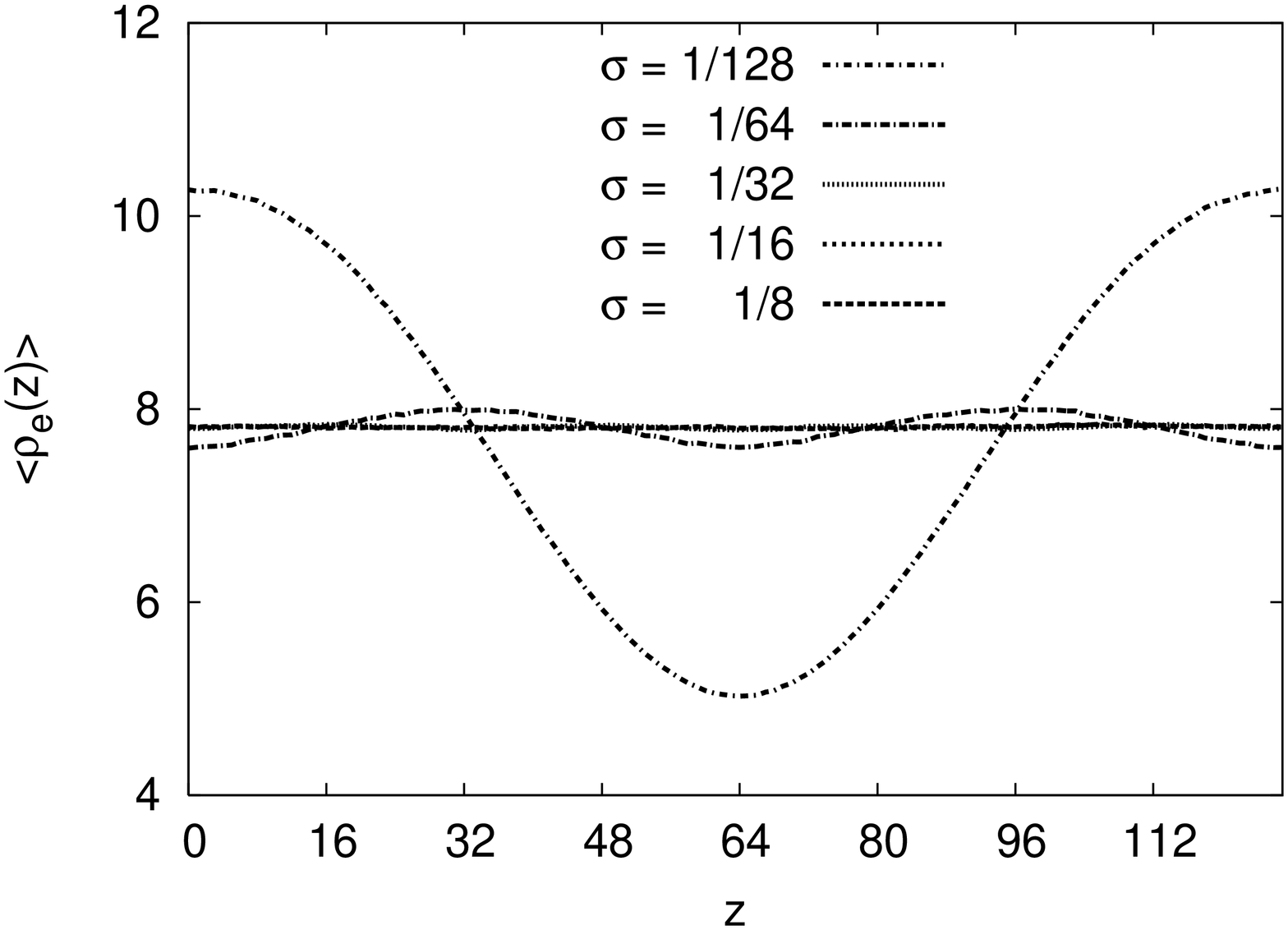, width=5.0cm, angle=270} \\
[0.05cm]\\
\multicolumn{2}{l}{\mbox{(c)}} \\ [-1.5cm] \\
&\psfig{file=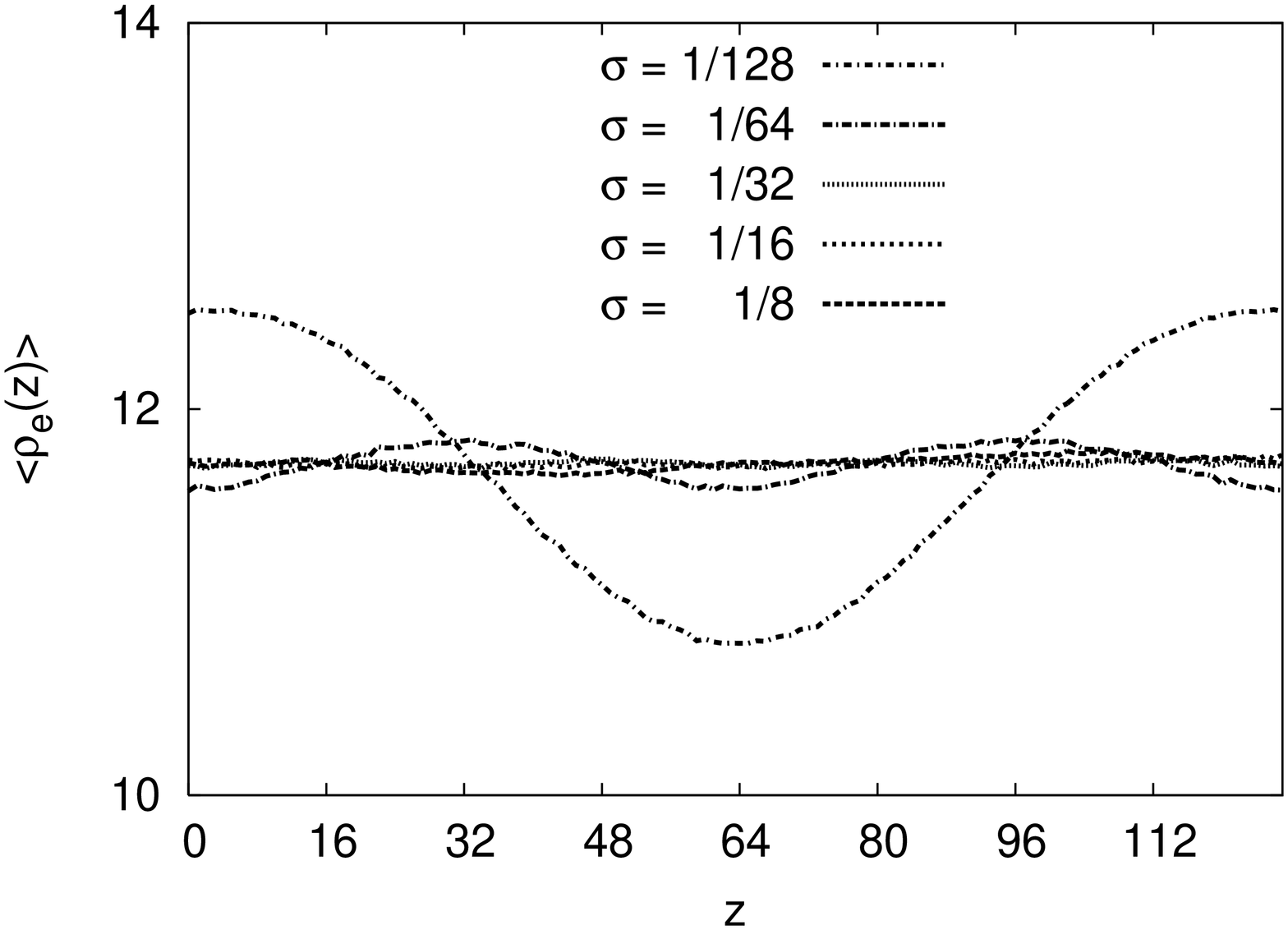, width=5.0cm, angle=270} \\
[1.0cm]
\end{array}$
\caption{Density distributions of chain ends $\rho_e(z)$ plotted
vs.~$z$, for the same choice of parameters as in
Figure~\ref{fig11}.}
\label{fig12}
\end{center}
\end{figure}

\begin{figure}
\begin{center}
$\begin{array}{c@{\hspace{0.2in}}c}
\multicolumn{2}{l}{\mbox{(a)}} \\ [-1.5cm] \\
&\psfig{file=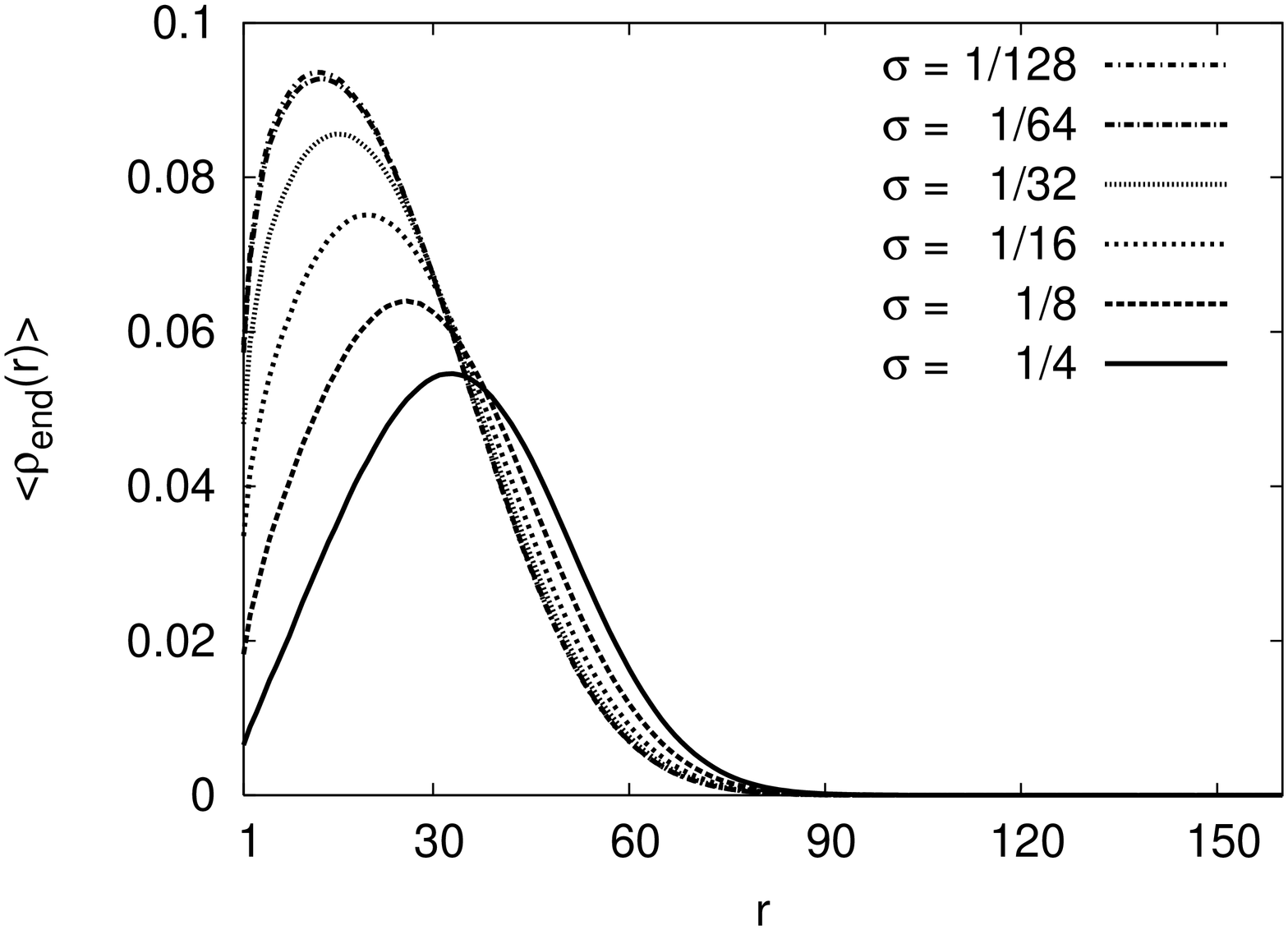,width=5.0cm,angle=270} \\
[0.05cm]\\
\multicolumn{2}{l}{\mbox{(b)}} \\ [-1.5cm] \\
&\psfig{file=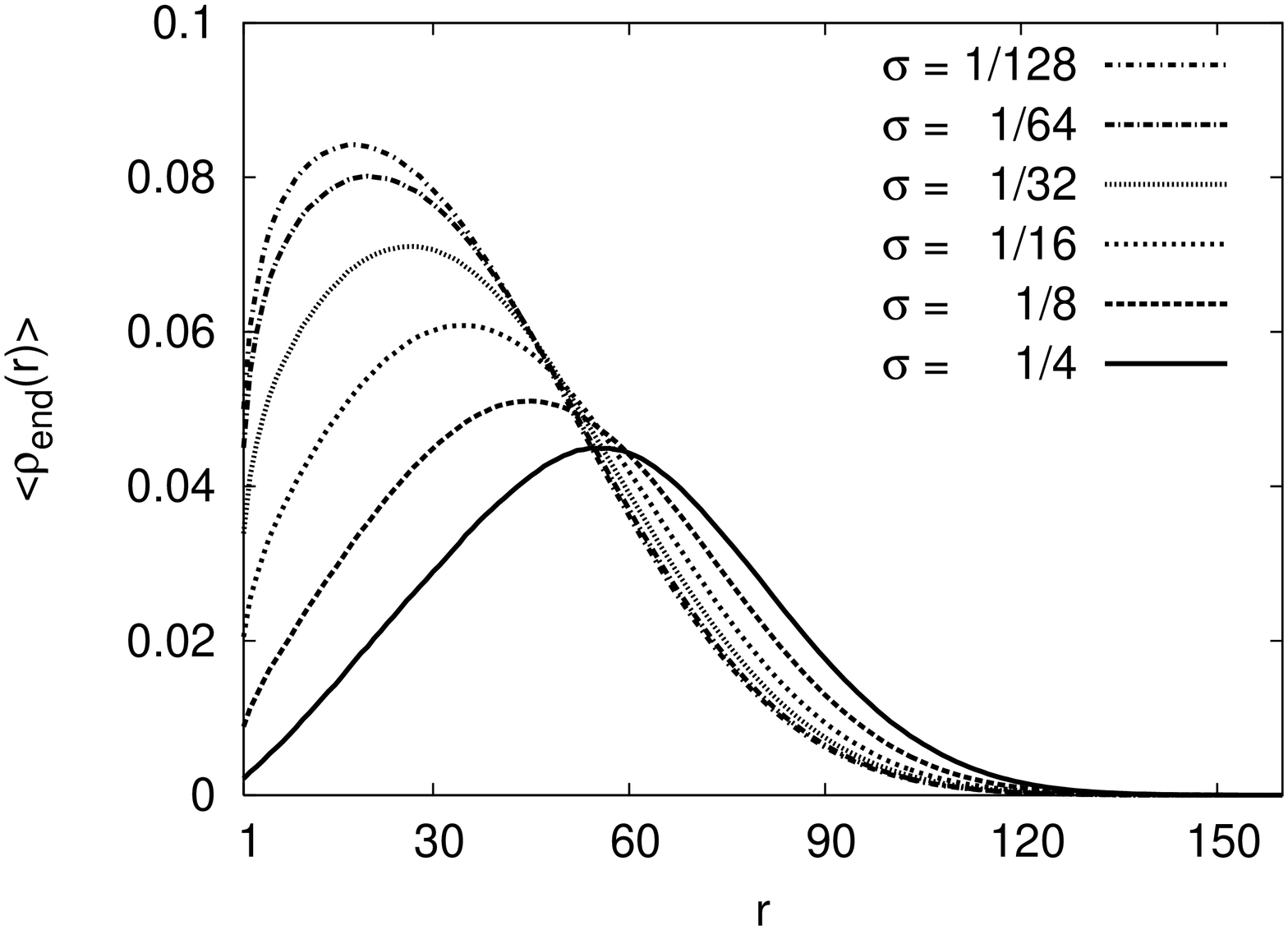, width=5.0cm, angle=270} \\
[0.05cm]\\
\multicolumn{2}{l}{\mbox{(c)}} \\ [-1.5cm] \\
&\psfig{file=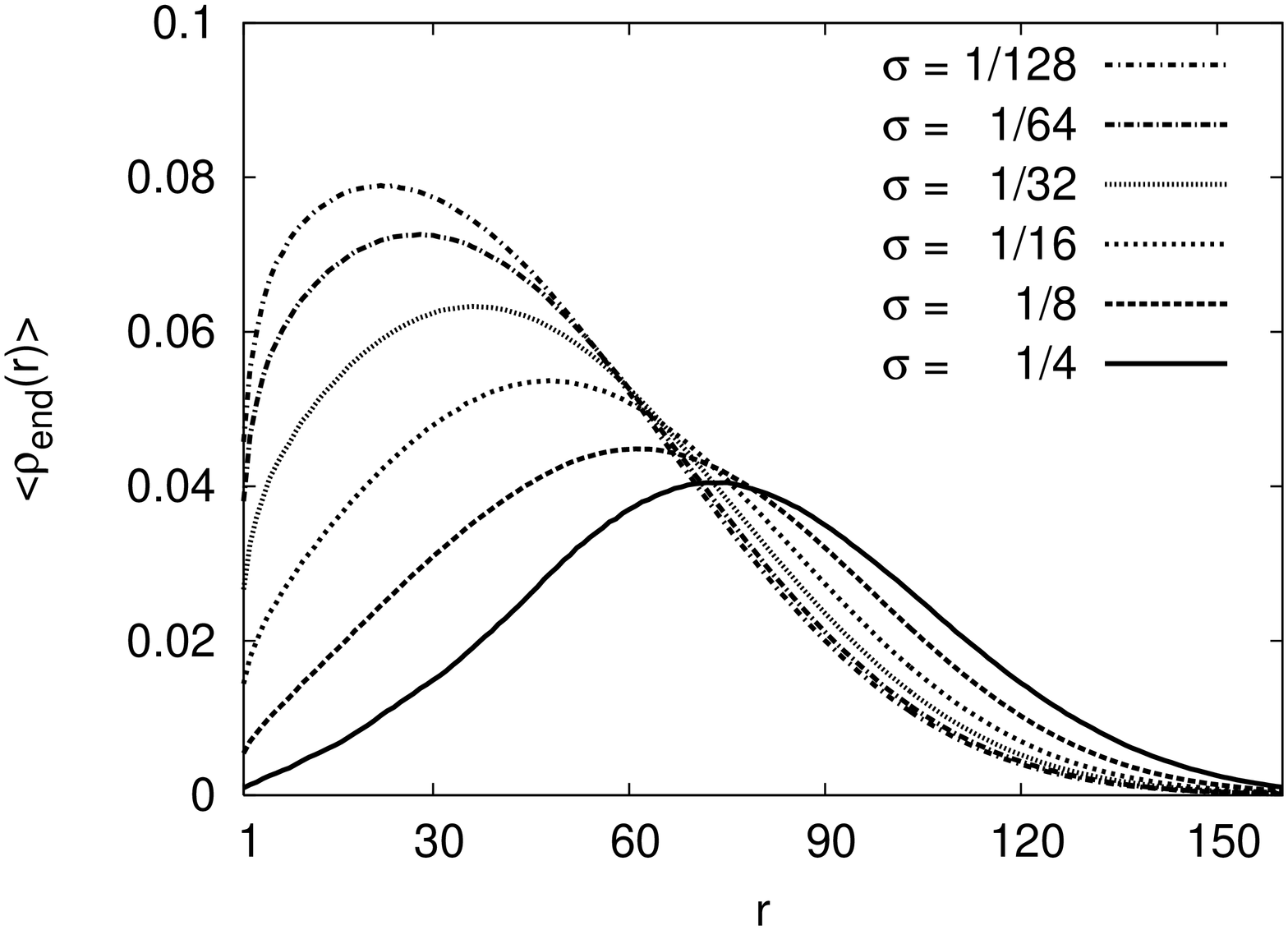, width=5.0cm, angle=270} \\
[1.0cm]
\end{array}$
\caption{Radial density distribution of chain ends $\rho _e(r)$
plotted vs.~$r$, for the same choice of parameters as in
Figures~\ref{fig10}-\ref{fig12}.}
\label{fig13}
\end{center}
\end{figure}

\vskip 1.0truecm
\noindent
{\large \bf IV. Monte Carlo Results for One-Component Bottle-Brush
Polymers}
\vskip 0.5truecm

In Figures~\ref{fig3} and \ref{fig4}, we present our data for
the mean square end-to-end radii and gyration radii components of
the side chains. Note that for each chain configuration a separate
local coordinate system for the analysis of the chain
configuration needs to be used; while the z-axis is always
oriented simply along the backbone, the x-axis is oriented
perpendicular to the z-axis and goes through the center of mass of
the chain in this particular configuration. The y-axis then also
is fixed simply from the requirement that it is perpendicular to
both the x- and z-axes.

For a polymer mushroom (which is obtained if the grafting density
$\sigma$ is sufficiently small) we expect that all chain linear
dimensions scale as $N^\nu$, for sufficiently long chains.
Therefore we have normalized all mean square linear dimensions in
Figures~\ref{fig3} and~\ref{fig4} by a factor $N^{2 \nu}$, using the
theoretical value$^{\textrm {\cite{64,65}}}$ $\nu = 0.588$. However, as we see
from Figures~\ref{fig3}, \ref{fig4} in the range $10 \leq N \leq
10^3$ displayed there, even for the smallest $\sigma$ presented,
where a single side chain is simulated, the shown ratios are not
constant. This fact indicates that corrections to 
scaling$^{\textrm {\cite{64,65}}}$ should not be disregarded, and 
this problem clearly
complicates the test of the scaling predictions derived above.
For the largest value of $\sigma$ included $(\sigma=1/4)$,
the irregular behavior of some of the data (Figure~\ref{fig3}b,c,
Figure~\ref{fig4}b) indicate a deterioration of statistical
accuracy. This problem gets worse for increasing number of
side chains $n_c$.

A further complication is due to the residual finite size effect.
Figure~\ref{fig5} shows a plot of $R^2_\bot = R_x^2+R_y^2$ vs.~the
scaling variable $\zeta = \sigma N^\nu$. One can recognize that
for small $L_b$ but large $N$ and not too large $\sigma$ 
systematic deviations from scaling occur (Figure~\ref{fig5}a), which
simply arise from the fact that an additional scaling variable
(related to $\langle R^2_{z}\rangle ^{1/2} / L_b$) comes into
play when $\langle R^2_{z}\rangle ^{1/2}$ no longer is negligibly
smaller than $L_b$. While for real bottle brush polymers with
stiff backbone effects due to the finite lengths of the backbone
are physically meaningful and hence of interest, the
situation is different in our simulation due to the use of
periodic boundary conditions. The choice of periodic
boundary conditions is motivated by the desire to study the
characteristic structure in a bottle brush polymer far away from
the backbone ends, not affected by any finite size effects.
However, if $\langle R_{gz}^2 \rangle ^{1/2}$ becomes comparable
to $L_b$, each chain interacts with its own periodic images, and
this is an unphysical, undesirable, finite size effect. Therefore
in Figure~\ref{fig5}b, the data affected by such finite size effects
are not included. 
One finds a reasonable data collapse of the scaled mean square 
end-to-end distance when one plots the data versus the scaling
variable $\zeta = \sigma N^\nu$. These results constitute the
first comprehensive test of the scaling relations for bottle brush
polymers, Equations~(\ref{eq21}) - (\ref{eq23}), and the consistency
between the data and the proposed scaling structure in terms of
the variable $\zeta$ is indeed very gratifying. On the other hand,
it is also evident from Figures~\ref{fig5} and \ref{fig6} that only a
mild stretching of the side chains away from the backbone is
observed, and in this region one is still far away from the region
of strong stretching, where the simple power laws
Equations~(\ref{eq8b}), (\ref{eq18}) and (\ref{eq16}) apply. Obviously,
the crossover from the simple mushroom behavior (observed for
$\zeta = \sigma N^\nu \ll 1$) to the strongly stretched
bottle brush (observed for $\zeta \gg 1$) takes at least one decade
of the scaling variable $\zeta$. There is a rather gradual and
smooth crossover rather than a kink-like behavior of the scaling
function. While in the $x$-component at least near
$\zeta=10$ a weak onset of stretching can be recognized,
hardly any evidence for the contraction of the $y$ and
$z$-components is seen. 

In order to test what one would expect if the picture of the
chains as quasi-two-dimensional self-avoiding walks were correct,
we have also studied single chains grafted to a straight backbone
of length $\sigma$, $\sigma =1,1/3,1/5,\ldots 1/63$ which are
confined between two parallel repulsive infinite walls. The
grafting site of the chains was chosen at the site located
symmetrically between the confining walls in the slit \{if the
grafting site is chosen to be the origin of the coordinate system,
the confining hard walls occur at $z=\pm(\sigma ^{-1}+1)/2$\}.
Figure~\ref{fig7} gives log-log plots of the gyration radii
components of such chains confined to such disk-like slits, which
we denote as $R_{g \alpha,\; \textrm{slit}}$, $\alpha =x,y$ and
$z$, in order to distinguish then from the actual gyration radii
components of the side chains in a bottle brush polymer. One can
clearly see that $\langle R^2_{gx,\textrm{slit}}\rangle$, $\langle
R^2_{gy,\textrm{slit}}\rangle$ scale as $N^{2 \nu_2} = N^{3/2}$ for
large $N$, as it must be. These data are very similar to data for
chains confined between repulsive walls without a grafting to a
piece of a rigid backbone,$^{\textrm {\cite{97}}}$ of course. In the
corresponding scaling plot (Figure~\ref{fig8}) one can see that both
$R_{gx,\textrm{slit}}^2$ and $R_{gy,\textrm{slit}}^2$ have the
simple crossover scaling behavior$^{\textrm {\cite{45,98}}}$
\begin{equation}\label{eq27}
\langle R_{gx,\textrm{slit}}^2\rangle = N^{2\nu} \tilde{f}_x
(\sigma N^\nu),\enspace \langle R_{gy,\textrm{slit}}^2\rangle =
N^{2\nu} \tilde{f}_y (\sigma N^\nu)
\end{equation}
with $\tilde{f}_x(\zeta) \propto \tilde{f}_y(\zeta) \propto \zeta
^{2(\nu_2-\nu)/\nu}$ with $\nu _2 =3/4$, as expected, and seen in
related previous work.$^{\textrm {\cite{97,99}}}$ Obviously, although we use in
Figures~\ref{fig7} and \ref{fig8} the same range of $N$ and $\sigma$ as
in Figures~\ref{fig4} and \ref{fig6}, the behavior is rather
different. As expected from our scaling analysis presented above,
the confinement that a chain experiences in a bottle brush due to
the presence of the other chains is much weaker than the
confinement of a chain in the equivalent disk-like sector between
confining walls. This fact is demonstrated very directly in
Figure~\ref{fig9}, where the ratios $\langle R_{g \alpha }^2\rangle
/ \langle R_{g \alpha, \textrm{slit}}^2 \rangle$ are plotted 
vs.~$N$ for the corresponding values of $\sigma$. If the hypothesis of
quasi-two dimensional confinement were valid, we would expect
these ratios to be constants. Obviously, this is not the case.

As a final point of this section, we discuss the distribution of
monomers (Figures~\ref{fig10} and \ref{fig11}) and chain ends
(Figures~\ref{fig12} and \ref{fig13}) in the
simulated model for the bottle brush polymer. Unlike corresponding
radial density distributions for the off-lattice bead-spring model
of Murat and Grest,$^{\textrm {\cite{22}}}$ where for small distances close to
the backbone a kind of ``layering'' was observed, we see a rather
smooth behavior also for small distances (Figure~\ref{fig10}). For
larger distances, the behavior is qualitatively very similar.
Again we fail to provide a clear-cut evidence for the predicted
power law decay, Equation~(\ref{eq6}). Note, however, that this power
law is supposed to hold only in the strong stretching limit, where
Equation~(\ref{eq8b}) is observable (which we do not verify either), and
in addition the stringent condition $1 \ll r \ll h$ needs to be
obeyed, to have this power law. Although our simulations encompass
much longer chains than every previous work on bottle brushes, we
clearly fail to satisfy this double inequality.

Turning to the distribution along the backbone (Figure~\ref{fig11}),
the periodicity due to the strictly periodic spacing of grafting
sites is clearly visible. If desired, one could also study a
random distribution of grafting sites along the backbone, of
course, but we have restricted attention to the simplest case of a
regular arrangement of grafting sites only. The distribution of
chain ends $\rho(z)$ exhibits an analogous periodicity for small
values of $\sigma$ only, while 
$\rho _e(z)$ is approximately constant (Figure~\ref{fig12})
for larger values of $\sigma$.

Most interesting is the radial distribution of chain ends
(Figure~\ref{fig13}). One can see an increasing depression of
$\rho_e(r)$ for small $r$ when $\sigma$ increases. Again these
data are similar to the off-lattice results of Murat and 
Grest.$^{\textrm {\cite{22}}}$ In no case do we see the 
``dead zone'' predicted by the
self-consistent field theory in the strong stretching 
limit,$^{\textrm {\cite{21,24}}}$ however.

\vskip 1.0truecm
\noindent
{\large \bf V. Phase Separation in Two-Component Bottle Brushes:
Theoretical Background}
\vskip 0.5truecm

We still consider a bottle brush polymer with a strictly rigid
straight backbone, where at regularly distributed grafting sites
(grafting density $\sigma$) side chains of length $N$ are
attached, but now we assume that there are two chemically
different monomeric species, A and B, composing these side chains
with $N_A=N_B=N$. These systems have found recent attention,
suggesting the possibility of intramolecular phase 
separation.$^{\textrm { \cite{56,57,58}}}$ In a binary system pairwise interaction energies
$\epsilon _{AA}$, $\epsilon _{AB}$, and $\epsilon _{BB}$ are expected, and
consequently phase separation between A and B could be driven by
the Flory-Huggins parameter

\begin{equation}\label{eq28}
\chi = z_c \epsilon / k_BT, \quad \epsilon =
\epsilon_{AB}-(\epsilon_{AA} + \epsilon_{BB})/2
\end{equation}
where $z_c$ is the coordination number of the lattice, like in the
Flory-Huggins lattice model of phase separation in a binary
polymer blend.$^{\textrm { \cite{44,45,50,51,52}}}$ However, since the side
chains are grafted to the backbone, only intramolecular phase
separation is possible here. Actually, the energy parameter
$\epsilon$ suffices for dense polymer blends or dense block
copolymer melts, where no solvent is present.$^{\textrm { \cite{44,45,50,51,52}}}$ 
For the problem of intramolecular phase
separation in a bottle brush, the solvent cannot be disregarded
and hence the enthalpy of mixing rather is written as.$^{\textrm {\cite{52,56}}}$

\begin{eqnarray}\label{eq29}
E_{\textrm{mix}}/k_BT= \int \frac {dV}{v} &[&\chi_{AS} \phi_A
(\vec{r}) \phi _S (\vec{r}) + \chi_{BS} \phi_B (\vec{r}) \phi_S
(\vec{r}) \nonumber \\
&+& \chi_{AB} \phi_A (\vec{r}) \phi_B (\vec{r})],
\end{eqnarray}
where $v$ is the volume per monomer, $\phi_A(\vec{r})$,
$\phi_B(\vec{r})$ and $\phi_S(\vec{r})$ are the local volume
fractions of monomers of types A and B, and the solvent density,
respectively, and three pairwise interaction parameters
$\chi_{AS},\chi_{BS}$, and $\chi_{AB}$ enter. The latter
$\chi_{AB}$ corresponds to the $\chi$-parameter written in
Equation~(\ref{eq28}), while the former two control the solvent quality for
polymers A and B, respectively. Of course, using the
incompressibility condition
\begin{equation}\label{eq30}
\phi_A(\vec{r})+ \phi_B(\vec{r})+\phi_S (\vec{r})=1
\end{equation}
the solvent density can be eliminated from the problem (but one
should keep in mind that in the free energy density there is an
entropy of mixing term $\phi_S(\vec{r}) \ln \phi_S(\vec{r})$
present$^{\textrm { \cite{56}}}$). 
In the framework of the lattice model studied
here, $v \equiv 1$ and solvent molecules are only implicitly
considered, identifying them with vacant sites. For simplicity,
the following discussion considers only the most symmetric case,
where $\chi_{AS}=\chi_{BS}$, and the numbers of A chains and B
chains are equal.
 
\begin{figure}
\begin{center}
\psfig{file=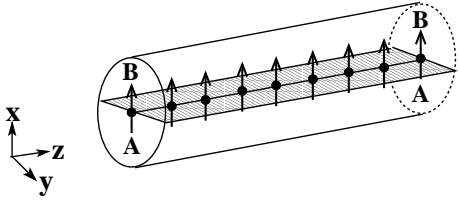,width=6.0cm,angle=0}
\caption{Schematic
description of perfect phase separation of side chains in a
cylindrical binary (A,B) bottle brush polymer into a ``Janus
cylinder'' structure, where the A-chains occupy the lower
hemicylinder and the B-chains occupy the upper hemicylinder. The
interface between B-rich and A-rich phases is assumed to be
oriented in the yz-plane. The figure indicates the
characterization of this order in terms of a local order
parameter, namely the vector $\vec{\psi}(z)$ oriented normal to the
interface at every grafting site. The absolute value of this vector
characterizes a suitable average of the local phase separation in
a disk of the cylinder located at z (see text). For perfect order
$\langle \mid \vec{\psi}(z) \mid \rangle = 1$ and $\vec{\psi}(z)$ is
oriented along the same axis for the whole bottle brush polymer.}
\label{fig14}
\end{center}
\end{figure}

Stepanyan et al.$^{\textrm { \cite{56}}}$ 
used Equations~(\ref{eq29}) and (\ref{eq30}),
as the starting point of a self-consistent field calculation, adding
conformational free energy contributions accounting for the
entropy associated with the stretching of Gaussian chains. It is
found that when $\chi_{AB}$ exceeds a critical value $\chi
^*_{AB}$, intramolecular phase separation of ``Janus
cylinder''-type occurs (Figure~\ref{fig14}). Then an interface is
formed, containing the backbone of the bottle brush, such that
there is an excess of A-monomers below the interface and an excess
of B-monomers above the interface (or vice versa). Assuming that
at the position $z$ of the backbone the interface is oriented 
in x-direction, we can describe this 
Janus-type phase separation in
terms of an one-dimensional order parameter density
\begin{equation}\label{eq31}
\psi(z)=(n_B^+ - n_A^++n_A^- - n_B^-)/(n_A^++n_A^- + n_B^+
+n_B^-),
\end{equation}
where $n_A^\pm$ and $n_B^\pm$ are the numbers of A(B) monomers in the
interval $[z,z+dz]$ with $x>0 (n_A^+,n_B^+)$ and
$x<0(n_A^-,n_B^-)$, respectively. Since we shall see below that
the orientation of the interface is an important degree of
freedom, we may consider $\psi(z)$ as the absolute value of a
vector order parameter $\vec{\psi}(z)$, and the direction of
$\vec{\psi}(z)$ is defined such that $\psi(z)$ takes a maximum.
However, Stepanyan et al.$^{\textrm { \cite{56}}}$ did not consider the
possibility of an inhomogeneity along the z-axis, and they also did
not derive how the order parameter depends on the parameters $N$,
$\sigma$, and the various $\chi$ parameters of the problem
\{Equation~(\ref{eq29})\}. Stepanyan et al.$^{\textrm {\cite{56}}}$ also assume that
the distribution of chain ends is a delta function at the radius
(or ``height'' $h$) respectively) of the bottle brush, where $h$
is given by $h/a=(\sigma a)^{1/4}N^{3/4}$ in the good solvent
regime. They then find that phase separation occurs for
\begin{equation}\label{eq32}
\chi_{AB}^* \propto 1/\sqrt{N},
\end{equation}
but argue that this result holds only for a ``marginal solvent''
rather than a good solvent, due to the mean-field approximation
used which neglects that inside a blob all binary contacts are
avoided, estimating the number of contacts simply proportional to
the product $\phi_A\phi _B$, cf. Equation~(\ref{eq29}), and neglecting
the correlations due to excluded volume. The regime of marginal
solvents is reached near the $\theta$-point (which occurs for $1-2
\chi_{AS}=1-2 \chi_{BS}=0$), and requires that$^{\textrm { \cite{56}}}$
\begin{equation}\label{eq33}
0<1-2\chi_{AS}<N^{-1/3}
\end{equation}
where again prefactors of order unity are omitted.

\begin{figure}
\begin{center}
\psfig{file=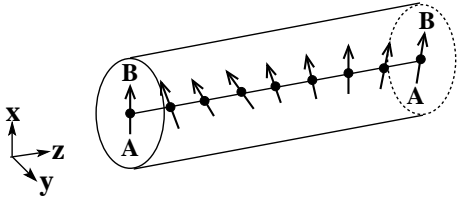,width=6.0cm,angle=0}
\caption{Same as
Figure~\ref{fig14}, but allowing for a long wavelength variation of
the vector $\vec{\psi}(z)$ characterizing the local interface
orientation in the Janus cylinder.}
\label{fig15}
\end{center}
\end{figure}

Stepanyan et al.$^{\textrm { \cite{56}}}$ extend Equation~(\ref{eq32}) by a simple
scaling-type argument, stating that the transition from the mixed
state to the separated state occurs when the energy of the A-B
contacts per side chain, $\Delta E$, is of the order of $k_BT$.
They estimate this energy as
\begin{equation}\label{eq34}
\Delta E / k_BT \approx Np(\bar{\phi}) \chi_{AB},
\end{equation}
where $p(\bar{\phi} )$ is the probability of the A-B contact, which
depends on the average volume fraction $\bar{\phi}$ of the monomers
inside the brush. According to the mean field theory, $p
(\bar{\phi})\propto \bar{\phi}$, where $\bar{\phi} = N/h^2$, 
$h$ being the radius of
the cylindrical brush. Using $h \propto N ^{3/4}$
(Equation~(\ref{eq8b})) one finds $p(\bar{\phi})\propto N^{-1/2}$ 
and using this result in Equation~(\ref{eq34}) implies
$\Delta E/k_BT \propto N^{1/2}\chi_{AB}$, and from $\Delta E /k_BT=1$
we recover Equation~(\ref{eq32}).

   The merit of this simple argument is that it is readily extended
to other cases: e.g., for a $\theta$-solvent, we have$^{\textrm {\cite{18}}}$
$h\propto N^{2/3}$ and hence
$p(\bar{\phi})\propto N^{-1/3}$, yielding$^{\textrm { \cite{56}}}$
\begin{equation}\label{eq35}
\chi^* _{AB} \propto N^{-2/3}\;, \quad \theta-{\textrm{solvent}}.
\end{equation}
For the poor solvent case, the bottle brush should collapse to a
cylinder densely filled with monomers, and hence $h \propto
N^{1/2}$, $\bar{\phi} =1$, and thus
\begin{equation}\label{eq36}
\chi^*_{AB} \propto N^{-1}\;,\quad {\textrm{poor solvent}} .
\end{equation}
Note that Equation~(\ref{eq36}) is the same relation as for a dense bulk
polymer blend or block copolymer melt, 
respectively.$^{\textrm {\cite{44,45,50,51,52}}}$

\begin{figure}
\begin{center}
\psfig{file=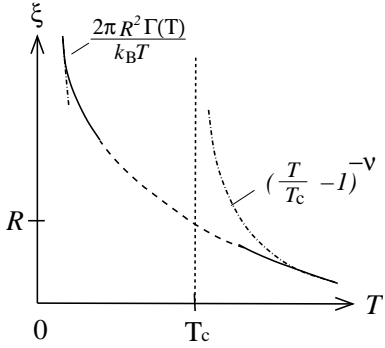,width=5.0cm,angle=0}
\caption{Schematic
sketch of the temperature dependence of the correlation length
$\xi$ of the ferromagnetic XY-model in a cylinder of radius R.}
\label{fig16}
\end{center}
\end{figure}

The situation is more subtle in the good solvent case, however,
since there the probability of contact is no longer given by the
mean field result $p(\bar{\phi}) \propto \bar{\phi}$ but 
rather by$^{\textrm { \cite{45}}}$
\begin{equation}\label{eq37}
p(\bar{\phi}) \propto \bar{\phi} ^{5/4} \propto N^{-5/8}\;,
\end{equation}
where the Flory approximation $\nu = 3/5$ was again used, with
$\bar{\phi} \propto N ^{-1/2}$ when $h \propto N^{3/4}$.
Equations~(\ref{eq34}) and (\ref{eq37}) now imply
\begin{equation}\label{eq38}
\chi _{AB}^* \propto N^{-3/8}\;, \quad {\textrm{good solvent}}
\end{equation}
However, these crude estimates do not suffice to prove that for
$\chi_{AB}>\chi_{AB}^*$ true long range order of this ``Janus
cylinder'' type (Figure~\ref{fig14}) is established. Hsu et
al.$^{\textrm {\cite{58}}}$ suggested that there is also the need to consider
variations of the direction of the order parameter $\vec{\psi}(z)$
along the z-direction (Figure~\ref{fig15}). It was argued that for
any finite side chain length $N$ also the cylinder radius (or
brush ``height'') $h$ is finite, and hence the system is
one-dimensional. In one-dimensional systems with short range
interactions at nonzero temperatures no long range order is
possible.$^{\textrm{\cite{100,101}}}$ The situation depicted in
Figure~\ref{fig15} is reminiscent of the one-dimensional XY-model of a
chain of spins on a one-dimensional lattice where each spin at
site $i$ is described by an angle $\varphi _i$ in the xy-plane, with
$0\leq i \leq 2 \pi$, and where neighboring spins are coupled. This
coupling is described by the Hamiltonian
\begin{equation}\label{eq39}
{\mathcal{H}} = - J \sum_i \cos (\varphi_{i+1}- \varphi
_i)=-J \sum \limits _i \vec{S}_{i+1}\cdot \vec{S}_i
\end{equation}
when $\vec{S}_i= (\cos \varphi _i,\sin \varphi _i)$ is a unit
vector in the xy-plane. While mean field theory predicts that
ferromagnetic order occurs along the chain, for ferromagnetic
exchange $J >0$ and temperatures $T$ less than the critical
temperature $T^{MF}_c$ which is of the order of $J/k_B$, the exact
solution of this model, Equation~(\ref{eq39}), shows$^{\textrm {\cite{100,101}}}$
that $T_c=0$, since ferromagnetic long range order is unstable
against long wavelength fluctuations. One can show that the
spin-spin correlation function for large $z=a(j-i) $ decays to
zero exponentially fast,
\begin{equation}\label{eq40}
\langle \vec{S}_i \cdot \vec{S}_j \rangle \propto \exp [-z/\xi],
\quad z \rightarrow \infty
\end{equation}
where $a$ is the lattice spacing of this spin chain. The correlation
length $\xi$ gradually grows as the temperature is lowered,
\begin{equation}\label{eq41}
\xi=2a(J/k_BT)
\end{equation}
Equation~(\ref{eq41}) is at variance with mean field theory, which
rather would predict$^{\textrm {\cite{100}}}$ (the index MF stands for
``mean field'' throughout)
\begin{equation}\label{eq42}
\xi_{MF}\propto (T/T_c^{MF}-1)^{- \nu_{MF}}\;,\quad \nu_{MF}=
1/2\quad .
\end{equation}
This critical divergence at a nonzero critical temperature
$T_c^{MF}$ is completely washed out by the fluctuations: rather
than diverging at $T_c^{MF}$, the actual correlation length $\xi$
\{Equation~(\ref{eq41})\} at $T_c^{MF}$ still is only of the order of
the lattice spacing.

This consideration can be generalized to spin systems on lattices
which have a large but finite size in $(d-1)$ dimensions and are
infinite in one space dimensions only.$^{\textrm {\cite{102,103}}}$ 
E.g., when
we consider an infinitely long cylinder of cross section $\pi R^2$
we expect that Equations~(\ref{eq41}), (\ref{eq42}) are replaced by the
finite size scaling relation$^{\textrm { \cite{104,105}}}$
\begin{equation}\label{eq43}
\xi = \xi _\infty \;\tilde{\xi}(R/\xi_\infty)\;, \enspace \tilde{\xi}(\zeta \gg
1)=1\;,\enspace \tilde{\xi}(\zeta \ll 1)\propto \zeta \;,
\end{equation}
where $\xi _\infty$ is the correlation length of the XY model
on a lattice which is infinitely large in all $d=3$ directions of space.

\begin{equation}\label{eq44}
\xi_\infty \propto (T/T_c-1)^{-\nu}\;,
\end{equation}
where again $k_BT_c \propto J$ (but with a smaller constant of
proportionality than that in the relation $k_BT_c^{MF} \propto
J)$, and $\nu$ is the correlation length exponent of the XY model
($\nu \approx 0.67$).$^{\textrm{\cite{106}}}$ Equation~(\ref{eq43}) describes a
smooth crossover of the ferromagnetic correlation length
describing spin correlations along the axis of the cylinder from
bulk, Equation~(\ref{eq44}), to a quasi-one-dimensional variation. For
$T \ll T_c$ the correlation length resembles Equation~(\ref{eq41}),
since$^{\textrm {\cite{102,103}}}$
\begin{equation}\label{eq45}
\xi= 2 \pi \Gamma (T) R^2/k_BT \approx 2 \pi (R^2/a)(J/k_BT),
\quad T \rightarrow 0.
\end{equation}
the ``helicity modulus'' (also called ``spin wave stiffness'') 
$\Gamma(T)$ is
of order $J$ for $T \rightarrow 0$ and shows a critical vanishing
as $T \rightarrow T_c$ from below, in the thermodynamic limit
$R\rightarrow \infty$. However, for finite $R$ there is a finite
size rounding of this singularity of $\Gamma (T \rightarrow T_c)$,
such that $\Gamma(T = T_c)$ is of order $1/R$, and hence a smooth
crossover to Equation~(\ref{eq43}) occurs near $T_c$. Figure~\ref{fig16}
gives a qualitative account of this behavior. For more details of
this finite size crossover we refer to the 
literature.$^{\textrm {\cite{102,103}}}$ 
But we suggest a qualitatively similar behavior for
the domain size $\xi$ of segregated A-rich and B-rich domains in
binary bottle brush polymers. So, when we study the correlation
function of the order parameter considered in Equation~(\ref{eq31})
\begin{equation}\label{eq46}
G_\psi(z)= \langle \vec{\psi}(z') \cdot \vec{\psi}(z'+z)\rangle
\propto \exp(-z/\xi _\psi)
\end{equation}
we expect that the correlation length $\xi_\psi$ describing the
local phase separation in the direction along the backbone of the
bottle brush polymer remains of order unity as long as
$\chi_{AB}^{-1}$ exceeds $\chi_{AB}^{*\;-1}$ distinctly. For
$\chi_{AB}$ near $\chi^*_{AB}$ we expect that $\xi_\psi $ becomes
of order $h$, the radius of the bottle brush. For $\chi_{AB}\gg
\chi_{AB}^{*}$, we expect $\xi_\psi \propto h^2\chi_{AB}$, by
analogy with Equation~(\ref{eq45}). Unfortunately, the test of those
predictions even in the poor solvent case where $\chi_{AB}^{*\; -1}$
is largest, is rather difficult, since the prefactor in the
relation $\chi_{AB}^* \propto 1/N$ is unknown.

\begin{figure*}
\begin{center}
$\begin{array}{c@{\hspace{0.4in}}c}
\multicolumn{2}{l}{\mbox{(a)}} \\ [-1.5cm] \\
&\psfig{file=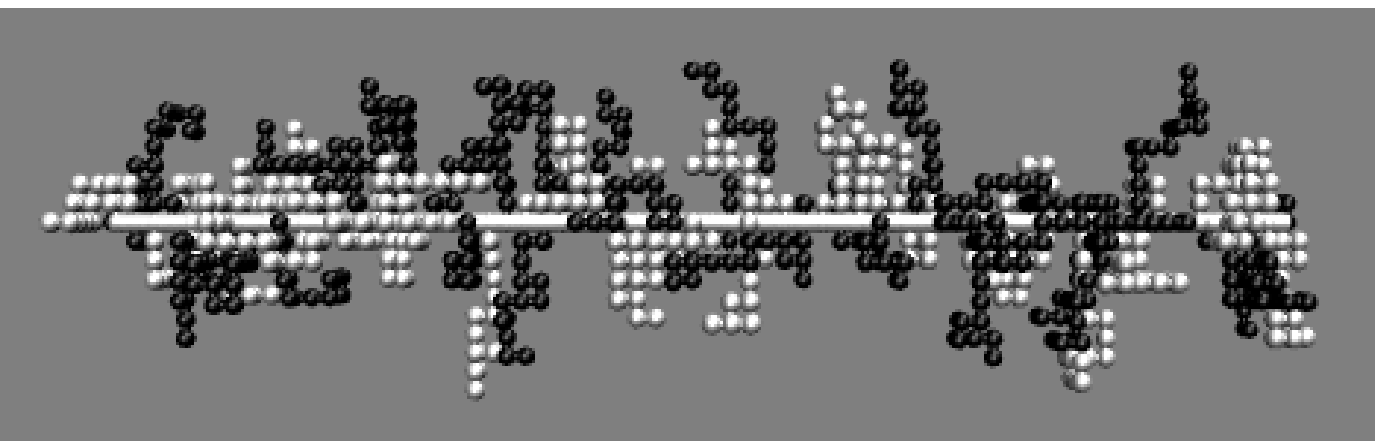,width=8.0cm,angle=0} \\
[0.5cm]\\
\multicolumn{2}{l}{\mbox{(b)}} \\ [-1.5cm] \\
&\psfig{file=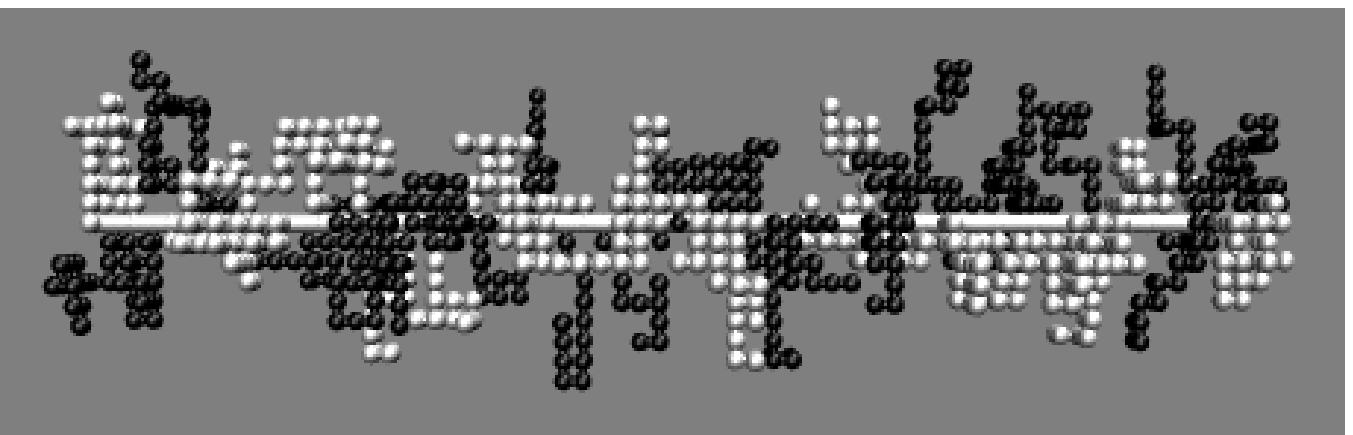, width=8.0cm, angle=0} \\
[0.5cm]\\
\multicolumn{2}{l}{\mbox{(c)}} \\ [-1.5cm] \\
&\psfig{file=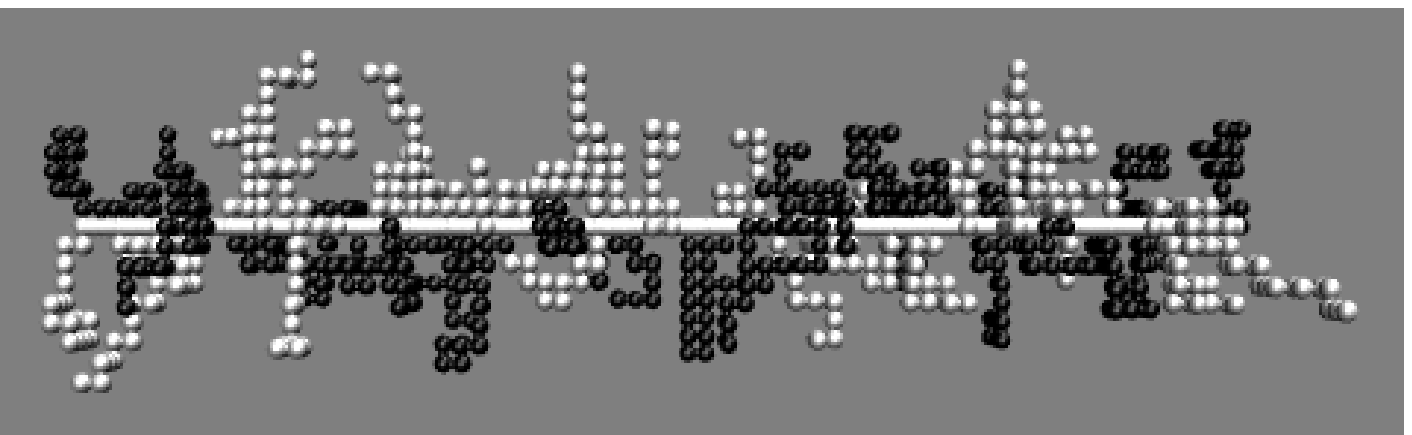, width=8.0cm, angle=0} \\
[1.0cm]
\end{array}$
\caption{Snapshot pictures of bottle brush configurations for
$L_b=64,\; q =1$, $N=18,\; \sigma = 1$ and three choices of
$q_{AB},\; q_{AB} = 1.0 $ (a), $q_{AB}=0.4$ (b) and $q_{AB}=0.1$
(c). Monomers A, monomers B, and a straight rigid backbone are shown
in black, light gray, and white colors, respectively.}
\label{fig17}
\end{center}
\end{figure*}

\begin{figure}
\begin{center}
$\begin{array}{c@{\hspace{0.2in}}c}
\multicolumn{2}{l}{\mbox{(a)}} \\ [-1.5cm] \\
&\psfig{file=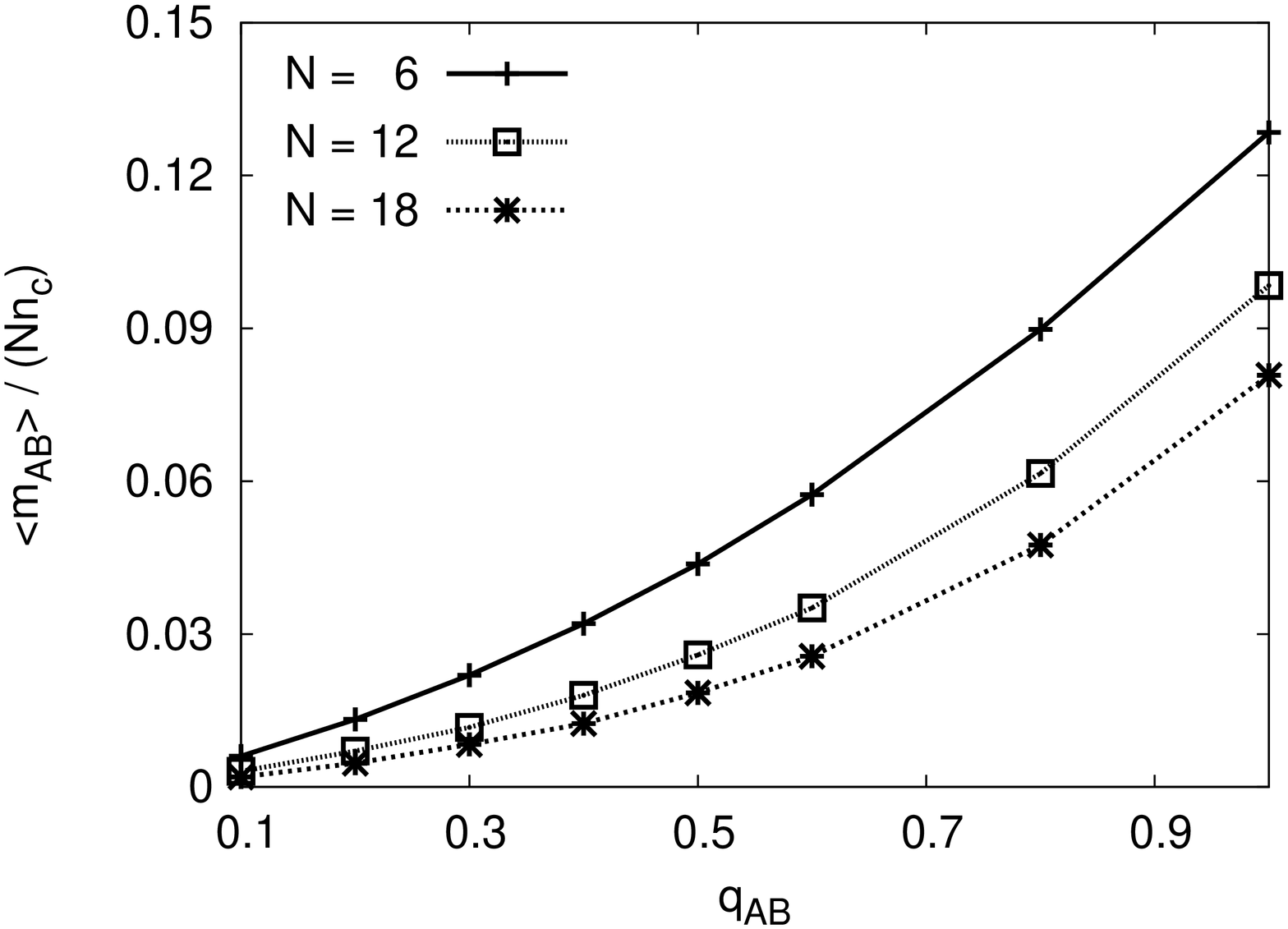,width=5.0cm,angle=270} \\
[0.05cm]\\
\multicolumn{2}{l}{\mbox{(b)}} \\ [-1.5cm] \\
&\psfig{file=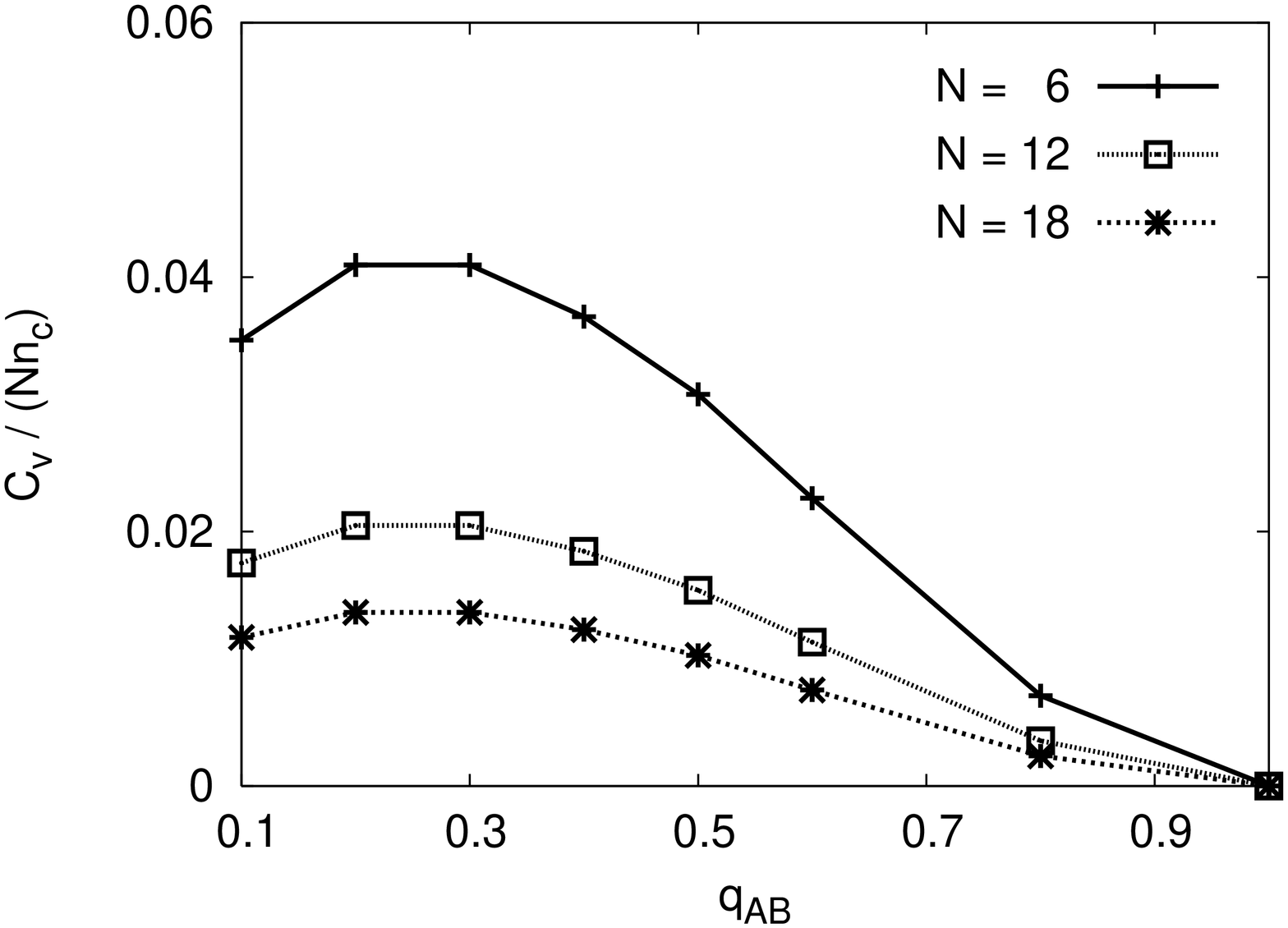, width=5.0cm, angle=270} \\
[1.0cm]
\end{array}$
\caption{(a) Average number of AB pairs per monomer $\langle
m_{AB}\rangle /(Nn_c)$ plotted vs.~$q_{AB}$ for side chain lengths
$N=6$, $12$, and $18$. All data refer to $\sigma = 1$, $L_b=64$. (b)
Specific heat per monomer, $c_v/(Nn_c)$ plotted vs.~$q_{AB}$ for
$N=6$, $12$ and $18$.}
\label{fig18}
\end{center}
\end{figure}

\begin{figure}
\begin{center}
$\begin{array}{c@{\hspace{0.2in}}c}
\multicolumn{2}{l}{\mbox{(a)}} \\ [-1.5cm] \\
&\psfig{file=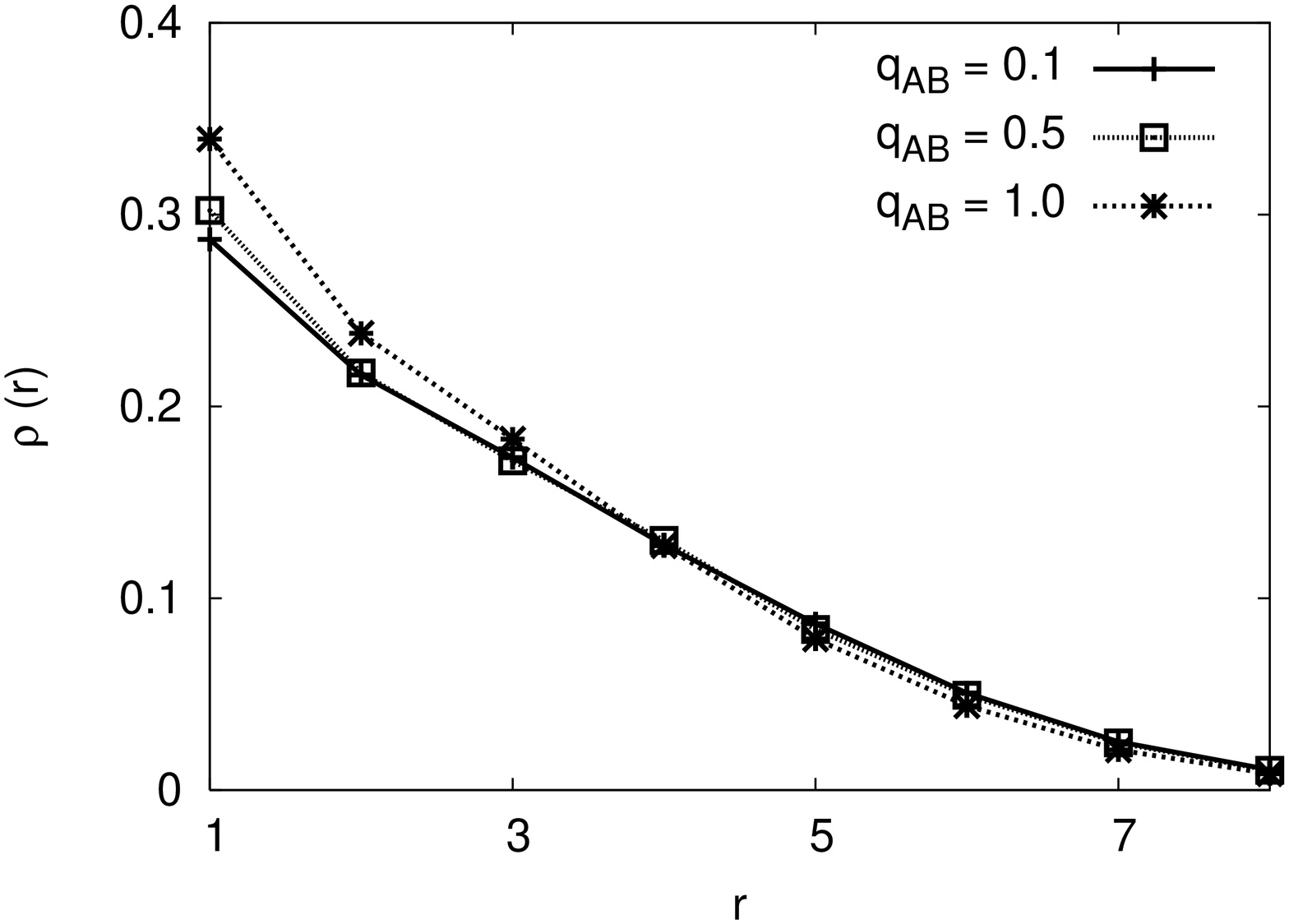,width=5.0cm,angle=270} \\
[1.0cm]\\
\multicolumn{2}{l}{\mbox{(b)}} \\ [-1.5cm] \\
&\psfig{file=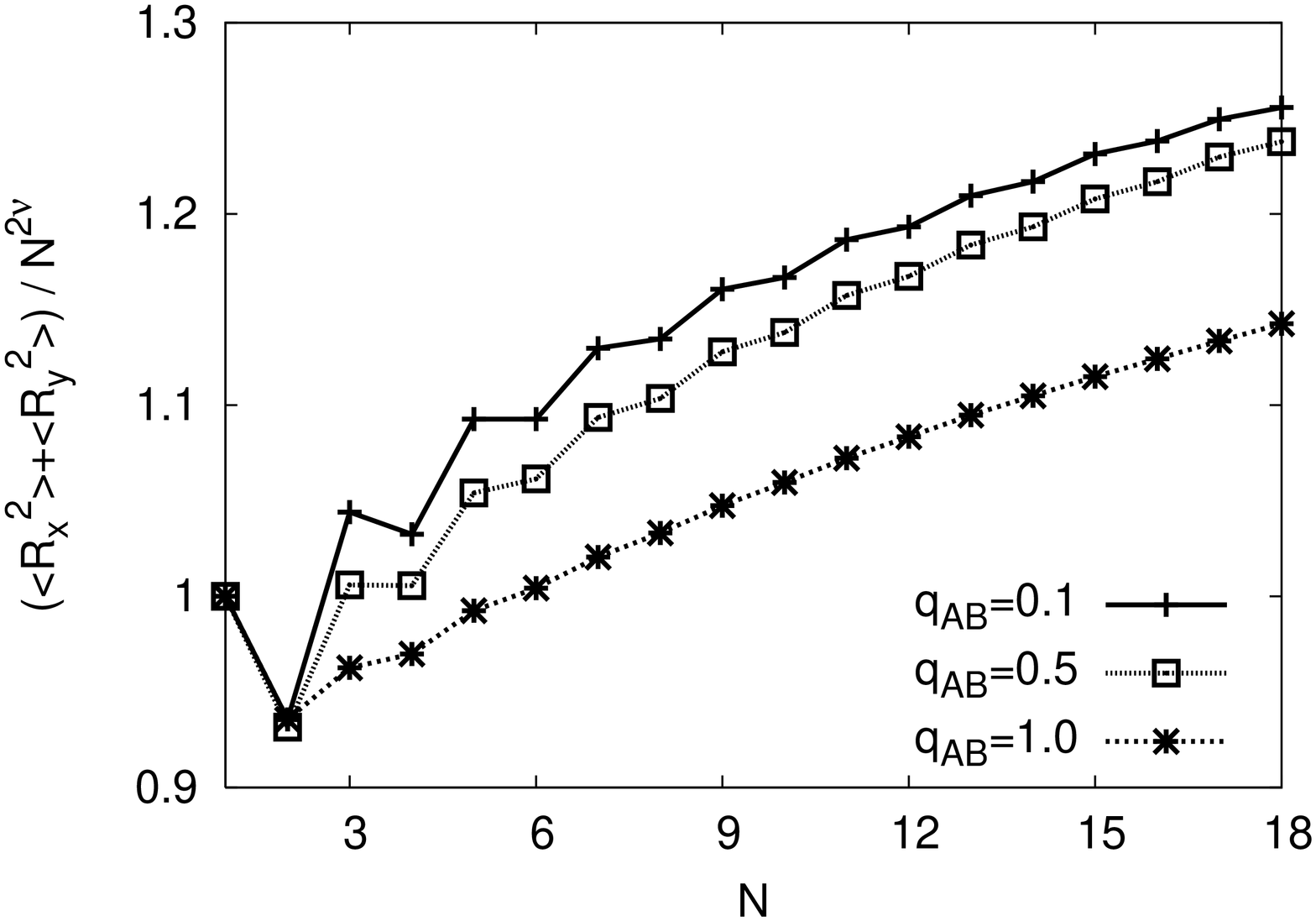, width=5.0cm, angle=270} \\
[1.0cm]
\end{array}$
\caption{(a) Radial density distribution $\rho(r)$ plotted vs.~$r$
for $L_b=64$, $N=18$, $q=1$, $\sigma = 1$, $f=1$ and various
$q_{AB}$. (b) Normalized transverse mean square end-to-end
distance $(\langle R_x^2\rangle + \langle R^2_y \rangle) /N^{2\nu}$
plotted vs.~$N$, for $L_b=64,\; q=1,\; \sigma = 1,\; f=1$, and
three choices of $q_{AB}$ as shown.}
\label{fig19}
\end{center}
\end{figure}

\begin{figure}
\begin{center}
\psfig{file=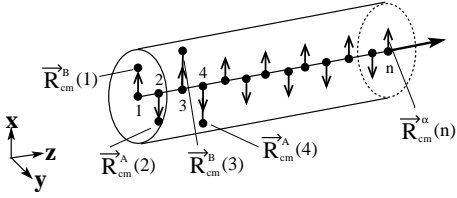,width=6.0cm,angle=0}
\caption{Vectors $\vec{R}_{\rm cm}^\alpha(n)$ from the grafting site $n$ to the
xy-component of the center of mass of the respective chain, and
corresponding unit vectors (denoted by arrows). For a perfectly
phase separated structure with the interface between A and B being
the yz plane, for $\alpha = A$ all unit vectors point along the
negative x-axis and for $\alpha = B$ all unit vectors point along
the positive x-axis.}
\label{fig20}
\end{center}
\end{figure}

\begin{figure}
\begin{center}
$\begin{array}{c@{\hspace{0.2in}}c}
\multicolumn{2}{l}{\mbox{(a)}} \\ [-1.5cm] \\
&\psfig{file=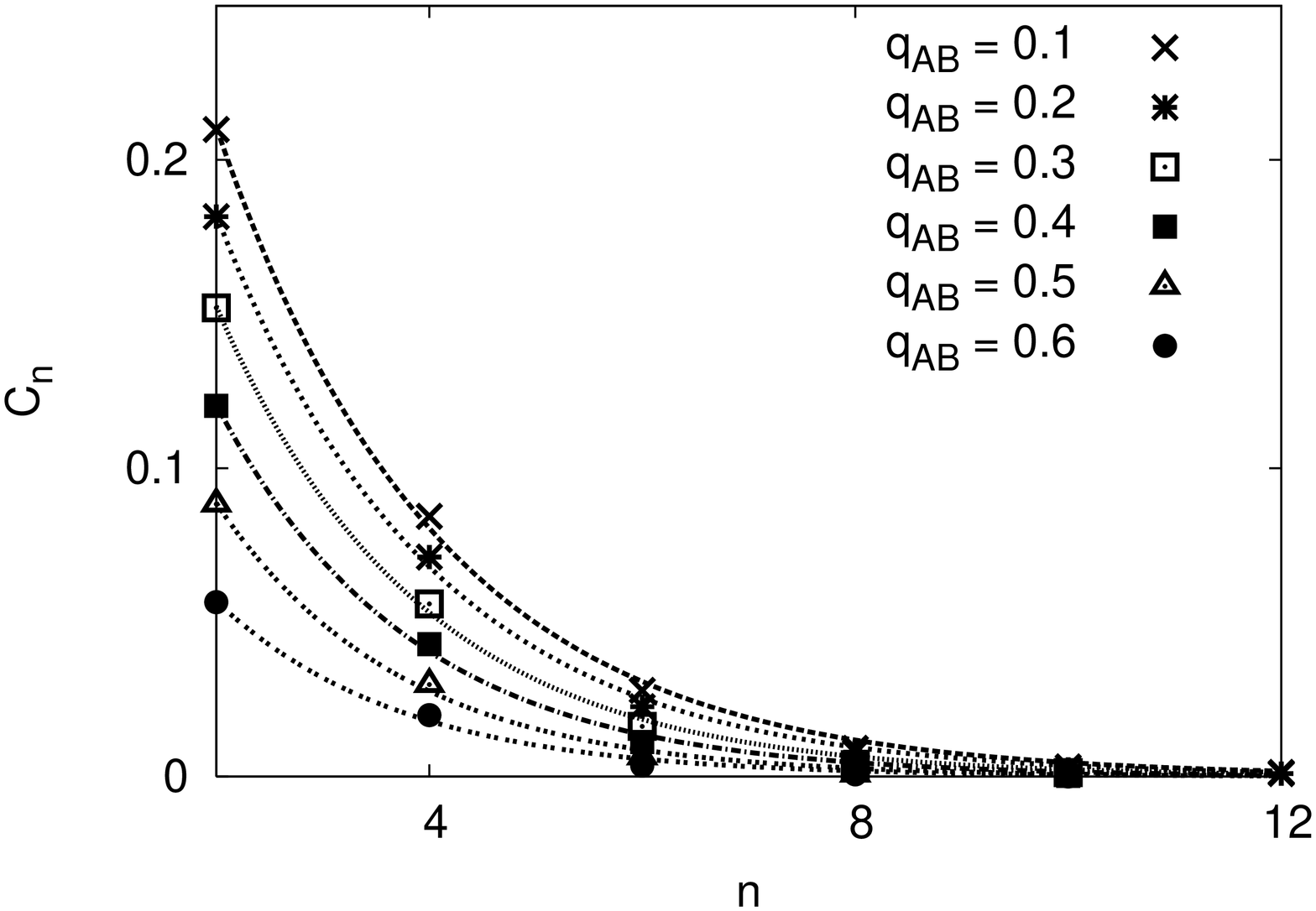,width=5.0cm,angle=270} \\
[0.05cm]\\
\multicolumn{2}{l}{\mbox{(b)}} \\ [-1.5cm] \\
&\psfig{file=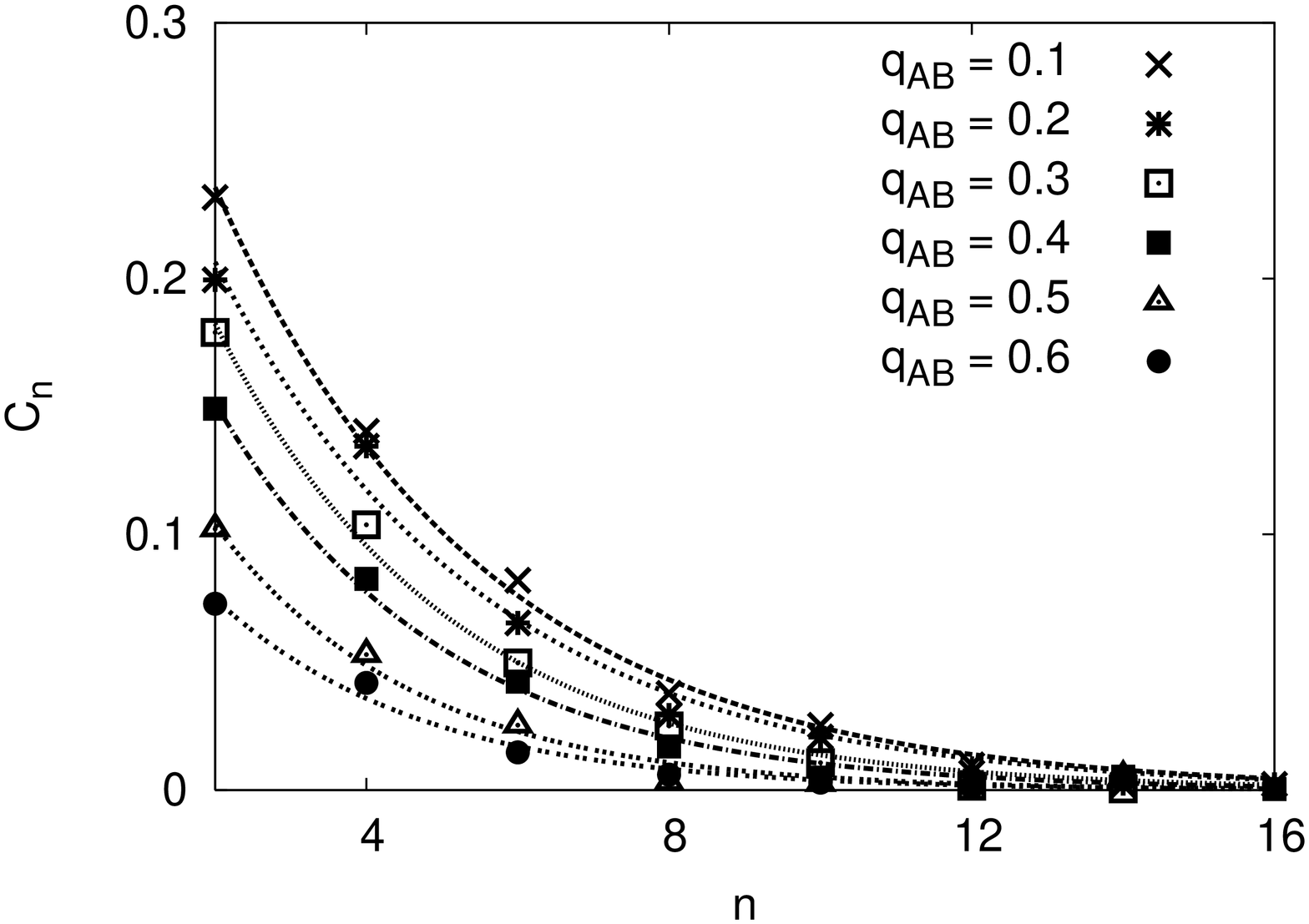, width=5.0cm, angle=270} \\
[0.05cm]\\
\multicolumn{2}{l}{\mbox{(c)}} \\ [-1.5cm] \\
&\psfig{file=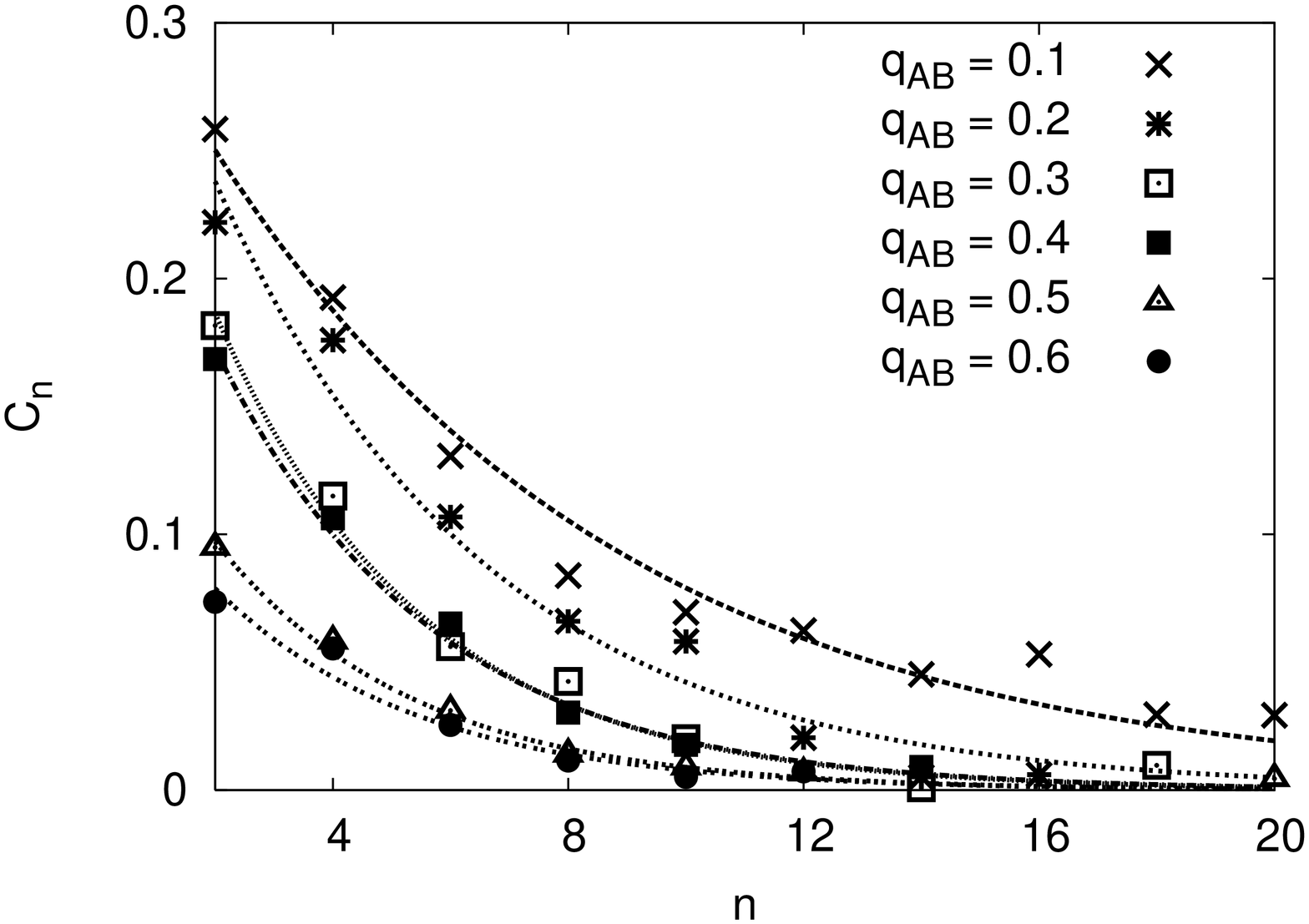, width=5.0cm, angle=270} \\
[1.0cm]
\end{array}$
\caption{Correlation function $C_n$ measuring local ``Janus cylinder''-type phase
separation plotted versus $n$, for $N=6$ (a), $12$ (b), and $18$ (c).
Various choices of $q_{AB}$ are included, as indicated in the figure.
Curves show fits to Equation~(\ref{eq50}). All data refer to good solvent
conditions $(q=1)$, and $L_b=64$, $\sigma=1$, $f=1$.}
\label{fig21}
\end{center}
\end{figure}

\begin{figure}
\begin{center}
$\begin{array}{c@{\hspace{0.2in}}c}
\multicolumn{2}{l}{\mbox{(a)}} \\ [-1.5cm] \\
&\psfig{file=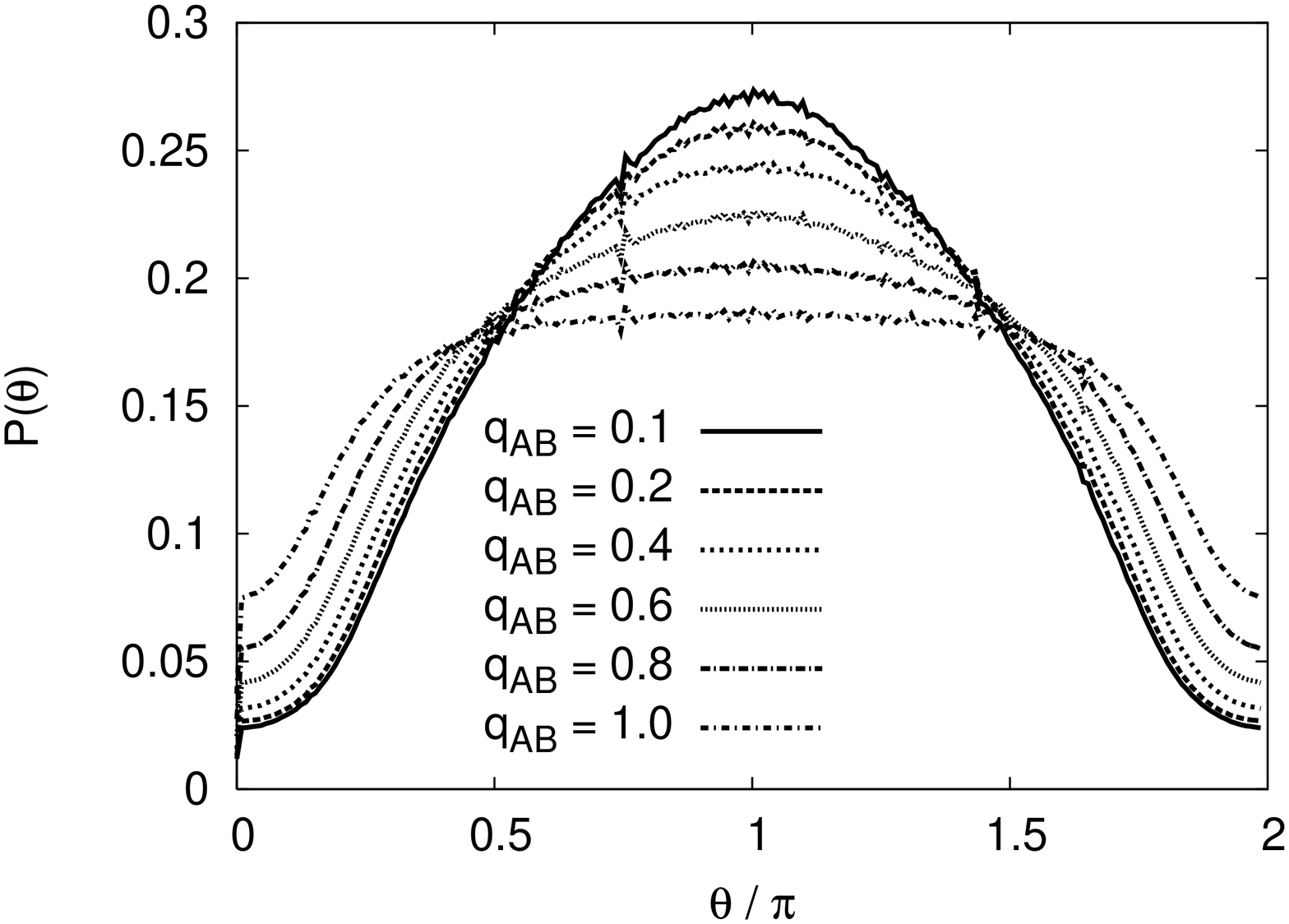,width=5.0cm,angle=270} \\
[0.05cm]\\
\multicolumn{2}{l}{\mbox{(b)}} \\ [-1.5cm] \\
&\psfig{file=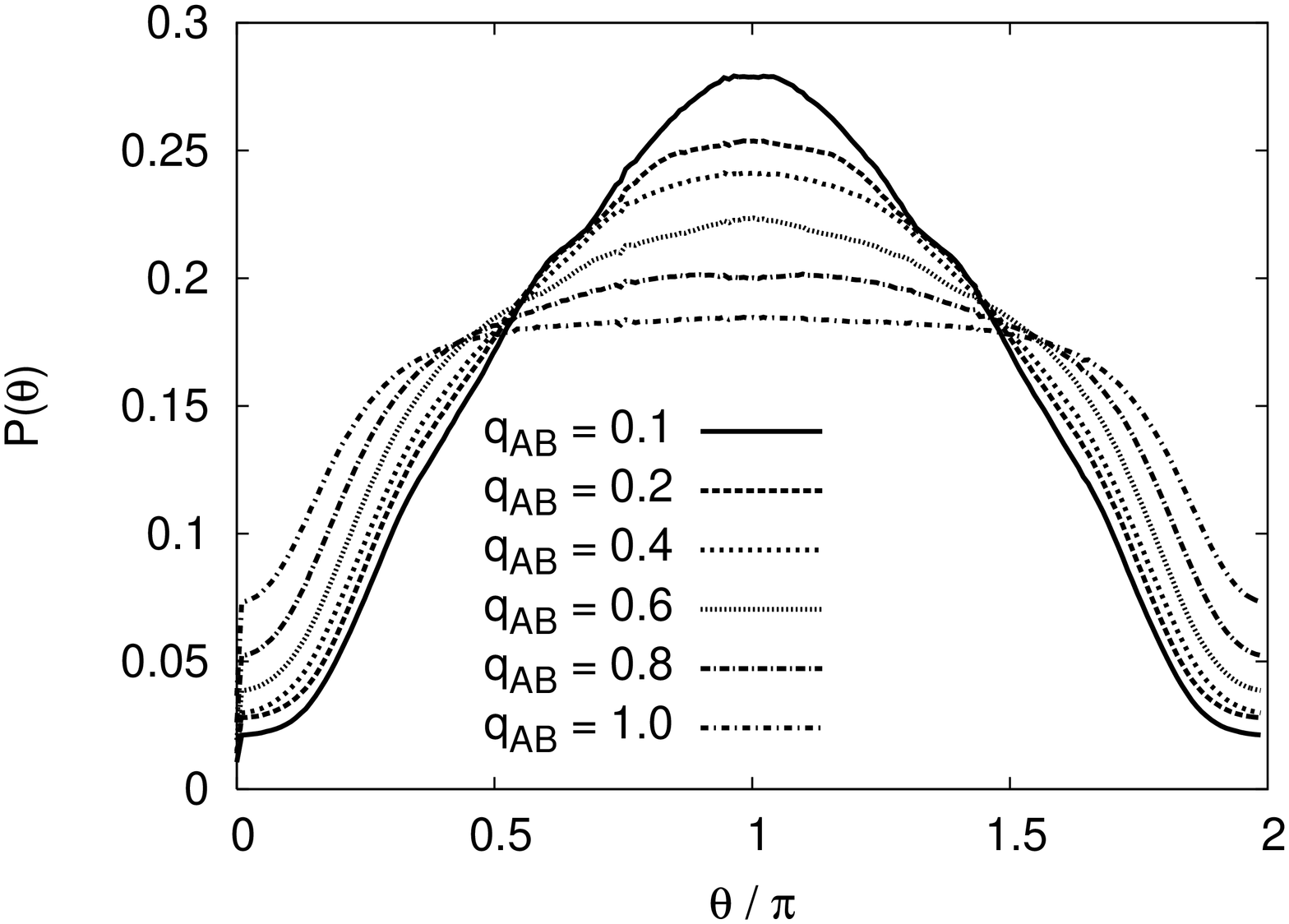, width=5.0cm, angle=270} \\
[1.0cm]
\end{array}$
\caption{Distribution $P(\theta)$ of the angle $\theta$ between the vectors
towards the centers of mass of subsequent (unlike) side chains plotted
vs.~$\theta / \pi$, for $N=12$ (a) and $N=18$ (b). Various choices of
$q_{AB}$ are included, as indicated.}
\label{fig22}
\end{center}
\end{figure}

\begin{figure}
\begin{center}
$\begin{array}{c@{\hspace{0.2in}}c}
\multicolumn{2}{l}{\mbox{(a)}} \\ [-1.5cm] \\
&\psfig{file=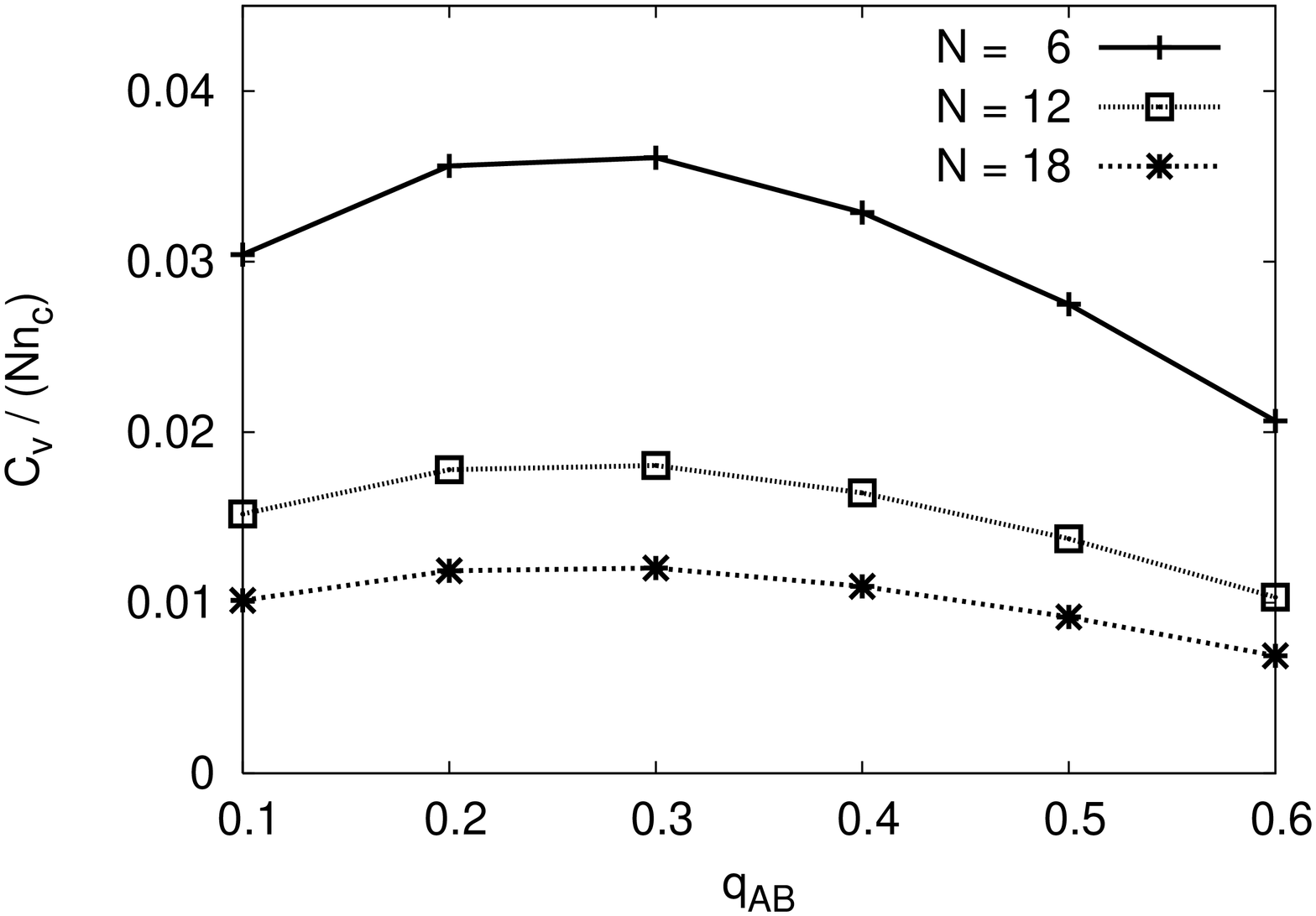,width=5.0cm,angle=270} \\
[0.05cm]\\
\multicolumn{2}{l}{\mbox{(b)}} \\ [-1.5cm] \\
&\psfig{file=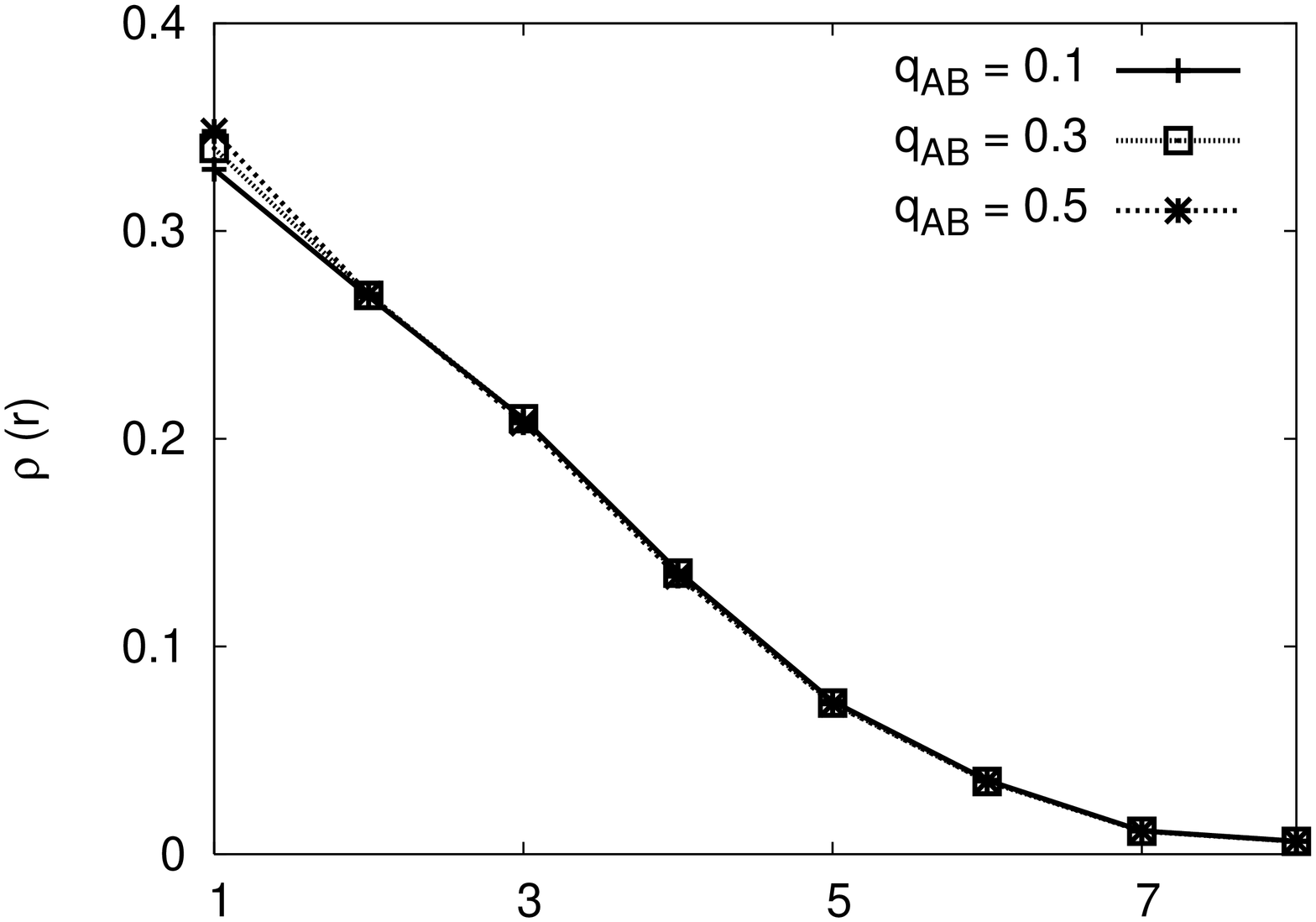, width=5.0cm, angle=270} \\
[1.0cm]
\end{array}$
\caption{ (a) Specific heat per monomer, $C_N / (Nn_c)$, plotted vs.~$q_{AB}$
for $N=6$, $12$, and $18$ for $\sigma=1$, $L_b=64$, and $q=1.3087$ (a Theta solvent).
(b) Radial density distribution $\rho(r)$ plotted vs.~$r$ for $L_b=64$, $N=18$,
$q=1.3087$, $\sigma=1$, $f=1$, and three choices of $q_{AB}$.}
\label{fig23}
\end{center}
\end{figure}

\vskip 1.0truecm
\noindent
{\large \bf VI. Monte Carlo Results for Binary Bottle Brush Polymers}
\vskip 0.5truecm

We use the same lattice model as considered before in our Monte
Carlo study of the chain conformations of one-component bottle
brush polymers, but now in the construction of the weights
$W_n(\alpha)$ we have to take into account that the partition
function now is
\begin{equation}\label{eq47}
Z = \sum_\alpha q^{m_{AA} + m_{BB}}  q_{AB}^{m_{AB}}
\end{equation}
with (remember that we restrict attention to the choice $\epsilon
_{AA} = \epsilon_{BB})$
\begin{equation}\label{eq48}
q=\exp(-\epsilon_{AA}/k_BT),\quad q_{AB}= \exp
(-\epsilon_{AB}/k_BT)\;.
\end{equation}
In Equation~(\ref{eq47}) the numbers of non-bonded occupied
nearest-neighbor monomer pairs AA, BB and AB are denoted as
$m_{AA}, m_{BB}$ and $m_{AB}$, respectively. Note that the sum in
Equation~(\ref{eq47}) extends over all possible configurations 
$\left\{\alpha\right\}$ of
the bottle brush polymer. The choice $q= q_{AB}=1$ corresponds to
the previously studied one-component bottle brush under good
solvent conditions, while the choice $q= q_{AB}>1$ corresponds
to variable solvent quality for the one-component brush (note that
$q=q_{AB}$ means $\chi=0$, Equation~(\ref{eq28}), and also $\chi_{AB}$
which is proportional to $\chi$ then vanishes: this means there is
no chemical incompatibility between A and B any longer, no physical
difference between A and B exists any more). From previous work on
single chains$^{\textrm {\cite{46}}}$ we know that the $\theta$-point occurs for
$q_\theta = 1.3087$. Therefore we varied $q$ in the range $1\leq q
\leq 1.5$; $q=1.5$ hence falls in the regime of poor solvent
quality already. Of course, in order to have rather compact
configurations of cylindrical bottle brushes a choice of much
larger $q$ would be desirable. However, the efficiency of the PERM
algorithm quickly deteriorates with increasing $q$: for $q=1.5$ we
encounter already for rather small values of the side chain length $N$
such as $N=18$ and a backbone length of $L_b=64$ huge
statistical fluctuations. The total size of the bottle brush
polymer under poor solvent conditions reached here,
$N_{\textrm{tot}}=N\sigma L_b+L_b=1216$, is almost two orders of
magnitude smaller than the maximal size studied under good solvent
conditions! However, all known simulation algorithms for polymers
suffer from difficulties of equilibration in the limit of very
dense configurations.$^{\textrm { \cite{84,85,86}}}$

Since we are mostly interested in the high grafting limit $(\sigma
=1$, so the number of side chains $n_c = \sigma L_b=L_b$) in the PERM
algorithm where all side chains grow simultaneously we use a bias
factor such that side chains are grown with higher probability in
the directions perpendicular to the backbone. This additional bias
(which is not present in the standard Rosenbluth$^{\textrm { \cite{91}}}$ 
and PERM$^{\textrm {\cite{46}}}$ methods) must be taken into account by suitable
weight factors. About 10$^6$ independent configurations were
typically generated.

Figures~\ref{fig17} - \ref{fig19} now show typical results for the
good solvent case $(q=1)$ but varying the parameter $q_{AB}$
controlling the chemical incompatibility. The visual inspection of
the configurations (Figure~\ref{fig17}) reveals little influence of
$q_{AB}$, however, and this observation is corroborated by the
more quantitative analysis: the average number of A-B pairs per
monomer is extremely small (Figure~\ref{fig18}a) even for
$q_{AB}=1$, and hence not much enthalpy could be won if A-chains
and B-chains avoid each other: due to the excluded volume
interaction, very few nearest neighbor contacts between any
non-bonded monomers occur in our bottle brush model. For $N=18$
and $q_{AB}=1$ the total number of AB contacts per chain is only
about 1.4, and increases with increasing $N$ only very slowly.
So for the range of side chain lengths accessible in our work, no
phase separation should be expected. In the specific heat one does
see a weak peak near $0.2 \leq q_{AB} \leq 0.4$, but the height of
this peak decreases very strongly with increasing $N$. Furthermore
does neither the peak position shift with increasing $N$ (as one
would expect from Equation~(\ref{eq38}), if this peak would be a
rounded precursor of the phase transition that should occur at
$\chi_{AB}^*$ in the limit $N \rightarrow \infty$) nor does the
peak width decrease with increasing $N$. Thus, it is clear that
this peak is not an indicator of a ``Janus-cylinder''-type phase
separation in the bottle brush: rather we interpret it as a
Schottky-type anomaly, expected from the fact that in our model of
alternatingly grafted A-and B-chains at a straight line backbone
with coordinates $(0,0,z)$ in the immediate environment of the
backbone (e.g. at lattice sites $(x,y,z)=(\pm1,0,z)$ or $(0,\pm
1,z)$) there is a nonzero a-priori probability of $1/4$ that
between the first monomer of the chain grafted at $z$ and the
first monomer of the chain grafted at $z+1$ a nearest-neighbor
contact occurs. The finite energy from this local contacts near
the backbone gives rise to the peak in the specific heat.

Also the radial density profile (Figure~\ref{fig19}a) shows 
little effects of varying $q_{AB}$, and there is also no effect
in the gyration radius component $\langle R_{g \bot}^2 \rangle =
\langle R_{gx}^2\rangle + \langle R_{gy}^2\rangle $ of the side
chains, although a weak increase occurs in the corresponding component of
the end-to-end distance (Figure~\ref{fig19}b). In the latter figure,
one can see for small $N$ an even-odd oscillation, but this
lattice effect clearly has died out for $N >10$.

In order to test for correlations measuring local phase separation
along the backbone, Equation~(\ref{eq46}) is somewhat cumbersome to
implement numerically, since for each $z$ one has to find the
direction of $\vec{\psi}(z)$ from the condition that
$|\vec{\psi}_(z)|$ is maximal (cf. Equation~(\ref{eq31})). A simpler
and similar correlation function has been defined from the vectors
pointing from the grafting site $i$ to the center of mass of the
respective side chains. Projecting this vector into the xy-plane
and defining an unit vector $\vec{S}_i^\alpha$ ($\alpha = A$ or $B$)
along this projection (Figure~\ref{fig20}) we define a correlation
function $C_n$ as follows
\begin{equation}\label{eq49}
C_n \equiv [\langle \vec{S}_i^A \cdot \vec{S}_{i+n}^A \rangle +
\langle \vec{S}_{i+1}^B \cdot \vec{S}_{i+1+n}^B \rangle]/2\;.
\end{equation}
Here we have explicitly incorporated the alternating grafting
ABAB... of side chains along the backbone. The average $\langle
\ldots \rangle$ in Equation~(\ref{eq49}) includes an averaging over all
sites $\left\{ i \right\}$ on which A chains are grafted, 
in order to improve the
statistics. If perfect long range order occurs, as implied in
Figure~\ref{fig20}, we clearly have $C_n=1$ independent of $n$,
while for the case of short range order, we expect $C_n \propto
\exp (-n/\xi)$. Actually, considering the fact that we use a
periodic boundary condition, we have analyzed our numerical data
in terms of the ansatz
\begin{equation}\label{eq50}
C_n\propto \{\exp (-n/\xi) + \exp [-(L_b-n)/\xi]\} \;.
\end{equation}
Figure~\ref{fig21} shows our data for $C_n$ for the three choices of
$N$: indeed we recognize that $C_n$ decays to zero with increasing
$n$, but the increase does get slower with increasing side chain
length $N$. The scale of this correlation effect clearly increases
with decreasing $q_{AB}$. While for $N=6$ the correlation length
$\xi$ hardly depends on $q_{AB}$, for large $N$ a slight
increase of $\xi$ with decreasing $q_{AB}$ is suggested.

\begin{figure}
\begin{center}
$\begin{array}{c@{\hspace{0.2in}}c}
\multicolumn{2}{l}{\mbox{(a)}} \\ [-1.5cm] \\
&\psfig{file=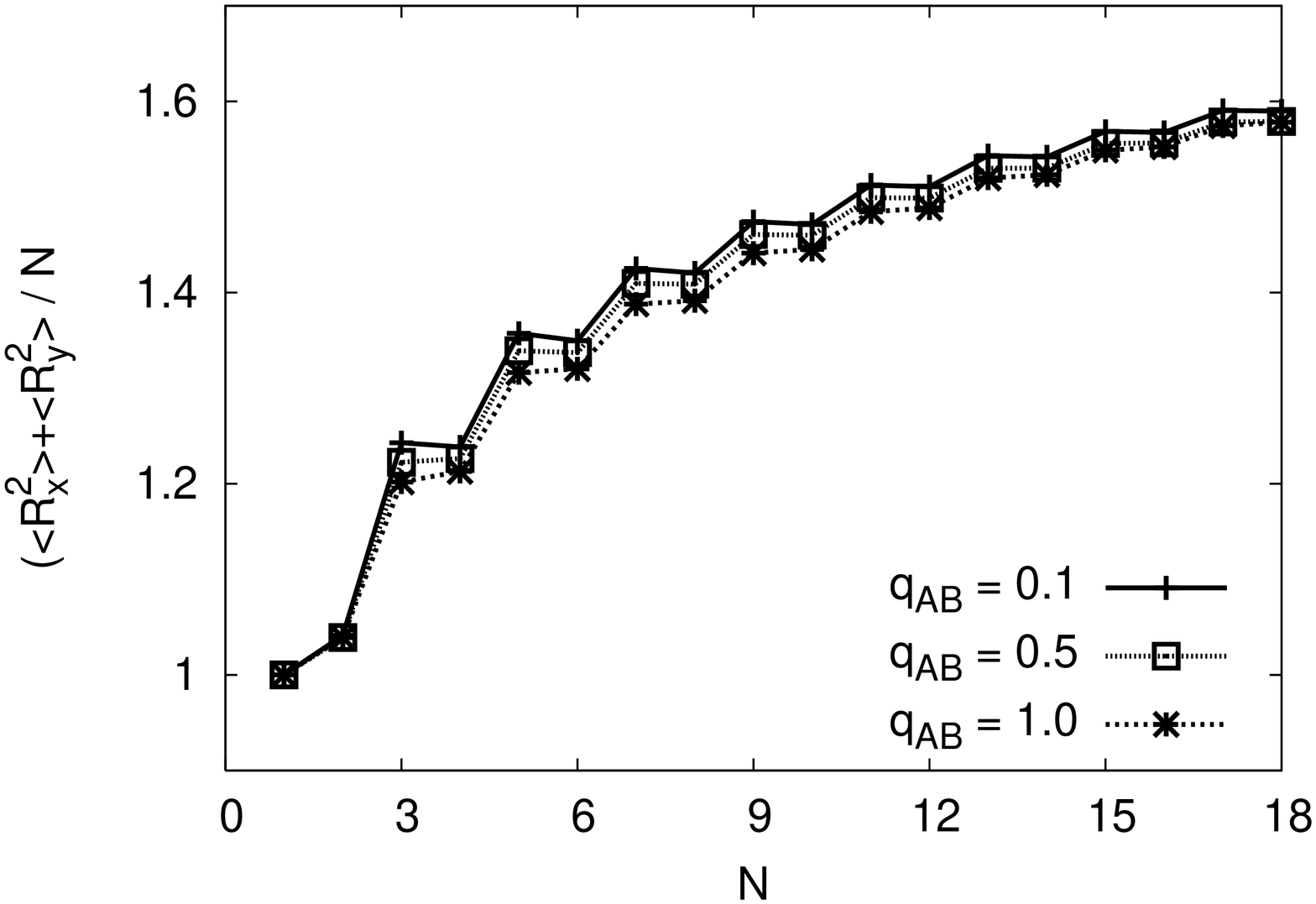,width=5.0cm,angle=270} \\
[0.05cm]\\
\multicolumn{2}{l}{\mbox{(b)}} \\ [-1.5cm] \\
&\psfig{file=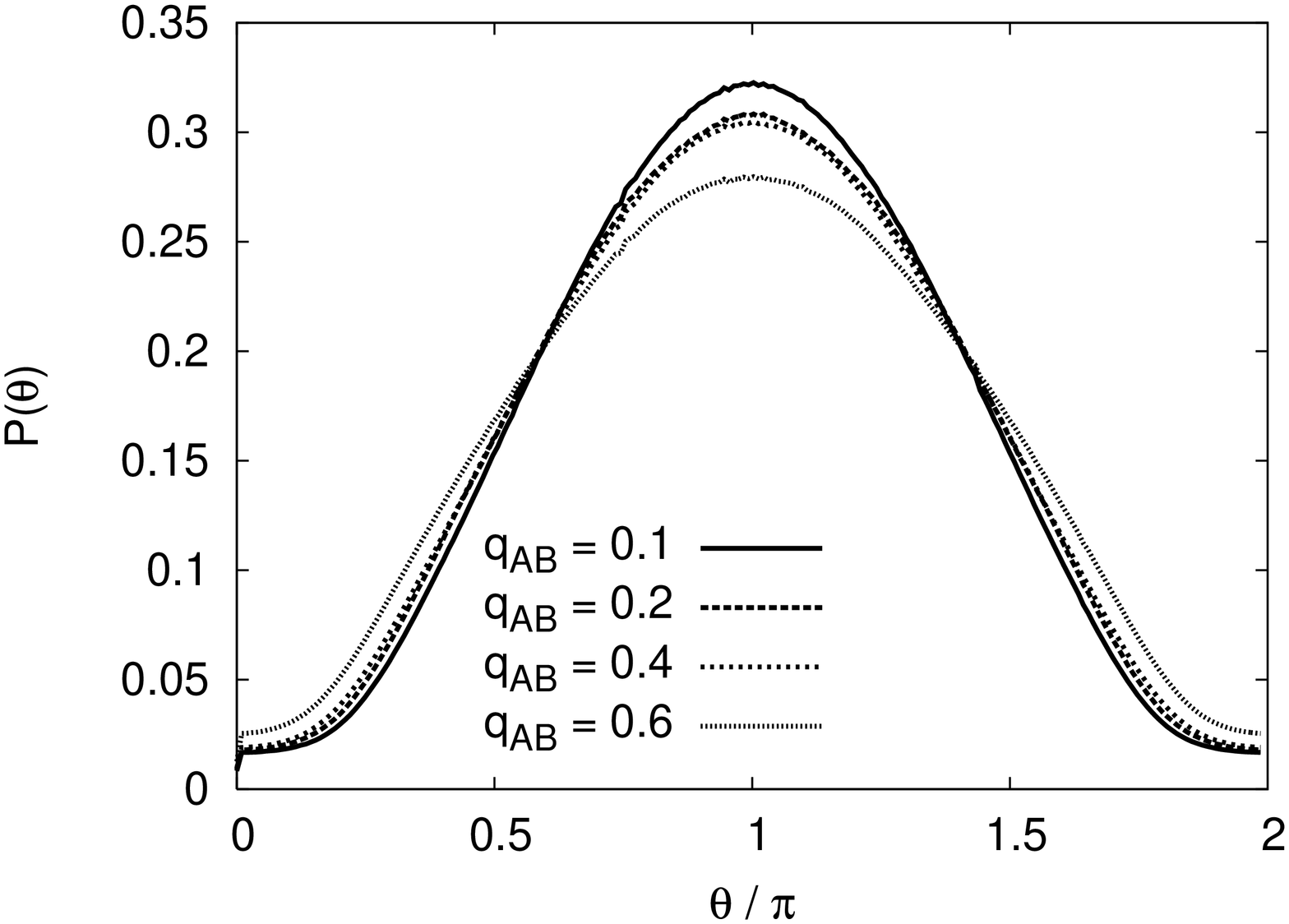, width=5.0cm, angle=270} \\
[1.0cm]
\end{array}$
\caption{(a) Normalized transverse mean square end-to-end distance
$(<R_x^2>+<R_y^2>)/N$ plotted vs.~$N$, for $L_b=64$, $\sigma=1$,
$f=1$, and $q=1.3087$ (Theta solvent). Three choices of $q_{AB}$
are included.
(b) Distribution $P(\theta)$ of the angle between the vectors
towards the centers of mass of subsequent (unlike) side chains
plotted vs.~$\theta / \pi$, for $N=18$, other parameters as in (a).}
\label{fig24}
\end{center}
\end{figure}

\begin{figure}
\begin{center}
$\begin{array}{c@{\hspace{0.2in}}c}
\multicolumn{2}{l}{\mbox{(a)}} \\ [-1.5cm] \\
&\psfig{file=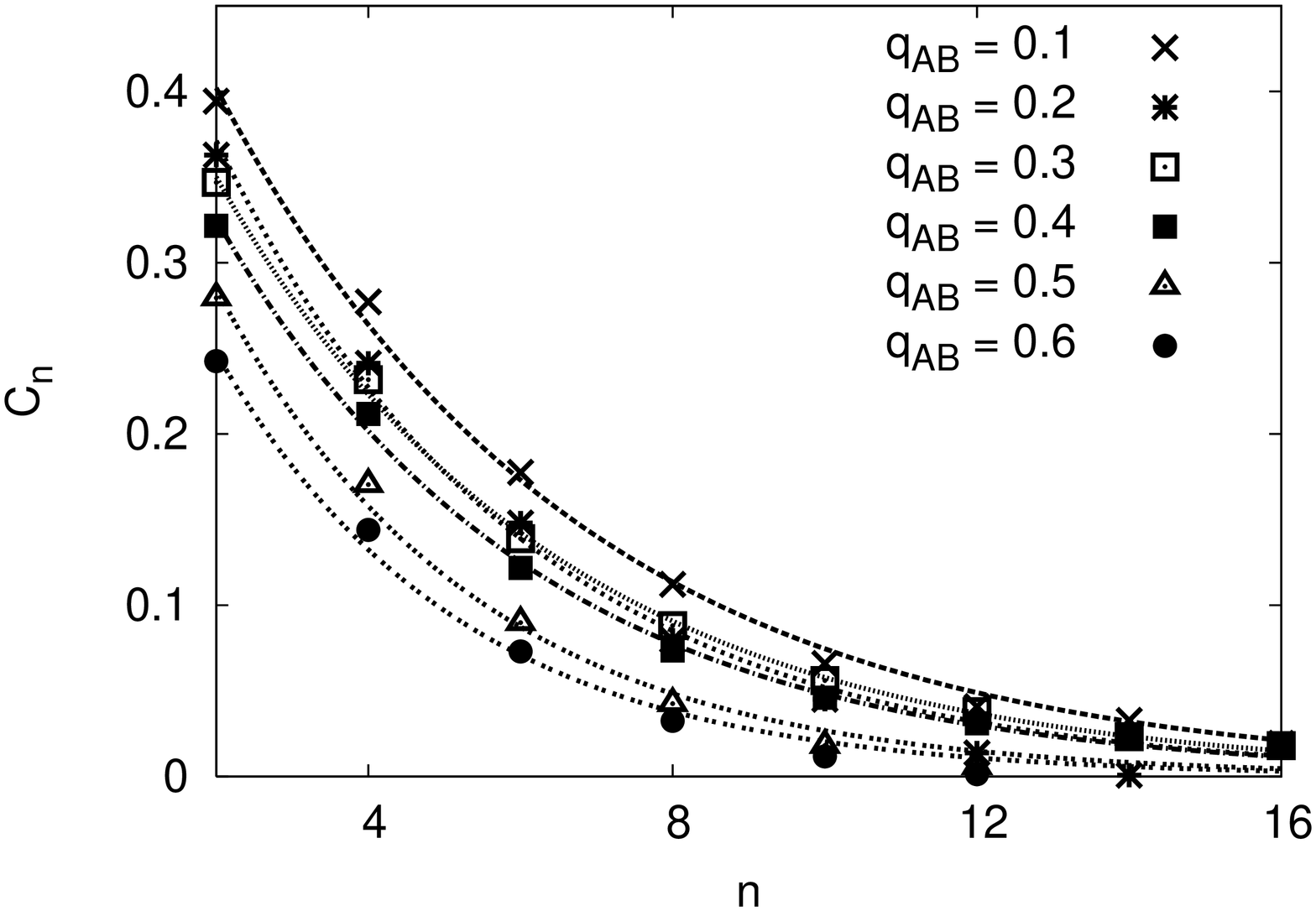,width=5.0cm,angle=270} \\
[0.05cm]\\
\multicolumn{2}{l}{\mbox{(b)}} \\ [-1.5cm] \\
&\psfig{file=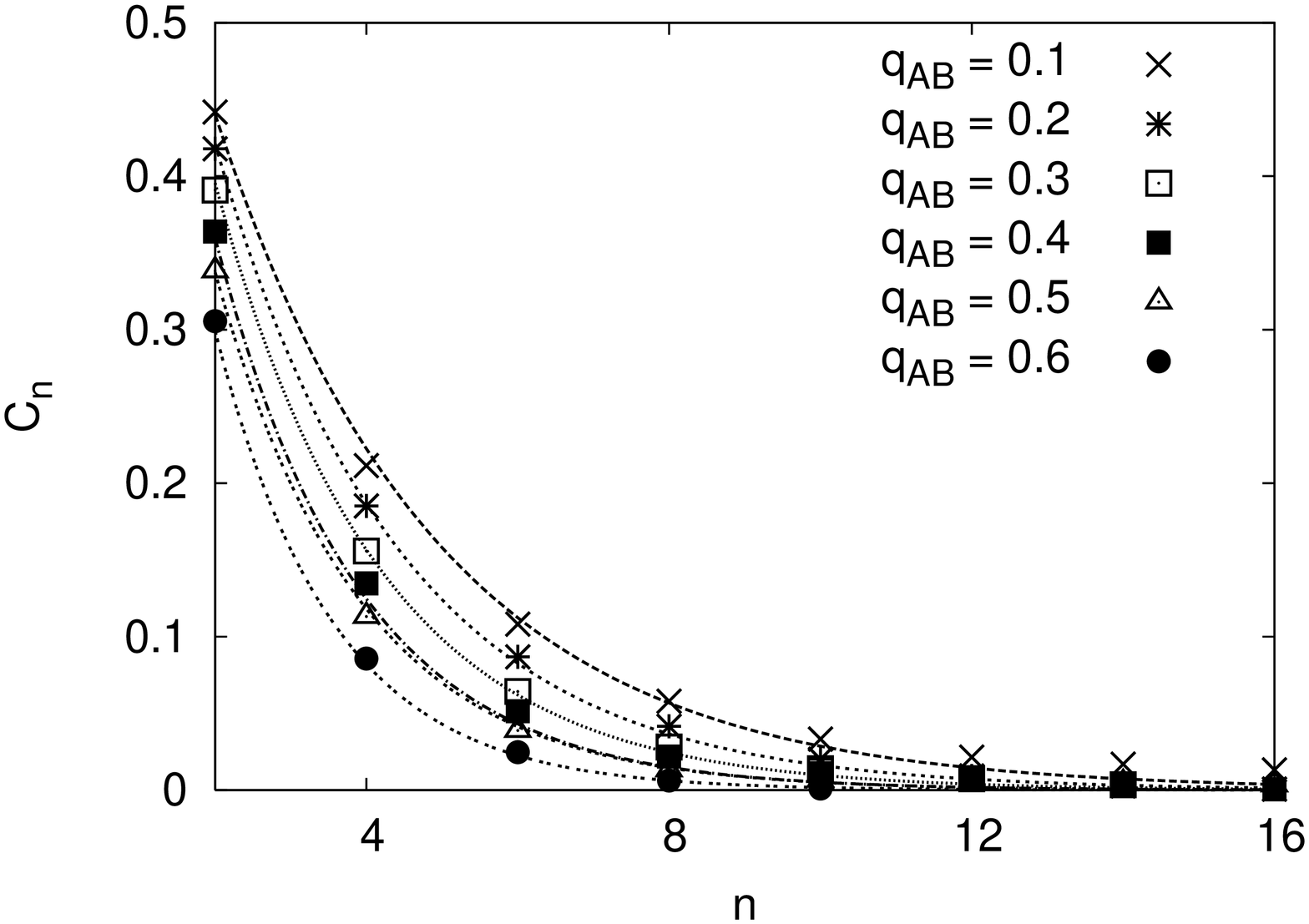, width=5.0cm, angle=270} \\
[1.0cm]
\end{array}$
\caption{Correlation function $C_n$
$\left\{ \right .$ Equation~(\ref{eq49}) $\left .\right\}$
plotted vs.~$n$, for $L_b=64$, $q=1.3087$, $\sigma=1$, $f=1$, $N=12$,
and various choices of $q_{AB}$, as indicated. Case (a) refers
to a choice of unit vectors from the $z$-axis to the center of
mass of the chain that is grafted at $z=i$, while case (b)
refers to a choice of unit vectors from the $z$-axis to
the center of mass of all monomers of type $\alpha$ in the xy-plane
at $z=i$. Curves are fits to Equation~(\ref{eq50}).}
\label{fig25}
\end{center}
\end{figure}

\begin{figure*}
\begin{center}
$\begin{array}{c@{\hspace{0.4in}}c}
\multicolumn{2}{l}{\mbox{(a)}} \\ [-1.5cm] \\
&\psfig{file=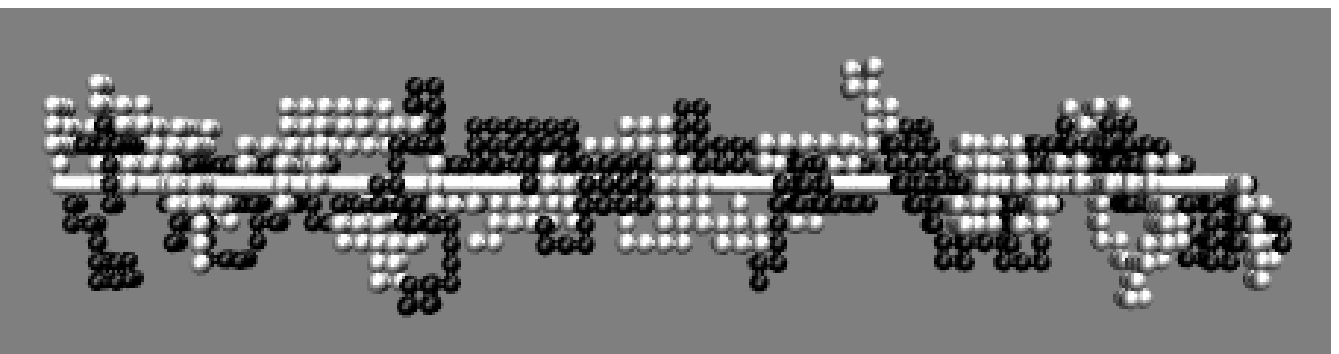,width=8.0cm,angle=0} \\
[0.5cm]\\
\multicolumn{2}{l}{\mbox{(b)}} \\ [-1.5cm] \\
&\psfig{file=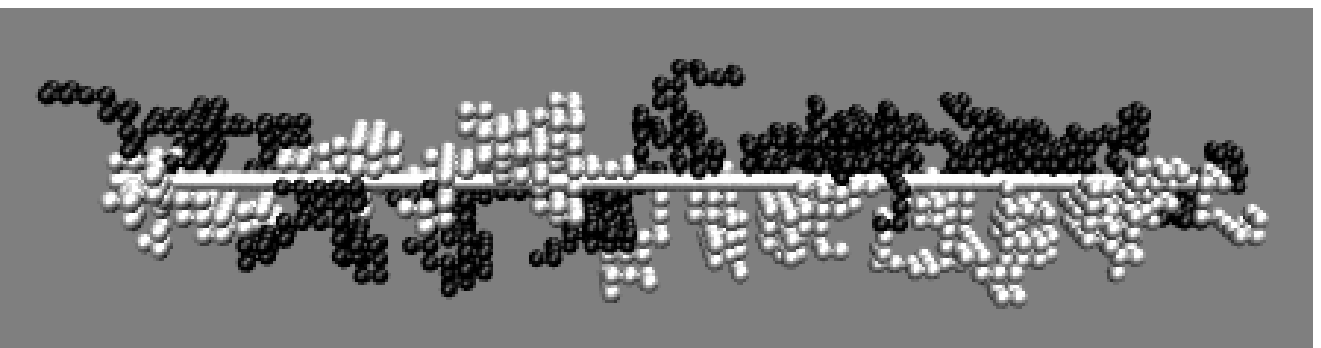, width=8.0cm, angle=0} \\
[0.5cm]\\
\multicolumn{2}{l}{\mbox{(c)}} \\ [-1.5cm] \\
&\psfig{file=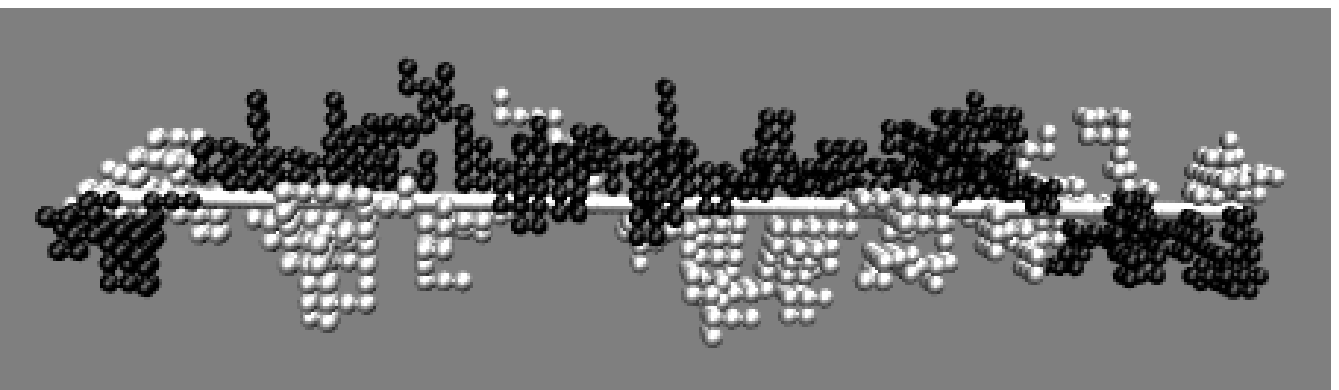, width=8.0cm, angle=0} \\
[1.0cm]
\end{array}$
\caption{Snapshot pictures of bottle brush conformations for
$L_b=64$, $q=1.5$, $N=18$, $\sigma=1$, and
three choices of $q_{AB}$, namely $q_{AB}=1.5$ (a),
$q_{AB}=0.6$ (b) and $q_{AB}=0.1$ (c).
Monomers A, monomers B, and a straight rigid backbone are shown
in black, light gray, and white colors, respectively.}
\label{fig26}
\end{center}
\end{figure*}

In an earlier simulation study of phase separation in binary
bottle brushes,$^{\textrm { \cite{57}}}$ it was suggested to quantify the degree
of separation by considering the distribution of the polar angle
$\varphi _ i$ from the axis to monomer $\ell$. Defining a variable
$\sigma _\ell^A$, which is $\sigma _\ell ^{A} = 1$ if monomer $\ell$
is of type A and zero otherwise, and similarly $\sigma_\ell^B$, de
Jong and ten Brinke$^{\textrm { \cite{57}}}$ introduced a function
\begin{equation}\label{eq51}
P(\phi) = \sum _ \ell \sum_{\ell '} (\sigma _\ell
^A \sigma _{\ell '}^B+ \sigma _\ell ^B \sigma ^A _{\ell '})
\delta (| \varphi _i - \varphi _j | - \phi)
\end{equation}
and studied $P (\phi)$ varying $\chi_{AB}$. However, it turns out
that $P(\phi)$ always has a rather complicated shape, and its
dependence on $\chi_{AB}$ is rather weak. Thus we investigated a
related but somewhat different function, namely the histogram
$P(\theta)$ of the angle $\theta$ between the vectors from the
axis to the centers of mass of subsequent unlike side chains
(Figure~\ref{fig22}). We see in this distribution $P(\theta)$ also
only a rather small dependence on $\chi_{AB}$, however: while for
$q_{AB}=1$ this distribution is essentially structureless for
$\pi/2 \leq \theta \leq 3 \pi /2$, for small $q_{AB}=1$ a rather
pronounced peak at $\theta \approx \pi$ develops, indicating a
preference of antiparallel orientation of subsequent side
chains. However, we feel that such indicators as $P(\phi)$ or
$P(\theta)$ are sensitive only to the presence of short range
order rather than long range order, and hence we shall focus on
the correlation function $C_n$ in the following.

Next we focus on binary bottle brushes in a Theta solvent
($q=1.3087)$. Figures~\ref{fig23} - \ref{fig25} show that the
results are not very different from the good solvent case: the
average number of AB pairs per monomer (not shown) and the
specific heat (Figure~\ref{fig23}a) are hardly distinguishable from
the good solvent case. Also the radial density profile
(Figure~\ref{fig23}b) exhibits only minor differences, slightly
larger densities occur inside the brush than in the good solvent
cases, and the transverse component of the end-to-end distance
does not exhibit much additional stretching, when $q_{AB}$
decreases (Figure~\ref{fig24}a). The distribution $P(\theta)$ is
somewhat less flat near $\theta = \pi$ (Figure~\ref{fig24}b) than in
the good solvent case (Figure~\ref{fig22}).

Also the correlation function $C_n$ shows again an exponential
decay with $n$ (Figure~\ref{fig25}a), the correlation length $\xi$
being rather similar to the good solvent case (Figure~\ref{fig21}).
Since one might argue that our definition of a correlation
function as given in Equation~(\ref{eq49}) is not the optimal choice, and
there might occur larger correlation lengths for a more clever
choice of a correlation function, we tried a different choice
which is also easy to compute. Namely, in the xy-plane at the
index $i$ of the z-coordinate we determine the center of mass of the
monomers of type $\alpha=A$ or B in that plane (the choice of
$\alpha$ is dictated by the type of chain grafted at $z=i$). Then
we introduce an unit vector $\vec{S}_i^\alpha$ from the z-axis in
the direction towards this center of mass. In terms of these unit
vectors, which then no longer distinguish from which chain the
monomers $\alpha$ in the $i$'th xy-plane are coming, we can
again apply the definition, Equation~(\ref{eq49}) to derive a correlation
function. This correlation function is shown in Figure~\ref{fig25}b.
One sees that the qualitative behavior of both types of
correlation functions is the same, but the decay of this second
type of correlation function even is slightly faster than that of
the correlation function used previously. Since normally, when one
studies problems involving a phase transition, the correlation
length of the order parameter is the largest correlation length
that one can find in the system, the previous definition (focusing
on the location of the center of mass of the individual side
chains) seems preferable to us. It is possible of course that such
differences between different ways of measuring correlations along
the backbone remain so pronounced for small $N$ only, where the
correlation lengths are only of the order of a few lattice
spacings, and hence are less universal and depend on the details
of the studied quantity.

\begin{figure}
\begin{center}
$\begin{array}{c@{\hspace{0.2in}}c}
\multicolumn{2}{l}{\mbox{(a)}} \\ [-1.5cm] \\
&\psfig{file=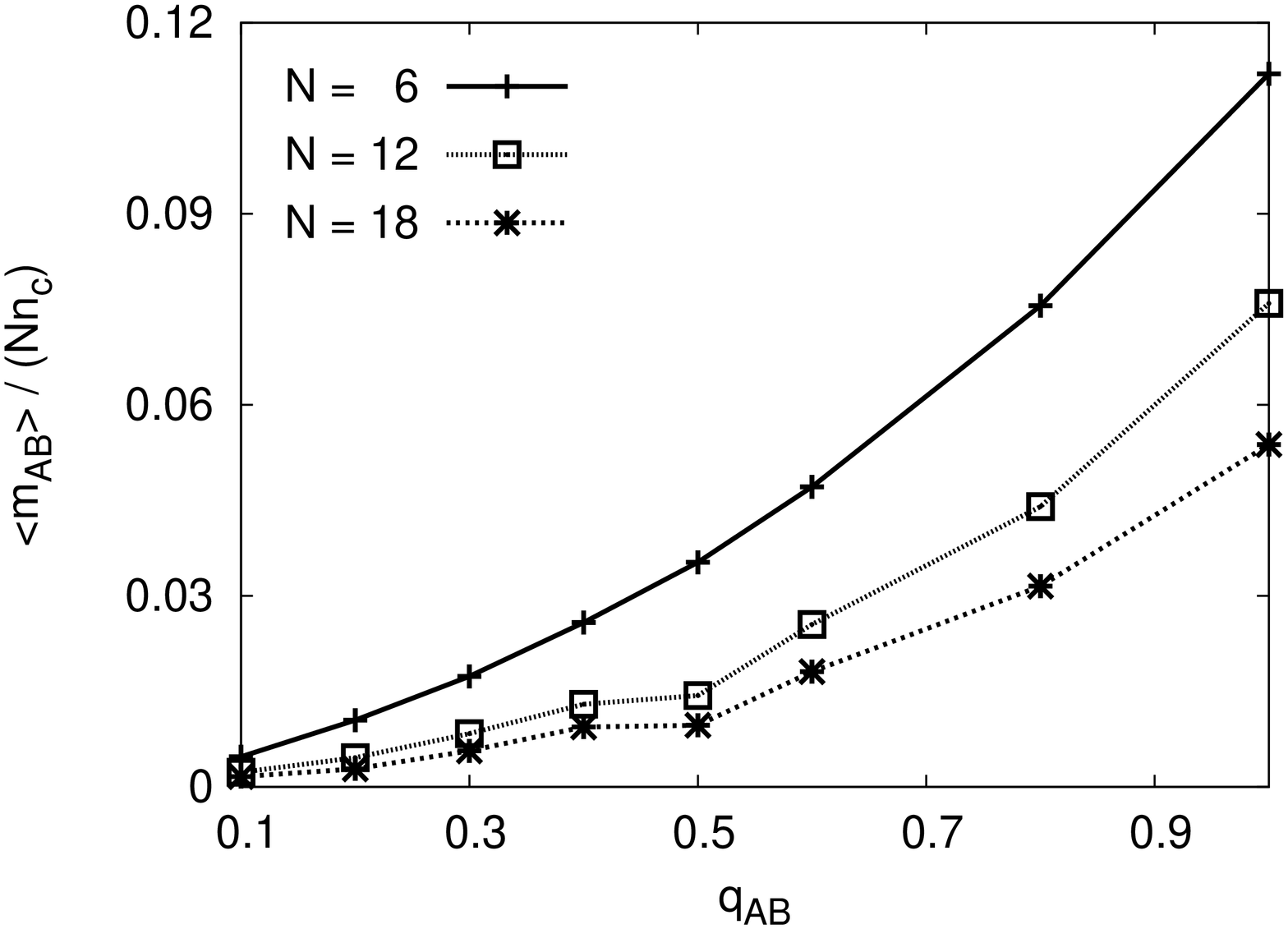,width=5.0cm,angle=270} \\
[0.05cm]\\
\multicolumn{2}{l}{\mbox{(b)}} \\ [-1.5cm] \\
&\psfig{file=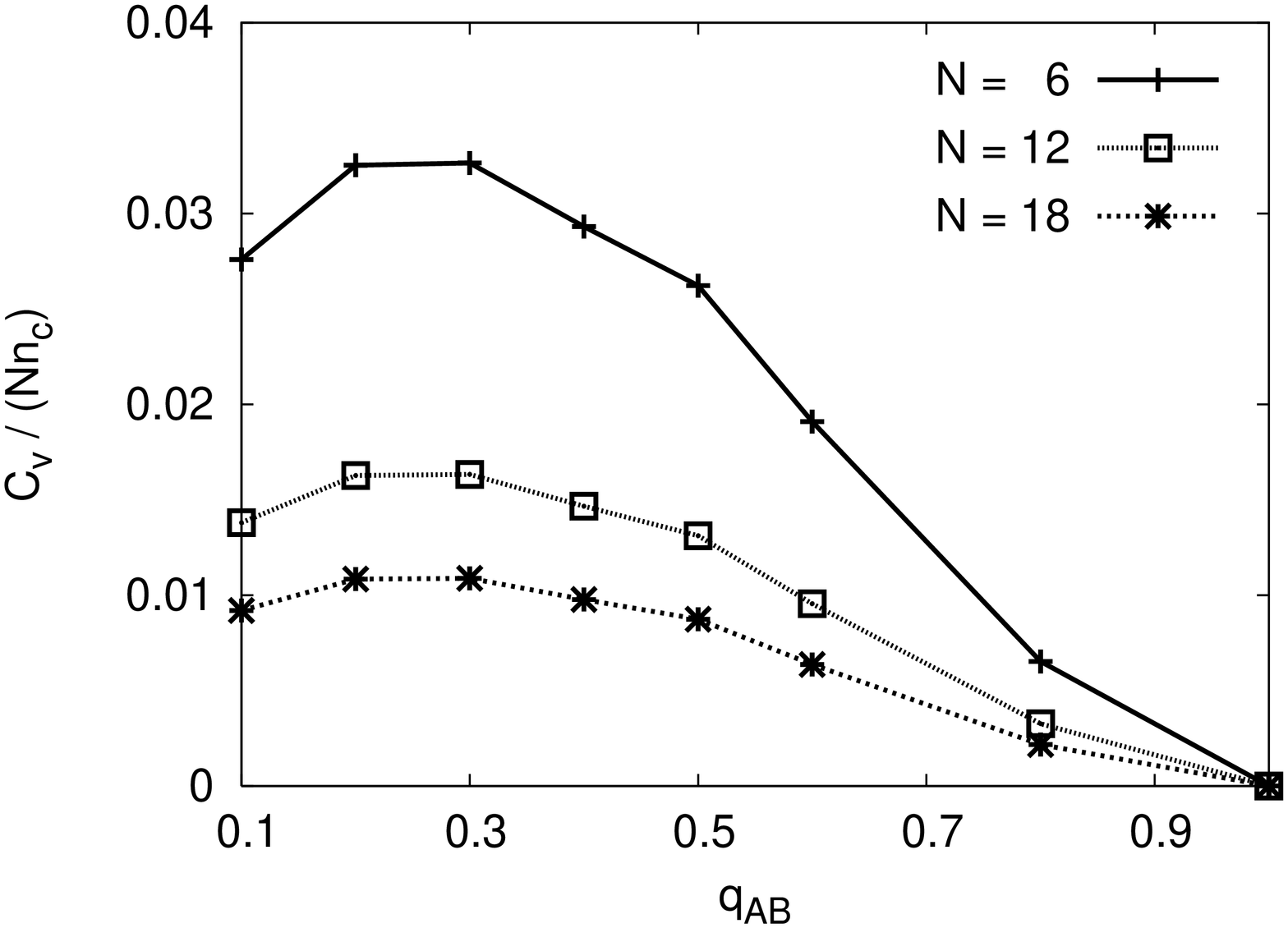, width=5.0cm, angle=270} \\
[1.0cm]
\end{array}$
\caption{ (a) Average number of AB pairs per monomers, $<m_{AB}>/(Nn_c)$,
plotted vs.~$q_{AB}$ for side chain lengths $N=6$, $12$, and $18$.
All data refer to $\sigma=1.5$, $L_b=64$. (b) Specific heat per
monomer, $C_v/(Nn_c)$, plotted vs.~$q_{AB}$ for $N=6$,
$12$, and $18$.}
\label{fig27}
\end{center}
\end{figure}

\begin{figure}
\begin{center}
\psfig{file=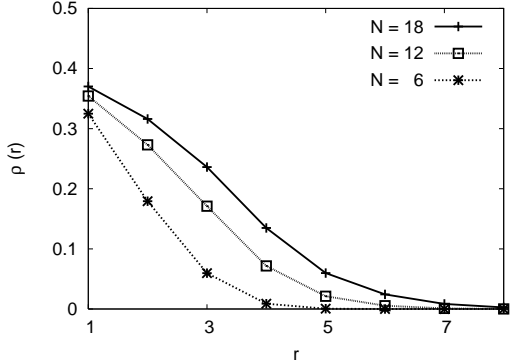,width=5.0cm,angle=270}
\caption{Radial density profile $\rho(r)$ plotted vs.~$r$ for $L_b=64$,
$q_{AB}=0.1$, $q=1.5$, $\sigma=1$, $f=1$, and three choices of
$N=6$, $12$, and $18$.}
\label{fig28}
\end{center}
\end{figure}

\begin{figure}
\begin{center}
$\begin{array}{c@{\hspace{0.2in}}c}
\multicolumn{2}{l}{\mbox{(a)}} \\ [-1.5cm] \\
&\psfig{file=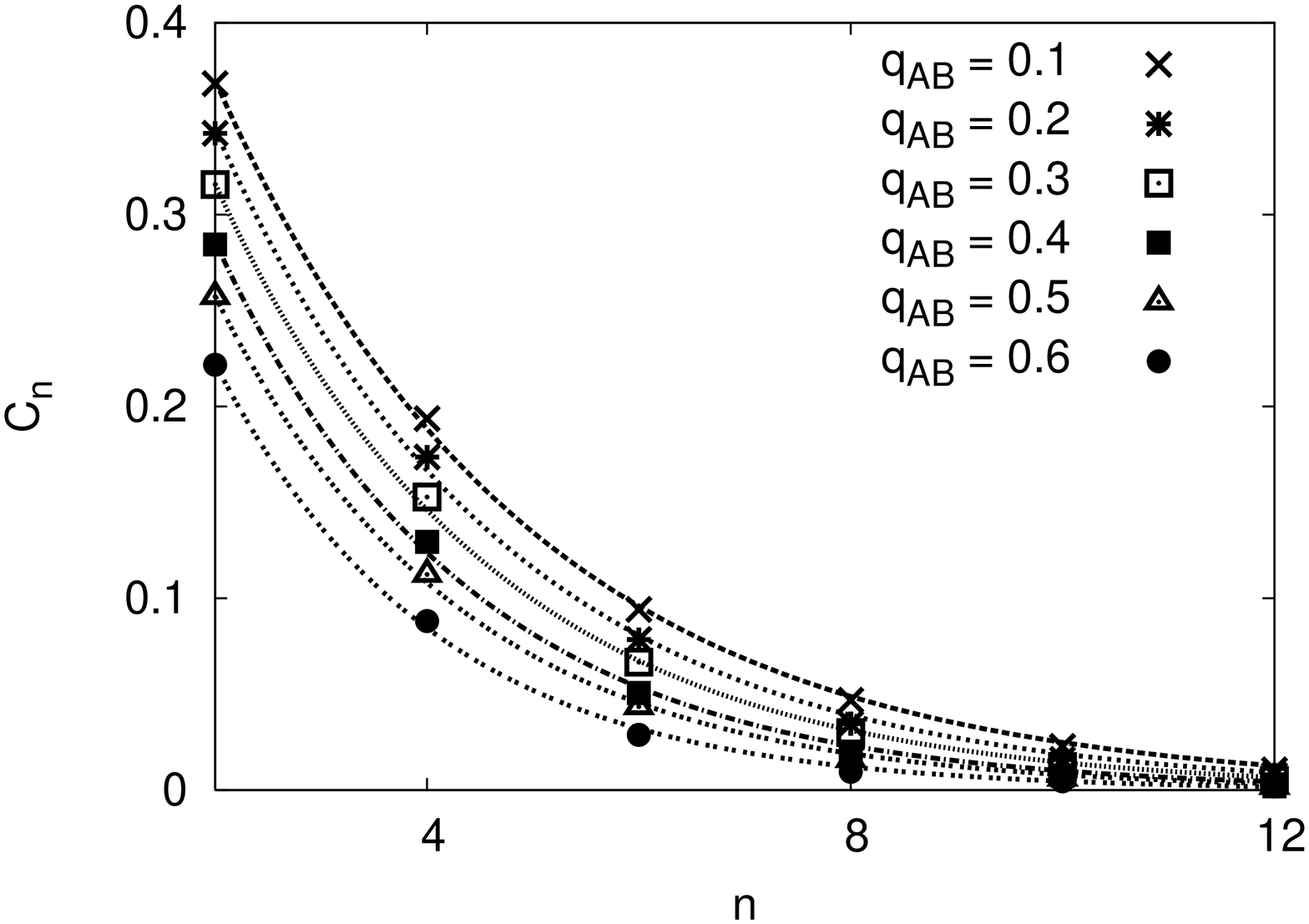,width=5.0cm,angle=270} \\
[0.05cm]\\
\multicolumn{2}{l}{\mbox{(b)}} \\ [-1.5cm] \\
&\psfig{file=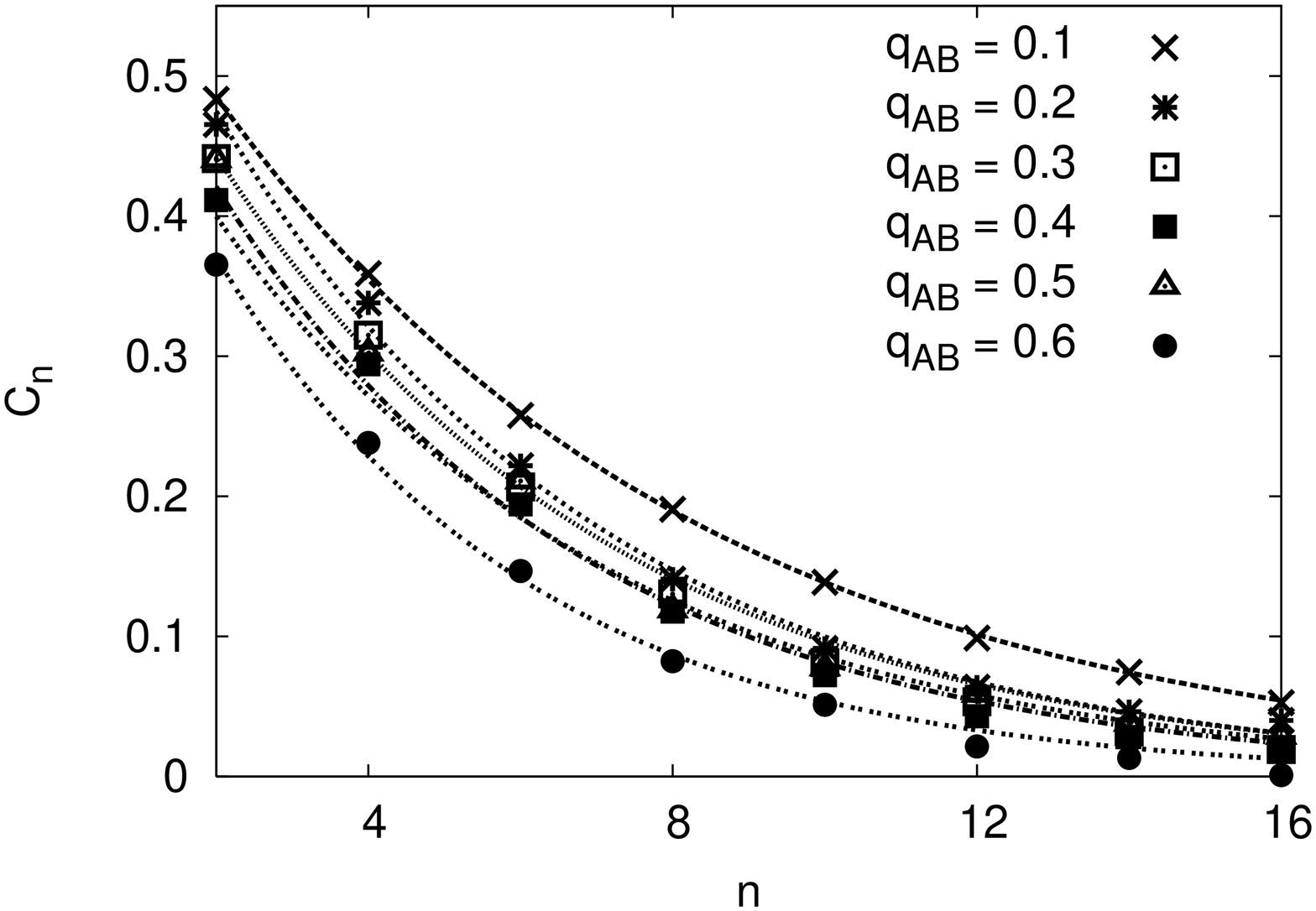, width=5.0cm, angle=270} \\
[0.05cm]\\
\multicolumn{2}{l}{\mbox{(c)}} \\ [-1.5cm] \\
&\psfig{file=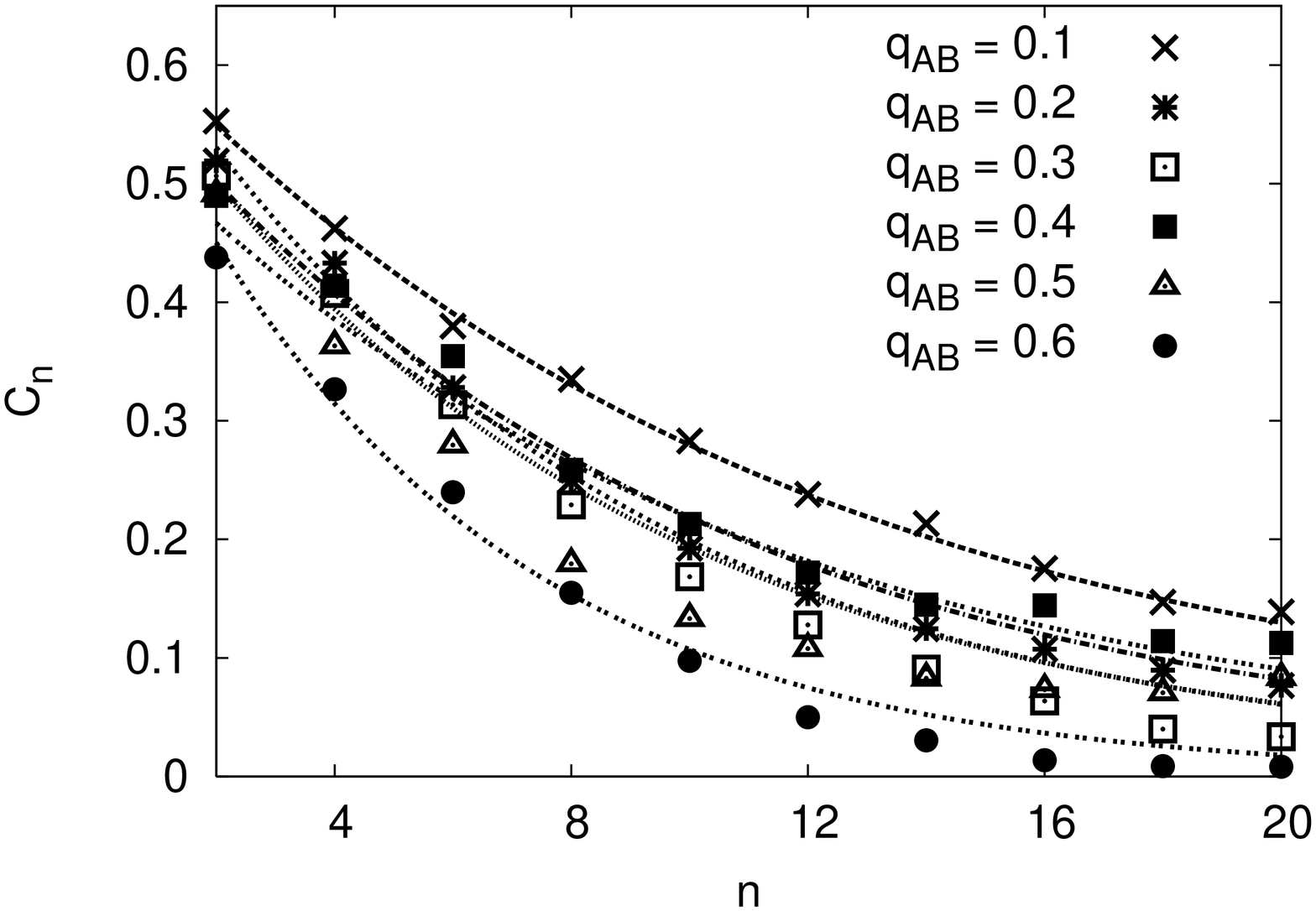, width=5.0cm, angle=270} \\
[1.0cm]
\end{array}$
\caption{Same as Figure~\ref{fig21}, but for poor solvent conditions $(q=1.5)$;
for $n=6$ (a), $n=12$ (b), and $n=18$ (c).
}
\label{fig29}
\end{center}
\end{figure}

We now turn to the poor solvent case (Figures~\ref{fig26} -
\ref{fig29}). Already the snapshot picture of the bottle brush
polymer conformations (Figure~\ref{fig26}) reveals that now the side
chains adopt much more compact configurations, due to the collapse
transition that very long single chains would experience in a poor
solvent.$^{\textrm { \cite{45}}}$ If now $q_{AB}$ becomes small, one recognizes
a more pronounced phase separation along the backbone of the
polymer, although there still is no long range order present.

Still, neither the variation of the average number of AB pairs per
monomer $\langle m _{AB} \rangle / (Nn_c)$ with $q_{AB}$
(Figure~\ref{fig27}a) nor the specific heat (Figure~\ref{fig27}b) give a
hint for the occurrence of a phase transition: in fact, these data
still look like in the good solvent case, Figure~\ref{fig18}! Also
the distribution $P(\theta)$ and the variation of $(\langle R_x^2
\rangle + \langle R_y^2 \rangle )/N^{2/3}$ with $N$ and $q_{AB}$
are qualitatively similar to the results found for good solvents
and Theta solvents, and therefore are not shown here. More
interesting is the monomer density profile (Figure~\ref{fig28}).
While for $N=6$ it still has the same character as in the previous
cases, for $N=12$ and $N=18$ we recognize an inflection point. In
fact, for a collapsed polymer bottle brush we expect a profile
exhibiting a flat interior region at melt densities ($\rho(r)$
near to $\rho = 1$), then an interfacial region where $\rho(r)$
rapidly decreases towards zero; in the ideal case reached for $q
\rightarrow \infty$, the profile should even be a Heaviside step
function, in the continuum limit, $\rho (r)= \theta (h-r)$.
Obviously, for $q=1.5$ we are still far from this behavior even
for $N=18$, but we can identify an interface location at about
$r_{\textrm{int}} \approx 3.5$, and in the interior of the
cylindrical bottle brush (for $r <r_{\textrm{int}}$) the monomer
density corresponds at least to a concentrated polymer solution.

Figure~\ref{fig29} then shows again plots of $C_n$ vs.~$n$, and the
resulting fits to Equation~(\ref{eq50}). The behavior is qualitatively
similar to the previous cases of good and Theta solvents again,
but now the decay of $C_n$ with $n$ is clearly much slower,
indicating a distinctly larger correlation length. This result
corroborates the qualitative observation made already on the basis
of the snapshot pictures, Figure~\ref{fig26}, and quantifies it.

\begin{figure}
\begin{center}
$\begin{array}{c@{\hspace{0.2in}}c}
\multicolumn{2}{l}{\mbox{(a)}} \\ [-1.5cm] \\
&\psfig{file=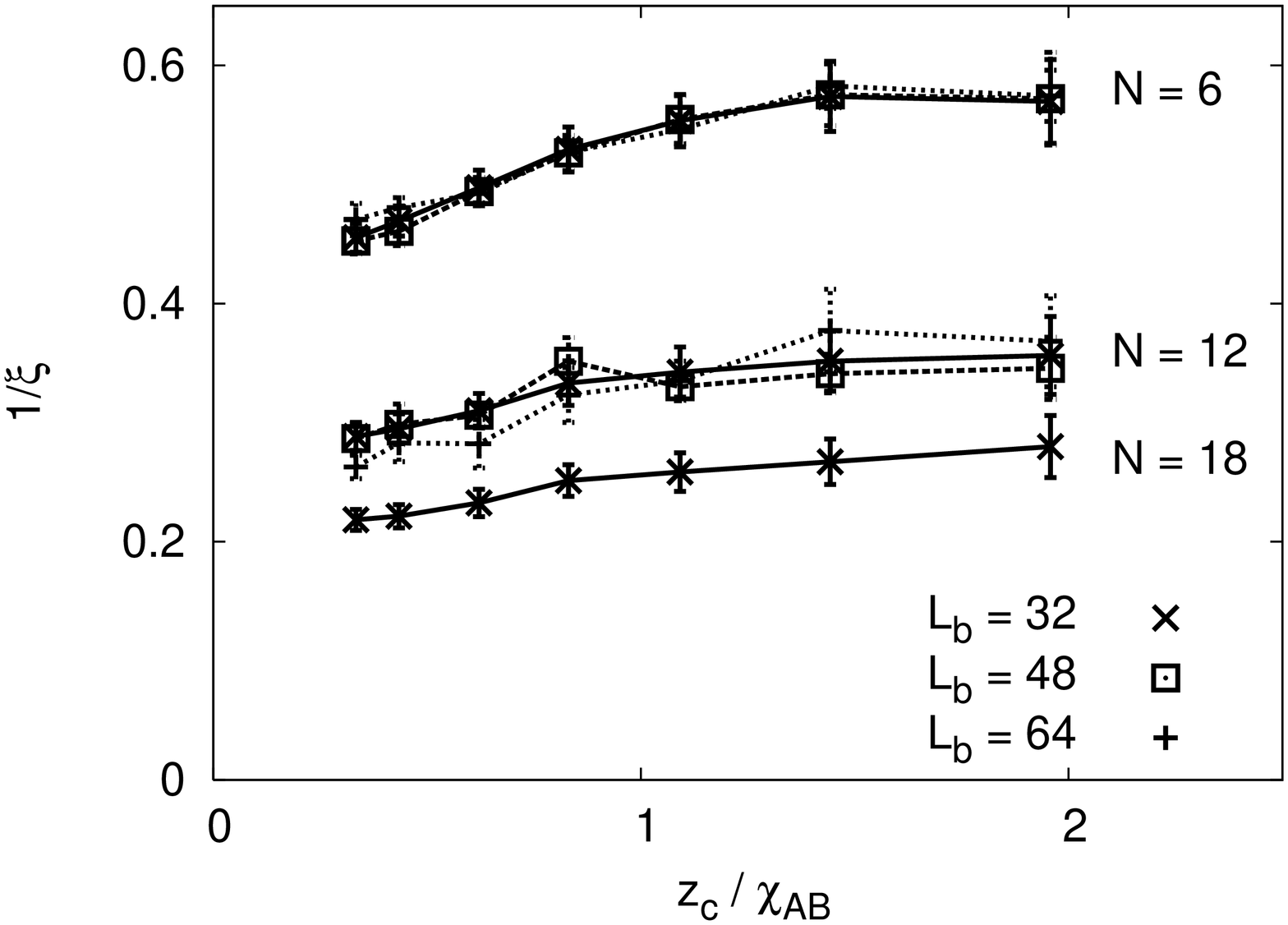,width=5.0cm,angle=270} \\
[0.05cm]\\
\multicolumn{2}{l}{\mbox{(b)}} \\ [-1.5cm] \\
&\psfig{file=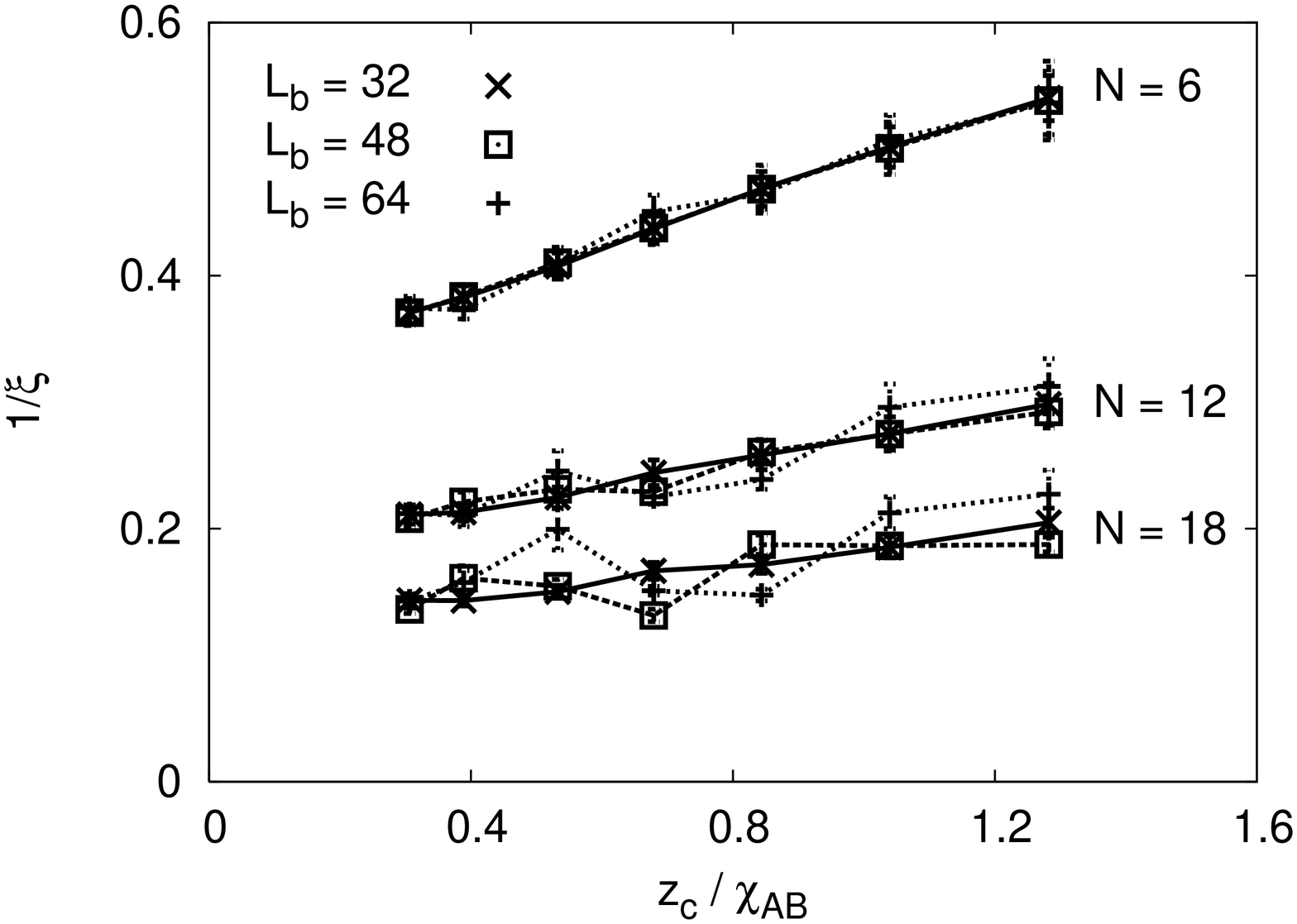, width=5.0cm, angle=270} \\
[0.05cm]\\
\multicolumn{2}{l}{\mbox{(c)}} \\ [-1.5cm] \\
&\psfig{file=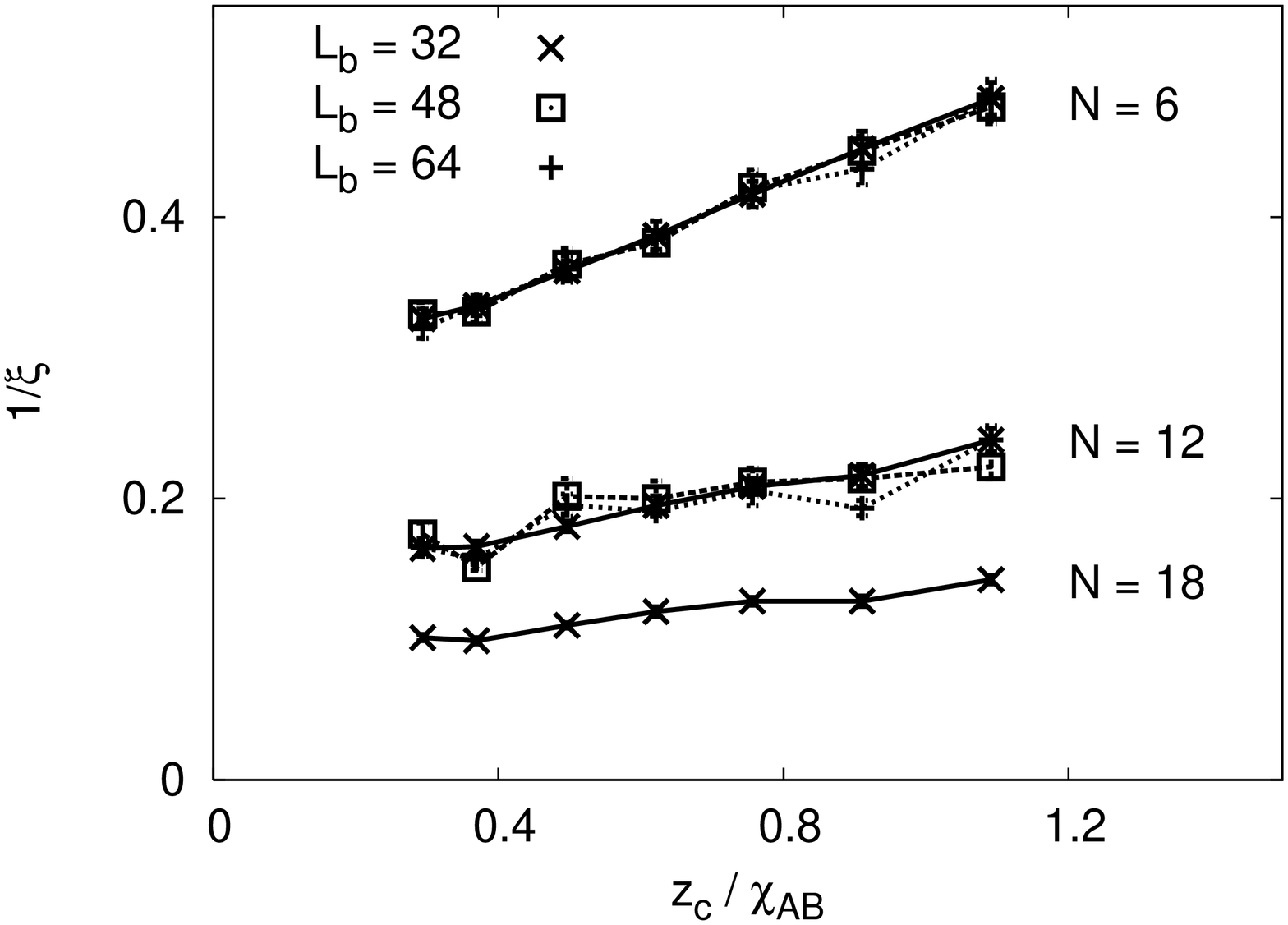, width=5.0cm, angle=270} \\
[1.0cm]
\end{array}$
\caption{Plot of the inverse correlation length $1/\xi$ versus the inverse
Flory Huggins parameters, $z_c/\chi$, for good solvent conditions
$(q=1)$, case (a), Theta solvent conditions $(q=1.3087)$, case (b),
and poor solvent conditions $(q=1.5)$, case (c). Data for three
chain lengths $N=6$, $12$, and $18$ are included throughout.
In several cases two or three choices of backbone lengths $L_b=32$, $48$
and $64$ are included as well. All data refer to the
choice of one grafted polymer per grafting site $(\sigma=1)$.}
\label{fig30}
\end{center}
\end{figure}

Our findings on the local character of phase separation in binary
bottle brush polymers are now summarized in Figure~\ref{fig30},
where plots of the inverse correlation length $1/\xi$ versus the
inverse Flory-Huggins parameter $z_c/\chi =
k_BT/(\epsilon_{AB}-\epsilon_{AA})=1/\ln [q/q_{AB}]$ are shown (cf.
Equation~(\ref{eq28}) and remember our choice
$\epsilon_{AA}=\epsilon_{BB}$). This choice of variables is
motivated by the analogy with the XY model in cylindrical
geometry, Equation~(\ref{eq45}); on the basis of this analogy, we would
expect that $\xi^{-1}\propto T$, i.e. we expect straight lines
that extrapolate through zero as $T \rightarrow 0$.

Figure~\ref{fig30} demonstrates that this analogy$^{\textrm { \cite{58}}}$ clearly
is not perfect. Rather the data are compatible with an
extrapolation
\begin{equation}\label{eq52}
\xi^{-1}(N,T)= \xi^{-1}(N,q,0)+C_N(q)T\quad , \quad T \rightarrow
0\;,
\end{equation}
where $C_N(q)$ is a coefficient that decreases with increasing $N$
and depends on solvent quality and with a nonzero intercept
$\xi^{-1}(N,q,0)$ implying that even in the ground state $(T=0)$
there is lack of long range order!

\begin{figure}
\begin{center}
\psfig{file=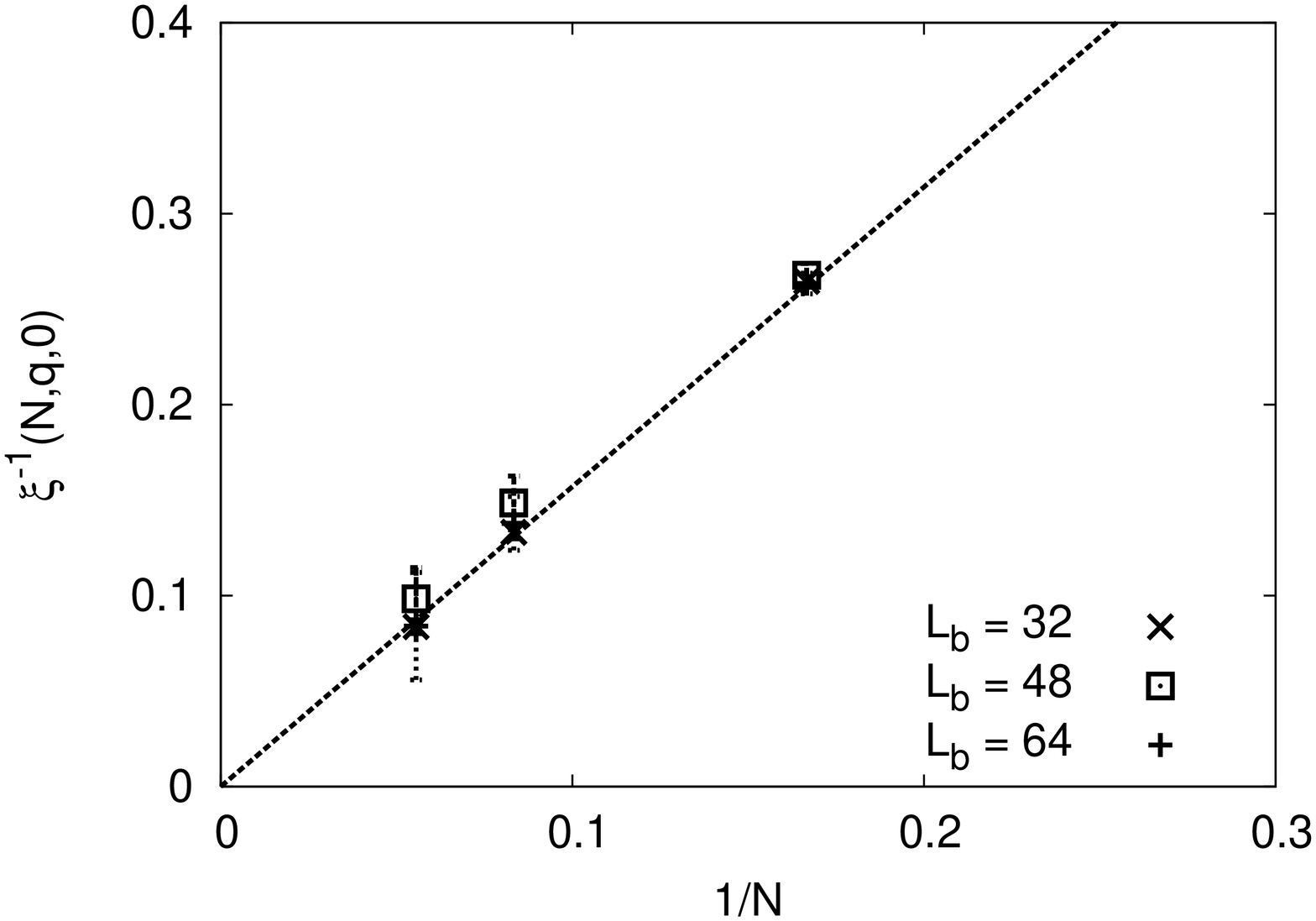,width=5.0cm,angle=270}
\caption{Inverse correlation length plotted vs.~$1/N$, in the
limit $q_{AB}\rightarrow 0$, for poor solvent conditions
$(q=1.5)$, and three choices of the backbone length $L_b$,
to exclude that the data are strongly affected by finite
size artifacts. The straight line indicates that the data
are compatible with a simple linear relation,
$\xi^{-1}(N,0) \propto 1/N$.}
\label{fig31}
\end{center}
\end{figure}

A very interesting question concerns the dependence of $\xi(N,T)$
on chain length. Figure~\ref{fig31} presents a plot of
$\xi^{-1}(N,q,0)$, as estimated from extrapolation of the data
shown in Figure~\ref{fig30} to $q_{AB}=0$, as a function of $1/N$.
Indeed the data are compatible with a relation $\xi^{-1}(N,q,0)
\propto 1/N$, implying long range order for $N \rightarrow \infty
$ in the ground state. In principle, carrying out the same
extrapolation at nonzero temperature and testing for which
temperature range $\xi^{-1}(N \rightarrow \infty ,T)$ starts to be
nonzero, one could obtain an estimate for the critical point of
intramolecular phase separation, that should occur (and then is
well defined) in the limit of infinite side chain length, $N
\rightarrow \infty$. Unfortunately the accuracy of our estimates
for $\xi (N,T)$ does not warrant such an analysis.

The reason for this surprising persistence of disorder at low
temperatures is configurational entropy, of course. As long as $q
<\infty$, the bottle brush polymer is not fully compact, and when
the density is small enough inside the bottle brush (as it still
is in the cases shown in Figure~\ref{fig28}), the two types of
chains A, B can avoid making binary contacts (as is evident from
Figures~\ref{fig18} and \ref{fig27}): 
$\langle m_{AB} \rangle \rightarrow 0$ as $q_{AB} \rightarrow 0$.
Hence there is a perfect avoidance of energetically unfavorable
contacts in the ground state, and the finite correlation length
$\xi(N,0)$ and nonzero entropy of the ground state
are not a consequence of ``frustrated interactions''$^{\textrm {\cite{107}}}$ as
in spin glasses, random field spin models, etc.$^{\textrm { \cite{107,108}}}$. 
In the present problem, if both $q$ and $N$ are not too large, a
ground state (for $q \rightarrow 0$) occurs where the structure of
the bottle brush is NOT a cylinder with an interface separating an
A-rich and a B-rich domain, as hypothesized in Figure~\ref{fig14}!
Thus a (coarse-grained) cross section of the bottle brush in the
xy-plane is NOT a circle, separating A-rich and B-rich regions by
a straight line, but rather looks like the number 8, i.e. a
dumbbell-like shape, where a (more or less circular and more or
less compact A-rich region) occurs for $x <0$, a similar B-rich
region occurs for $x >0$, but no monomers (apart from the backbone
monomer) occur at the y-axis at $x=0$. The orientation of the
x-axis can fluctuate as one moves along the z-axis, and unlike
Figure~\ref{fig15} (twisting an interface in a cylinder along the
z-axis clearly involves an energy cost, and this is described by
the helicity modulus in the XY-model analogy, Equation~(\ref{eq45})) the
free energy cost of this structural distortion is outweighed by
the configurational entropy gain, at least for $q$ and $N$ not too
large. It is an unresolved question whether some critical value
$q_c(N)$ exists, where $\xi^{-1}(N,q,0)$ vanishes, and a long
range ordered ground state occurs. Another unresolved question is,
whether (another?) critical value $q'_c(N)$ exists, where the
character of the ground state changes such that the local cross
section of the binary bottle brush changes from an 8-shaped to a
circular density distribution. Actually, the problem of
alternatingly grafting A-chains and B-chains along a line is
equivalent to grafting symmetric AB diblock copolymers along a
line, such that the junction points of the diblocks form a
straight line. This problem is the lower-dimensional analog of
diblocks grafted with their junction points to the flat interface
between an unmixed binary (A,B) homopolymer blend: for this
problem it is well-known (see Werner et al.$^{\textrm { \cite{109}}}$ for
references) that the shape of the diblocks is dumbbell like, the
A-block being stretched away from the interface, in order to be
embedded in the A-rich phase underneath of the interface, and the
B-rich block also being stretched away from the interface, in
order to be embedded in the B-rich phase above the interface.

\begin{figure*}
\begin{center}
$\begin{array}{c@{\hspace{0.4in}}c@{\hspace{0.4in}}c@{\hspace{0.4in}}c}
\multicolumn{2}{l}{\mbox{(a)}} &
        \multicolumn{2}{l}{\mbox{(b)}} \\ [0.05cm]\\
&\epsfig{file=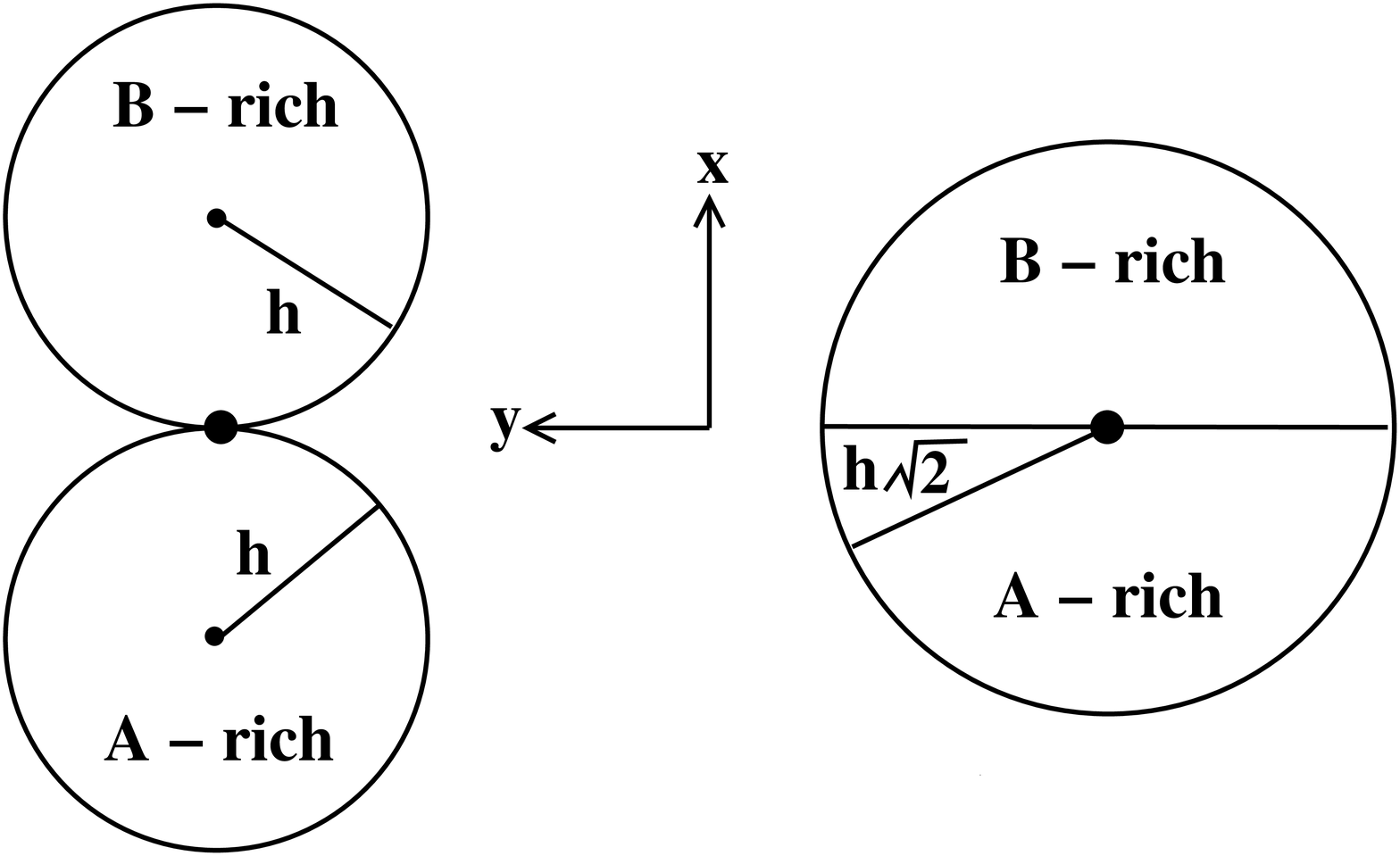, width=5.0cm, angle=0} &
& \epsfig{file=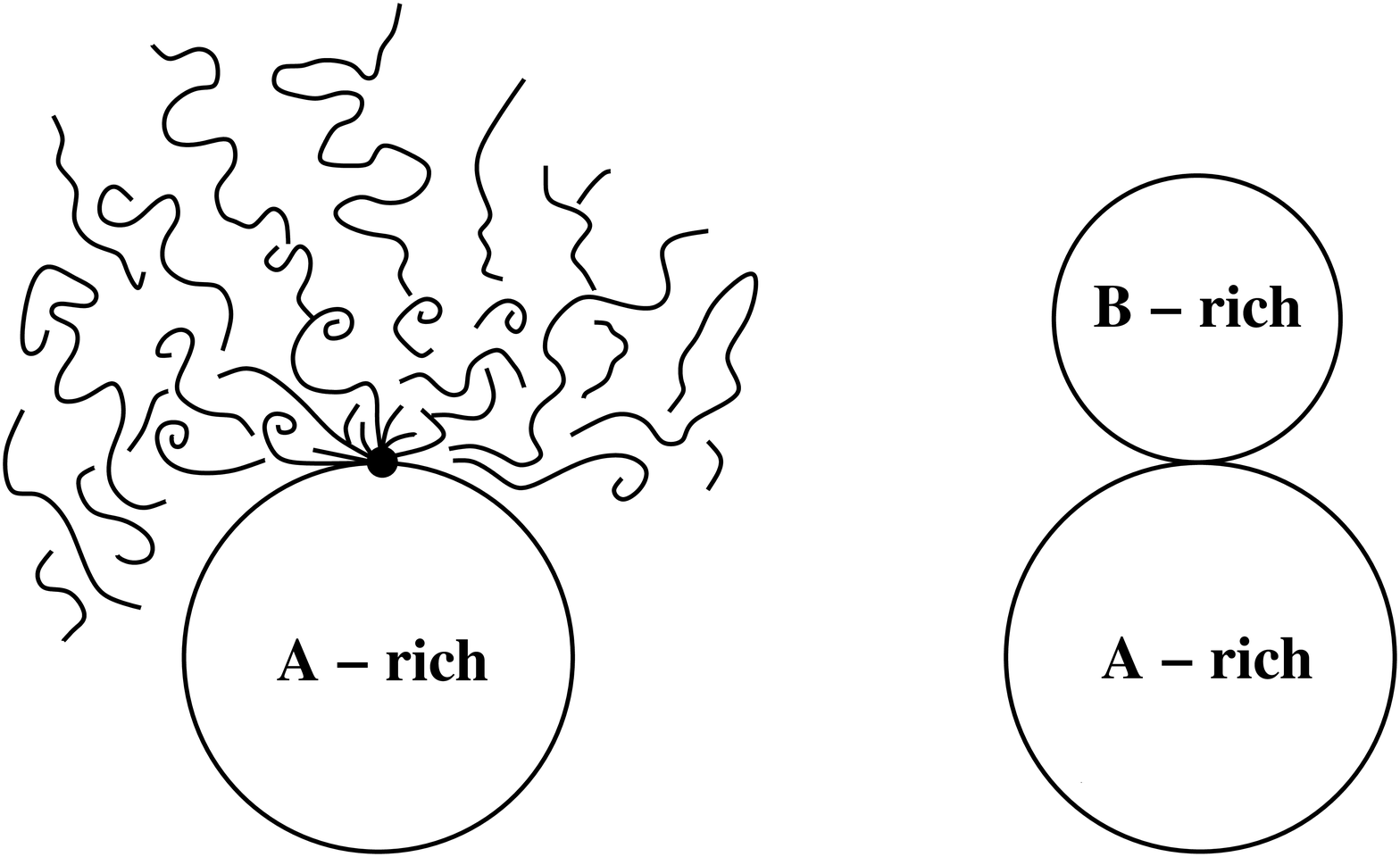, width=5.0cm, angle=0} \\
[1.0cm]
\end{array}$
\caption{Possible ground structures of binary bottle brush polymers,
showing the cross section perpendicular to the backbone (dot) in the
continuum limit, for the symmetric case $\epsilon_{AA}=\epsilon_{BB}$.
Left part shows a double cylinder, each cylinder has a radius $h$.
Right part shows a Janus cylinder (with the same volume this requires a
radius $h'=\sqrt{2}h$.
The surface energy cost of the double cylinder is
$4h\pi L_b \mid \epsilon_{AA} \mid$, while the surface energy cost of the
Janus cylinder is $2\sqrt{2}hL_b(\pi \mid \epsilon_{AA} \mid + \mid \epsilon_{AB}\mid)$.
(b) Possible generalizations of the double cylinder structure to the asymmetric case
$\epsilon_{AA} \neq \epsilon_{BB}$. The left part assumes
$\epsilon_{BB}=0$, so the B chains
are in good solvent, only the A-chains collapse. The right part assumes
$\mid \epsilon_{BB} \mid > \mid \epsilon_{AA} \mid$, so that
the collapsed B-rich cylinder is denser than the A-rich cylinder.}
\label{fig32}
\end{center}
\end{figure*}

As a consequence of our discussion, we call into question the idea
of the ``Janus cylinder''-type phase 
separation$^{\textrm { \cite{56,57,58}}}$
and propose as an alternative possibility (Figure~\ref{fig32}) the
``double cylinder'' (with cross section resembling the number 8).
Which of these cross-sectional structures occur will depend on the
interaction parameters $\epsilon_{AA}= \epsilon_{BB}$, and
$\epsilon_{AB}$, of course. If the strength of the attractive
interactions $|\epsilon_{AA}|=|\epsilon_{BB}|$ exceeds the
strength of the repulsion $|\epsilon_{AB}|$ sufficiently much, it
is clear that for $T\rightarrow 0$ the Janus cylinder type phase
separation will occur, while in the opposite limit, when
$|\epsilon_{AB}|$ exceeds $|\epsilon_{AA}|=|\epsilon_{BB}|$
sufficiently much, the double cylinder geometry will win. In the
macroscopic continuum limit a comparison of the respective surface
energies would imply that the double cylinder geometry, which
avoids an AB interface, but requires for the same volume taken by
A and B monomers a surface area of pure A and B that is larger by
a factor of $\sqrt{2}$ than for the ``Janus cylinder''
(Figure~\ref{fig32}), becomes energetically preferable for
\begin{equation}\label{eq53}
\epsilon_{AB} > (\sqrt{2}-1)\pi |\epsilon_{AA}|=1.301 |
\epsilon_{AA}|
\end{equation}
Since in the present work we consider the limit $q_{AB}=
\exp(-\epsilon_{AB}/k_BT) \rightarrow 0$ at fixed $q=\exp
(-\epsilon _{AA}/k_BT)$, it is clear that in our case
Equation~(\ref{eq53}) is fulfilled. However, for the 
''most symmetric''$^{\textrm { \cite{52}}}$ choice of interaction 
parameters, $\epsilon_{AB}=
-\epsilon_{AA} = - \epsilon_{BB}$, the conclusion would be
different. For lattice models Equation~(\ref{eq53}) is
questionable since very compact configurations of collapsed
polymers must respect the lattice structure, and different
geometrical factors, depending on the type of the lattice, in the
inequality Equation~(\ref{eq53}) may occur.

\vskip 1.0truecm
\noindent
{\large \bf VII. Conclusions and Outlook}
\vskip 0.5truecm

In this article, we have restricted attention exclusively to
static conformational properties of very long bottle brush
polymers with a rigid backbone. We feel that this ``simple''
limiting case needs to be understood first, before the very
interesting extension to the case of flexible or semiflexible
backbones,$^{\textrm{ \cite{28,29,30,31,32,33,34,35,36,37,38,39,40}}}$ 
or the question of the crossover between a bottle brush polymer and a
star polymer,$^{\textrm { \cite{110}}}$ can be correctly addressed. Thus the
latter two problems were not at all considered in this paper, and
hence we also do not wish to comment on the recent 
controversy$^{\textrm { \cite{14,15,16}}}$ concerning the correct 
interpretation of
experiments on the overall linear dimensions of bottle brush
polymers. Thus, the focus of the present article, as far as
one-component bottle brush polymers are concerned, is the
conformation of the side chains. We recall that this information
is experimentally accessible, if one prepares bottle brushes with
a single arm being deuterated, while all remaining arms of the
polymer remain protonated, to allow a study of the static
structure factor $S(q)$ of a single arm by elastic coherent
neutron scattering. It is clear that such experiments are very
difficult, and hence no such experiment of this type is known to
the present authors yet, but clearly it would be highly desirable
to obtain such experimental information.

As discussed in the first part of the paper, one can find in the
literature rather diverse concepts about the conformations of the
side chains of a bottle brush polymer under good solvent
conditions and high grafting density. One concept assumes that the
cylindrical volume that the bottle brush occupies can be
partitioned into disks, such that each disk contains just one
polymer chain confined into it, no other chain participating in
the same disk. A simple geometric consideration, reviewed in the
first part of the present paper, then yields the following
predictions for the linear dimensions of the chain, as a function
of grafting density $\sigma$ and side chain length $N$
\begin{equation}\label{eq54}
R_{gx}\propto \sigma ^{1/4}N^{3/4}, \; R_{gy}\propto \sigma ^{1/4}
N^{3/4},\quad R_{gz} \propto \sigma ^{-1}\;.
\end{equation}
Remember that we choose the z-axis as the direction of the rigid
backbone, the x-axis is oriented normal to the z-axis towards the
center of mass of the chain, and the y-axis is perpendicular to
both x- and z-axes. Thus the chain conformation has a quasi-two
dimensional character, and there is no stronger stretching of a
chain in radial direction than in tangential direction, since
$R_{gy}/R_{gx}= \textrm{const}$ (independent of both $N$ and
$\sigma$).

Clearly, this quasi-two-dimensional picture is not very plausible,
and more popular is an extension of the 
Daoud-Cotton$^{\textrm{ \cite{8}}}$ blob
picture for star polymers to the present case. While for a star
polymer the blob radius simply increases proportional to the
distance from the center, $r$, and there is no geometrical
difficulty to densely pack the conical compartments resulting from
dividing a sphere into $f$ equal sectors, such that each sector
contains a single arm of the star, with a sequence of spheres of
increasing size, it has been argued in the literature that for
cylindrical geometry the blob radius scales as $\xi(r) \propto
(r/\sigma)^{1/2}$. While in the literature no mentioning of a
non-spherical blob shape (or, equivalently anisotropic local
screening of the excluded volume interaction) is found, we have
emphasized here the geometrically obvious fact that a dense
filling of space with blobs in the cylindrical geometry
appropriate for a bottle brush requires that the blobs have the
shape of ellipsoids with three different axes, proportional to
$r$, $(r/\sigma)^{1/2}$, and $\sigma ^{-1}$, respectively. While
this picture does not alter the prediction for the stretching of
the chain in the radial x-direction found in the literature,
$R_{gx}\propto \sigma^{(1-\nu)/(1+\nu)}N^{2\nu/(1+\nu)}\approx
\sigma ^{1/4}N^{3/4}$, different predictions result for the other
two linear dimensions. While the regular Daoud-Cotton picture
would yield for $R_{gy}$ the size of the last blob, $R_{gy}=
\xi(r=h)=(h/\sigma)^{1/2} \propto \sigma ^{-\nu / (1+\nu)} N^{\nu
/(1+\nu )}\approx \sigma ^{-3/8}N^{3/8}$, we find instead
\begin{equation}\label{eq55}
R_{gy}\propto \sigma ^{-(2\nu -1)/(1+\nu)}N^{3\nu /(2\nu
+2)}\approx \sigma ^{-1/16}N^{9/16}\;,
\end{equation}
\begin{equation}\label{eq56}
R_{gz}\propto \sigma ^{-(2\nu -1)/(1+\nu)}N^{\nu /(2\nu
+2)}\approx \sigma ^{5/8}N^{3/8}\;.
\end{equation}
Equations~(\ref{eq55}) and (\ref{eq56}) make use of both the anisotropic blob
linear dimensions and random walk-type arguments (cf.
Equations~(\ref{eq13})-(\ref{eq28})) and are new results, to the best of
our knowledge.

Also the crossover scaling towards mushroom behavior has been
considered, Equations~(\ref{eq20})-(\ref{eq23}), and Monte Carlo evidence
for this crossover scaling description was obtained, from
extensive work using the PERM algorithm. Although rather large
bottle brushes were simulated (backbone length $L_b$ up to
$L_b=128$, avoiding free end effects by periodic boundary
conditions, side chain length up to $N=2000$ for grafting density
$\sigma = 1$, see Figure~\ref{fig2}), it was not possible to reach
the asymptotic regime of strong side chain stretching where
Equations~(\ref{eq55}) and (\ref{eq56}) hold, however. One reason why for
moderate grafting densities even fairly long side chains can avoid
each other with small significant amount of stretching is the fact
that the natural shape of a self avoiding walk configuration is an
elongated ellipsoid, with three rather different eigenvalues of
the gyration tensor. So side chains can to a large extent avoid each
other$^{\textrm {\cite{36}}}$ by orienting themselves such that the longest axis
of the ellipsoid is oriented in the radial x-direction and the
smallest axis in the backbone z-direction. This implies that very
large values of $\sigma a N^\nu$ are needed to obtain significant
stretching. This difficulty to verify any such scaling laws in our
simulations, which were only able to explore the onset of
stretching away from simple mushroom-type behavior of the side
chains and not the strongly stretched behavior, suggests that no
such scaling behavior should be observable in the experiments as
well: given the empirical fact that one bond of a coarse-grained
lattice model corresponds to $n=3-5$ chemical 
monomers,$^{\textrm{\cite{84,85,86}}}$ it is clear that the 
experimentally accessible
side chain lengths do not exceed those available in our
simulation. Thus, scaling theories of bottle brushes are
unfortunately of very restricted usefulness for the interpretation
of either experiments or simulations. The only firm conclusion
about the theories mentioned above that we like to make is that
there is no evidence whatsoever for the quasi-two-dimensional
picture, since we do see a decrease of $R_{gy}/R_{gx}$ with
increasing $N$.

Turning to the problem of intramolecular phase separation in
binary (AB) bottle brush polymers, we have examined the proposal
that a ``Janus cylinder''-type phase separation 
occurs.$^{\textrm {\cite{56,57,58}}}$ This idea is questionable 
for several reasons (i)
In a quasi-one-dimensional system, no sharp phase transition to a
state with true long range order can occur; at most one can see a
smooth increase in the corresponding correlation length, as the
temperature is lowered (or the incompatibility between A and B is
enhanced, respectively). Thus we have defined suitable correlation
functions and studied the variation of the corresponding
correlation length as function of the chain length $N$ and the
parameter $q_{AB}=\exp[-\epsilon _{AB}/k_BT]$ controlling the
incompatibility ($\epsilon_{AB}$ is the repulsive energy
encountered when two neighboring lattice sites are occupied by
monomers of different kind). Three choices of solvent quality
(taken symmetric for both A and B through the choice
$\epsilon_{AA}=\epsilon_{BB}$) were considered,
$q=\exp(-\epsilon_{AA}/k_BT)=1$ (good solvent), $q=1.3087$ (Theta
solvent) and $q=1.5$ (poor solvent). It was found that in all
cases the correlation length suited to detect Janus-cylinder type
ordering increases rather weakly as $q_{AB}\rightarrow 0$,
approaching finite values even for $q_{AB}=0$, for finite side
chain length $N$, while for $N \rightarrow \infty$ an infinite
correlation length is compatible with the data. No evidence for
the predicted critical points$^{\textrm {\cite{56}}}$ and their scaling behavior
with $N$ (Equations~\ref{eq32}-\ref{eq38}) could be detected, however.
The gradual establishment of a local phase separation lacking a
sharp transition is compatible with observations from a previous
simulation,$^{\textrm{ \cite{57}}}$ however. (ii) Depending on the relation
between the energy parameters $\epsilon_{AA}$ and $\epsilon_{AB}$
(cf. Equation~(\ref{eq53})), the ``local'' phase separation (considering
a slice of suitable thickness perpendicular to the backbone of the
bottle brush to obtain suitable coarse-grained densities
$\rho_A,\rho_B$ on mesoscopic scales) for temperatures $T
\rightarrow 0$ may have two different characters
(Figure~\ref{fig32}a): the ``Janus cylinder'', which contains an
interface between A-rich and B-rich phases, competes with the
``double cylinder''. In the latter structure, there is no extended
AB interface, A-chains and B-chains meet only in the immediate
vicinity of the backbone where they are grafted. Thus, the
coarse-grained density distribution under poor solvent conditions
in a slice has the shape of the number 8, where in the upper part
of the 8 we have the B-rich phase and in the lower part we have the
A-rich phase. Of course, when we orient the x-axis
(Figure~\ref{fig14}) such that it is parallel to the vector
connecting the center of mass of the A-rich region to the center
of mass of the B-rich region in the slice, the orientation of the
x-axis for both structures in Figure~\ref{fig32} can randomly rotate
when we move along the backbone (z-direction) due to long
wavelength fluctuations, cf. Figure~\ref{fig15}, and hence the
comments about finite correlation lengths of this phase separation
apply here as well. Actually, the numerical results of the present
Monte Carlo simulations give clear evidence that local phase
separation of ``double cylinder''-type rather than ``Janus
cylinder'' type is observed, since the number of AB-contacts tends
to zero.

\begin{figure}
\begin{center}
$\begin{array}{c@{\hspace{0.2in}}c}
\multicolumn{2}{l}{\mbox{(a)}} \\ [-1.5cm] \\
&\psfig{file=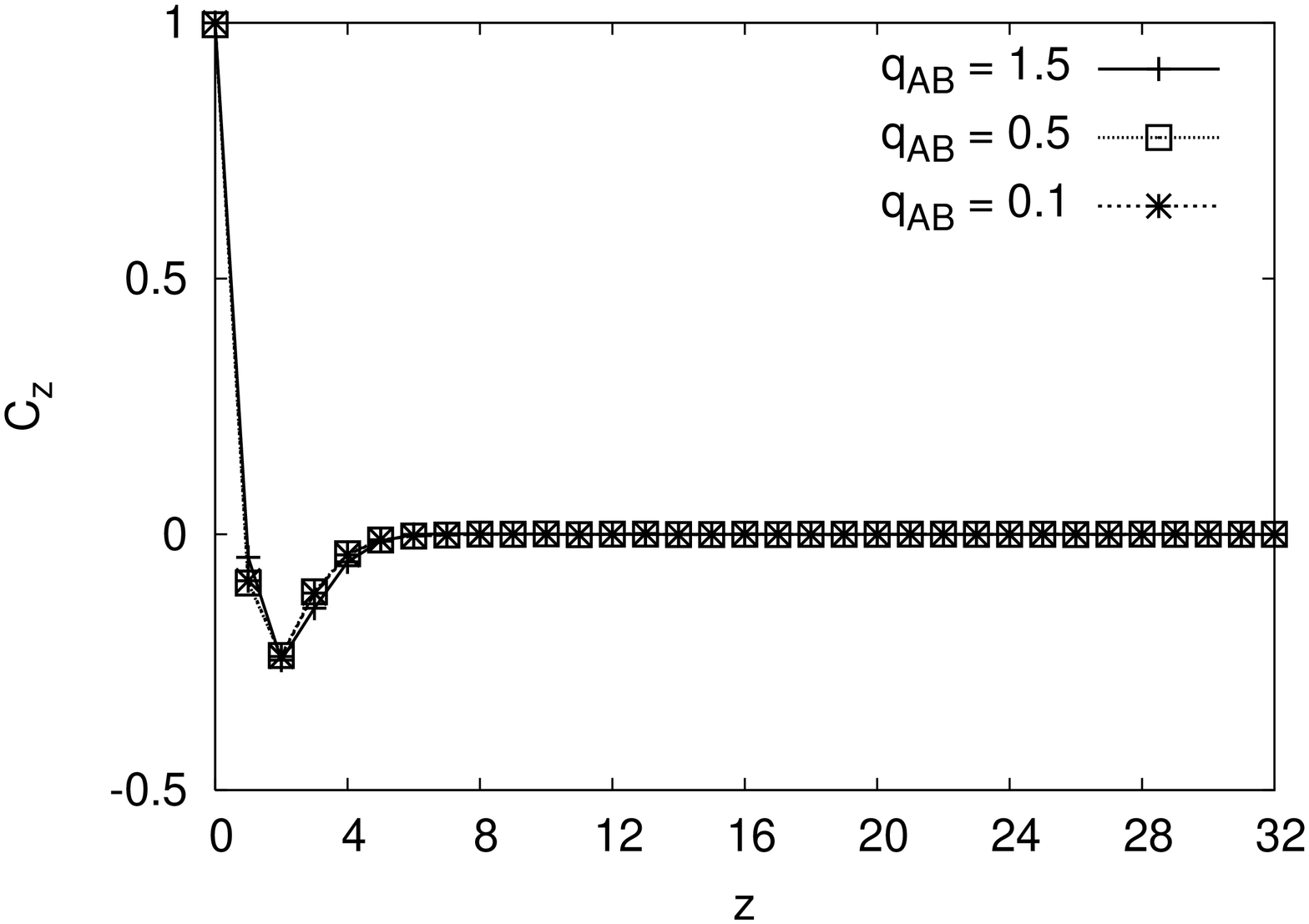,width=5.0cm,angle=270} \\
[0.05cm]\\
\multicolumn{2}{l}{\mbox{(b)}} \\ [-1.5cm] \\
&\psfig{file=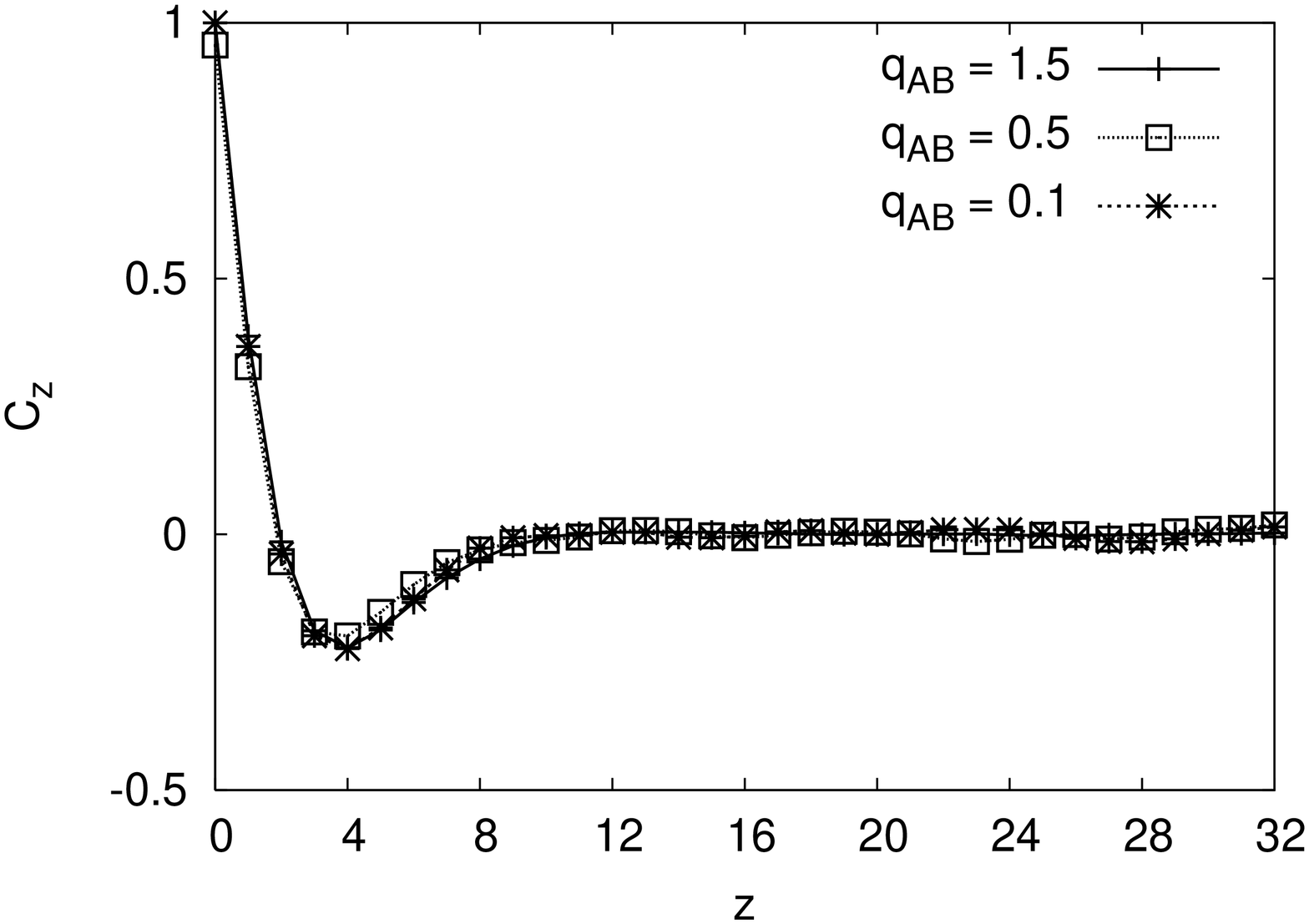, width=5.0cm, angle=270} \\
[1.0cm]
\end{array}$
\caption{Correlation function $C_z$ (Equation~(\ref{eq57}))
plotted vs.~$z$, for $L_b=64$, $q=1.5$, $\sigma=1$, $f=1$,
$N=6$ (a) and $N=18$ (b), and various choices of $q_{AB}$, as indicated.}
\label{fig33}
\end{center}
\end{figure}

Thus, varying the solvent quality and chemical incompatibility,
one can influence the character of local intramolecular phase
separation, and one can control the correlation length over which
the vector characterizing either interface orientation or the axis
of the local dumbbell is oriented in the same way along the
z-axis. Of course, in real systems one must expect that the
solvent quality for the two types of chains will differ
(Figure~\ref{fig32}b). Then asymmetric 8-shaped local structures
will result: e.g., if the solvent is a good solvent for B but a
poor one for A, but the incompatibility between A and B is very
high, we expect a structure where the A-chains are collapsed in a
cylinder with the backbone on the cylinder surface, and from there
the B-chains extend into the solution like a ``flower'' in the
cross section (Figure~\ref{fig32}b, left part). Conversely, it may
happen that the solvent quality is poor for both A and B, but
nevertheless different, so the densities of both cylinders and
hence their radii will differ (Figure~\ref{fig32}b, right part).
Similar asymmetries are also of interest if the local character of
the phase separation is of the Janus cylinder type: then in the
cross section the A-rich and B-rich regions will take unequal
areas, rather than equal areas as shown in Figure~\ref{fig32}a, and
the interface will be bent rather than straight. Similar effects
will occur when the chain lengths $N_A,N_B$ differ from each
other, or the flexibilities of both types of chains are different,
etc.

An aspect which has not been discussed here but which is important
for the binary bottle brush under poor solvent conditions is the
question whether or not the density in the collapsed bottle brush
is homogeneous along the z-direction (so one really can speak
about ``Janus cylinders'' or ``double cylinders'',
Figure~\ref{fig32}), or whether it is inhomogeneous so the state of
the system rather is a chain of ``pinned clusters''. 
For the one-component bottle brush, this question was discussed by Sheiko
et al.,$^{\textrm {\cite{60}}}$ from a scaling point of view, but (unlike the
related problem of brush-cluster transition of planar brushes in
good solvents$^{\textrm {\cite{115,116,117,118}}}$) we are not aware of any
simulation studies of this problem yet.
In order to understand the structures of the binary bottle brush
in the poor solvent, we count the number of monomers $M(z)$ 
(irrespective of A or B) in each xy-plane, and calculate the 
normalized correlation function $C_z$ of $\delta M(z)=M(z)-N$
along the backbone,  
\begin{equation}
  C_z =<\delta M(i+z) \delta M(i)> / <\delta M(i)^2> \; . \label{eq57}
\end{equation}
Here the average $<\ldots>$ includes an averaging over
all sites $\{i\}$ on which chains are grafted, in order to improve the statistics. 
In Figure~\ref{fig33}, we see that this correlation actually is 
negative for $2 \leq z \leq 8$, indicating some tendency to
pinned cluster formation.  

Before one can try to understand the real two-component bottle
brush polymers studied in the laboratory,$^{\textrm{ \cite{111,112}}}$ two more
complications need to be considered as well: (i) the flexibility
of the backbone; (ii) random rather than regular grafting along
the sequence. While the effects due to (i) were already addressed
to some extent in an earlier simulation,$^{\textrm{ \cite{57}}}$ randomness of
the grafting sequence has not been explored at all. However, in
the study of binary brushes on flat substrates$^{\textrm{\cite{113}}}$ it has
been found that randomness in the grafting sites destroys the long
range order of the micro-phase separated structure, that is
predicted to occur$^{\textrm{\cite{114}}}$ for a perfectly periodic arrangement
of grafting sites. In the one-dimensional case, we expect the
effects of quenched disorder in the grafting sites to be even more
dominant than in these two-dimensional mixed polymer brushes.
Hence, it is clear that the explanation of the structure of one-
and two-component bottle brush polymers still is far from being
complete, and it is hoped that the present article will motivate
further research on this topic, from the point of view of
theory, simulation and experiment.

\underline{Acknowledgement} This work was financially supported by
the Deutsche Forschungsgemeinschaft (DFG), SFB 625/A3. K. B.
thanks S. Rathgeber and M. Schmidt for stimulating discussions and
for early information about recent papers (Refs.~\cite{14,15,16}).
H.-P. H. thanks Prof. Peter Grassberger and Dr. Walter Nadler for
very useful discussions. We are grateful to the NIC J\"ulich
for providing access to the JUMP parallel processor.

\end{document}